\documentclass[twocolumn,floatfix,prl]{revtex4}%
\usepackage[dvipdfmx]{graphicx}%
\usepackage{amsmath,amssymb}%
\usepackage{amsfonts}
\usepackage{amssymb}
\usepackage{mathrsfs}
\usepackage{color}
\usepackage{bm}

\def\s{{\sigma}}
\def\e{{\epsilon}}
\def\k{{ {\bf k} }}
\def\p{{ {\bf p} }}
\def\q{{ {\bf q} }}

\def\B{{ {\bf B} }}

\def\r{{ {\bf r} }}
\def\aa{{ {\bf a} }}
\def\ee{{ {\bf e} }}
\def\bmphi{{ {\bm \phi} }}
\def\bmeta{{ {\bm \eta} }}
\def\bmsig{{ {\bm \sigma} }}
\def\w{{\omega}}
\def\a{{\alpha}}
\def\b{{\beta}}

\def\g{{\gamma}}

\def\A{{ {\rm A} }}
\def\B{{ {\rm B} }}
\def\C{{ {\rm C} }}
\allowdisplaybreaks[4]

\begin{document}
\title{
Charge-loop current order and $Z_3$ nematicity mediated by bond order fluctuations in kagome metals 
}
\author{
Rina Tazai$^1$, Youichi Yamakawa$^2$, and Hiroshi Kontani$^2$
}
\date{\today }

\begin{abstract}
Recent experiments on geometrically frustrated kagome metal $A$V$_3$Sb$_5$ (A$=$K, Rb, Cs)
have revealed the emergence of the charge loop current (cLC) order near the bond order (BO) phase.
However, the origin of the cLC and its interplay with other phases have been uncovered.
Here, we propose a novel mechanism of the cLC state,
by focusing on the BO phase common in kagome metals.
The BO fluctuations in kagome metals,
which emerges due to the Coulomb interaction and the electron-phonon coupling,
mediate the odd-parity particle-hole condensation
that gives rise to the topological current order.
Furthermore, the predicted cLC+BO phase gives rise to 
the $Z_3$-nematic state in addition to the giant anomalous Hall effect.
The present theory predicts the
close relationship between the cLC, the BO, and the nematicity,
which is significant to understand the cascade of quantum electron states
in kagome metals.
%
\end{abstract}

\address{
$^1$Yukawa Institute for Theoretical Physics, Kyoto University, 
Kyoto 606-8502, Japan \\
$^2$ Department of Physics, Nagoya University,
Furo-cho, Nagoya 464-8602, Japan. 
}
\sloppy

\maketitle
\subsection{Introduction}



\begin{figure}[htb]
\includegraphics[width=.99\linewidth]{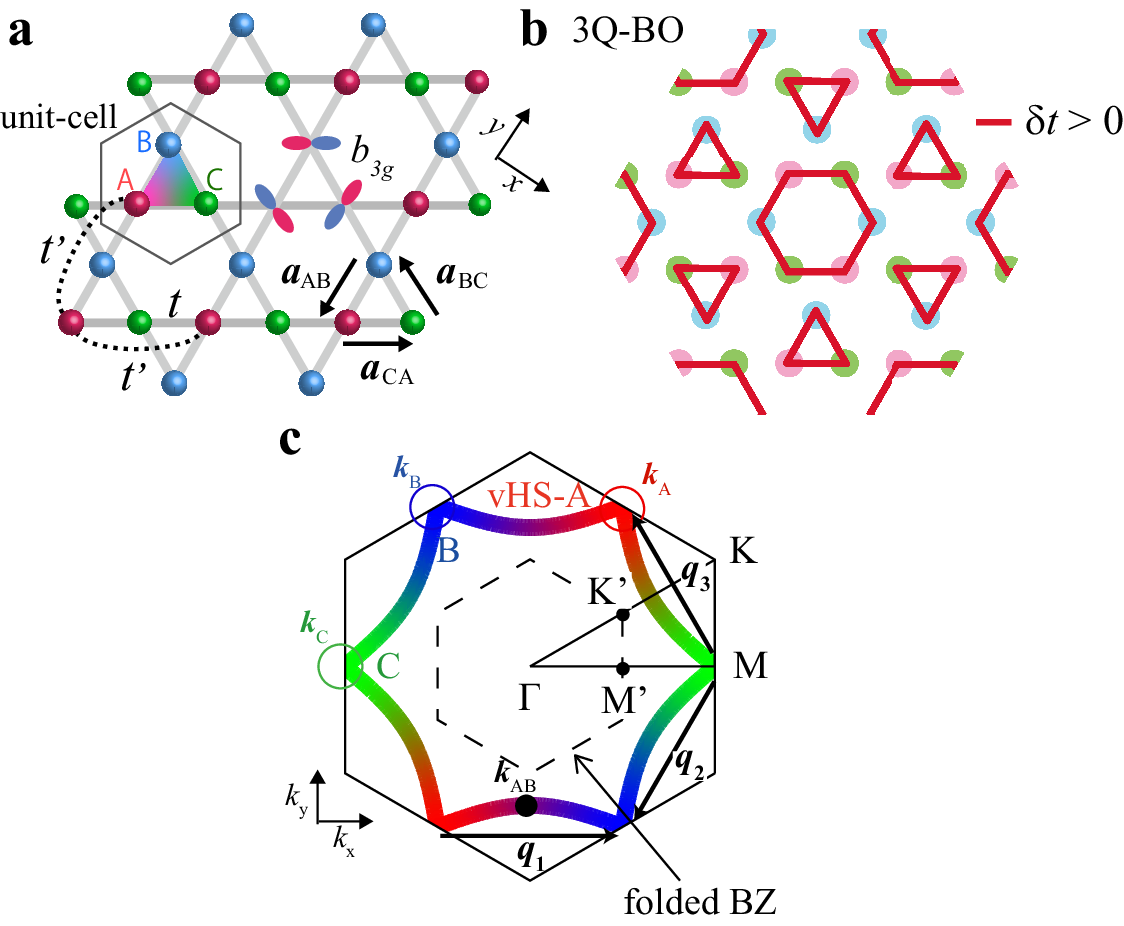}
\caption{
{\bf Lattice structure, Fermi surface, and BO form factor in kagome metal}. \ 
{\bf a}. Kagome-lattice structure 
composed of the sublattices A, B and C.
$2{\aa}_{lm}$ is the minimum translation vector, and we set $|2{\aa}_{lm}|=1$.
The relation 
${\aa}_{\rm AB}+{\aa}_{\rm BC}+{\aa}_{\rm CA}={\bm0}$ holds.
{\bf b}. $3Q$ Tri-Hexagonal bond order (BO) state. 
{\bf c}. Fermi surface (FS) at $n=0.917$ with the nesting vectors $\q_1,\q_2,\q_3$.
The color of the FS represents the weight of the sublattice
(A = red, B = blue, C = green).
The FS has single sublattice character
near the van-Hove singularity (vHS) points.
In kagome metals, $\q_1$ connects vHS-A and vHS-B.
It is given as 
$\q_1=(2{\aa}_{\rm AB})\times(2\pi/\sqrt{3}){\ee}_z$,
where ${\ee}_z$ is the unit vector perpendicular to the $xy$-plane.
}
\label{fig:fig1}
\end{figure}

Recent discovery of the kagome-lattice metal 
$A$V$_3$Sb$_5$ ($A$=K, Rb, Cs) shown in Fig. \ref{fig:fig1} {\bf a} 
has opened the way to study the unique physics of 
geometrically frustrated metals with strong correlation
\cite{kagome-exp1,kagome-exp2,kagome-P-Tc1}.
In CsV$_3$Sb$_5$, the formation of the 2$\times$2 
Star-of-David or Tri-Hexagonal density wave (DW)
was detected by scanning tunneling microscopy (STM) 
at $T\approx 90$ K at ambient pressure
\cite{STM1,STM2}.
It is presumably the triple-$\q$ ($3Q$) bond order (BO)
shown in Fig. \ref{fig:fig1} {\bf b}, 
which is the even-parity modulation in the hopping integral 
$\delta t_{ij}^{\rm b}$ (=real)
\cite{Thomale2013,SMFRG,Thomale2021,Neupert2021,Balents2021,Nandkishore,Tazai-kagome}.
Below the BO transition temperature $T_{\rm BO}$, 
superconductivity (SC) with highly anisotropic gap 
emerges for $A$=Cs
\cite{Roppongi,SC2},
and the gap structure changes to isotropic by introducing impurities.
Also, nodal to nodeless crossover is induced by
the external pressure in $A$=Rb,K \cite{muSR5-Rb}.
These results are naturally understood
based on the BO fluctuation mechanism \cite{Tazai-kagome}.



More recently, the non-trivial time reversal symmetry breaking (TRSB) order at $T_{\rm TRSB}$ 
attracts considerable attention.
It has been reported by $\mu$SR study 
\cite{muSR3-Cs,muSR2-K,muSR4-Cs,muSR5-Rb},
Kerr rotation analysis \cite{birefringence-kagome},
field-tuned chiral transport study \cite{eMChA}
and STM measurements \cite{STM1,eMChA}.
The transition temperature $T_{\rm TRSB}$ 
is close to $T_{\rm BO}$ in many experiments,
while the TRSB order parameter is strongly magnified at 
$T^* \approx 35$K for $A$=Cs
\cite{eMChA,muSR3-Cs,muSR4-Cs} and
$T^* \approx 50$K for $A$=Rb
\cite{muSR5-Rb}.
Recently, magnetic torque measurement reveals the 
TRSB order associated with the rotational symmetry breaking, 
which is called the nematic order,
at $T^*\approx130$K \cite{Asaba}.
In contrast,
TRSB was not reported by different experimental groups
using the Kerr rotation 
\cite{Kapitulnik} and STM \cite{nematic-SC2} measurements.
Thus, the TRSB onset temperature is still under debate.
The chiral cLC is driven by the additional odd-parity hopping integral 
$\delta t_{ij}^{\rm c}$ (=imaginary), and the accompanied 
topological charge-current \cite{Haldane}
gives the giant anomalous Hall effect (AHE) below $T \approx 35$ K
\cite{AHE1,AHE2}.
The correlation-driven topological phase in kagome metals is very unique,
while its mechanism is still unknown.

%

In addition to the cascade of quantum phase transitions,
the emergent nematic order inside the BO and the cLC phases
attracts great attention.
The nematic transition is clearly observed by the elastoresistance
\cite{elastoresistance-kagome},
the scanning birefringence \cite{birefringence-kagome},
and the STM \cite{STM2} studies.
In addition, nematic SC states have been reported
\cite{nematic-SC1,nematic-SC2}.
Thus, kagome metals provide a promising platform for exploring the interplay 
between electron correlations and topological nature.

To understand the rich quantum phases in kagome metals,
lots of theoretical studies have been performed
\cite{Thomale2013,SMFRG,Thomale2021,Neupert2021,Balents2021,Nandkishore,Tazai-kagome,Zhou-cLC-RPA,Fernandes-coexistence,Kennes-coexistence}.
Each BO and cLC order is explained 
by introducing various off-site interactions
in the mean-field approximation (MFA)
\cite{Varma,Nersesyan,Neupert2021,Zhou-cLC-RPA},
while a fine-tuning of off-site interactions is necessary
to explain the cascade of phase transitions.
On the other hand, beyond-MFA mechanisms have been developed
to explain the rich phase transitions 
\cite{Fradkin-rev2012,Davis-rev2013,Onari-SCVC,Tsuchiizu1,Tsuchiizu4,Yamakawa-Cu,Yamakawa-FeSe,Onari-FeSe,Chubukov-PRX2016,Fernandes-rev2018,Onari-TBG,Kontani-AdvPhys}.
For example, strong interplay between the magnetism, 
nematicity and SC in Fe-based superconductors
and other strongly correlated metals
were understood by beyond-MFA mechanisms
\cite{Onari-SCVC,Yamakawa-FeSe,Onari-FeSe,Chubukov-PRX2016,Fernandes-rev2018,Kontani-AdvPhys,Tsuchiizu1,Tsuchiizu4,Yamakawa-Cu,Onari-TBG}.
It is urgent and important to elucidate why the 
BO and cLC orders/fluctuations coexist 
in the study of kagome metals.
For example, these fluctuations will mediate 
non-BCS SC \cite{Tazai-kagome}
and exotic pair-density-wave states
\cite{PDW-theory,Raghu-PDW,Wu-6e}.



In this paper, we reveal that the cLC order is
mediated by the BO fluctuations that are abundant above $T_{\rm BO}$
in kagome metals \cite{X-ray160K,He-BOfluc}.
The sizable off-site Umklapp scattering by the BO fluctuations
induces the odd-parity and TRSB current order.
(=imaginary $\delta t_{ij}^{\rm c}$).
This cLC mechanism is universal because it is 
irrelevant to the origin of the BO.
Furthermore, we discover that the coexistence of 
the BO and the cLC order gives rise to the novel
$Z_3$ nematicity along the three lattice directions
reported in Refs.
\cite{elastoresistance-kagome,birefringence-kagome,STM2}.
The present theory reveals the close relationship 
between the cLC, BO, nematicity and SC state,
which is significant to understand the unsolved
quantum phase transitions in kagome metals.

The phase transitions in metals are
described as the symmetry breaking of the normal self-energy;
$\Delta\Sigma\equiv \Sigma-\Sigma_{A_{1g}}$
\cite{Tazai-Matsubara,Kontani-AdvPhys}.
$\Delta\Sigma$ is determined by the stationary condition of 
the free energy; $\delta F[\Delta\Sigma]/\delta(\Delta\Sigma)=0$.
The DW equation enables us to derive the solution 
that satisfies the stationary condition, 
as we proved based on the Luttinger-Ward theory \cite{Tazai-Matsubara}.
Based on the DW equation, we discover that the odd-parity 
and TRSB $\Delta\Sigma$ 
is driven by the BO fluctuation exchange processes. 
(Note that the DW equation for $\Delta\Sigma$
is analogous to the Eliashberg equation for the SC gap $\Delta$.) 


\subsection{Results}
\vspace{5mm}
{\bf BO form factor and fluctuations}. \

Here, we introduce the kagome-lattice tight-binding model 
with a single $d$-orbital of each vanadium site (A, B, or C) shown in Fig. \ref{fig:fig1} {\bf a}.
(The $d$-orbital belongs to $b_{3g}$ of the $D_{2h}$ point group at V site,
while its representation is not essential here.)
The kinetic term is given by
$\hat{H}_0 =\sum_{\k,l,m,\sigma} h_{lm}^0(\k) c^{\dagger}_{\k,l,\sigma} c_{\k,m,\sigma}$,
%
%
where $l,m$ denote the sublattices A,B,C, and
$h_{lm}^0 (\k)\ (=h_{ml}^{0} (\k)^*)$ is the Fourier 
transform of the nearest-neighbor hopping integral $t$
in Ref. \cite{Frantz}
in addition to the inter-sublattice hopping $t'$
shown in Fig. \ref{fig:fig1} {\bf a}.
We set $t \ (=-0.5{\rm eV})$ to fit the bandwidth, and
$t' (=-0.08{\rm eV})$ to reproduce the shape of the Fermi surface (FS).
Numerical results are insensitive to the presence of $t'$.
Hereafter, the unit of energy is eV unless otherwise noted.
The FS around the van-Hove singularity (vHS) point
($\k\approx \k_{\A}$, $\k_{\B}$, or $\k_{\C}$)
is composed of a single $3d$-orbital on V ion,
which is called the sublattice interference \cite{Thomale2013}.
This simple three-site model 
well captures the main pure-type FS in kagome metals
\cite{STM1,ARPES-VHS,ARPES-band,ARPES-CDWgap,ARPES-Lifshitz,ARPES-CDWgap2}.
The FS at the vHS filling 
($n_{\rm vHS}=0.917$ per site and both spins) 
is shown in Fig. \ref{fig:fig1} {\bf c}.
The wavevectors of the BO correspond to the inter-sublattice 
nesting vectors $\q_n$ ($n=1,2,3$) in Fig. \ref{fig:fig1} {\bf c}.
(The equivalent square lattice kagome model is
convenient for the numerical study; see Supplementary Note 1.)
The good inter-sublattice nesting of the FS naturally
triggers the observed inter-sublattice BO at $\q=\q_n$,
as shown in previous theoretical studies
\cite{Thomale2013,SMFRG,Tazai-kagome}.

\begin{figure}[htb]
\includegraphics[width=.99\linewidth]{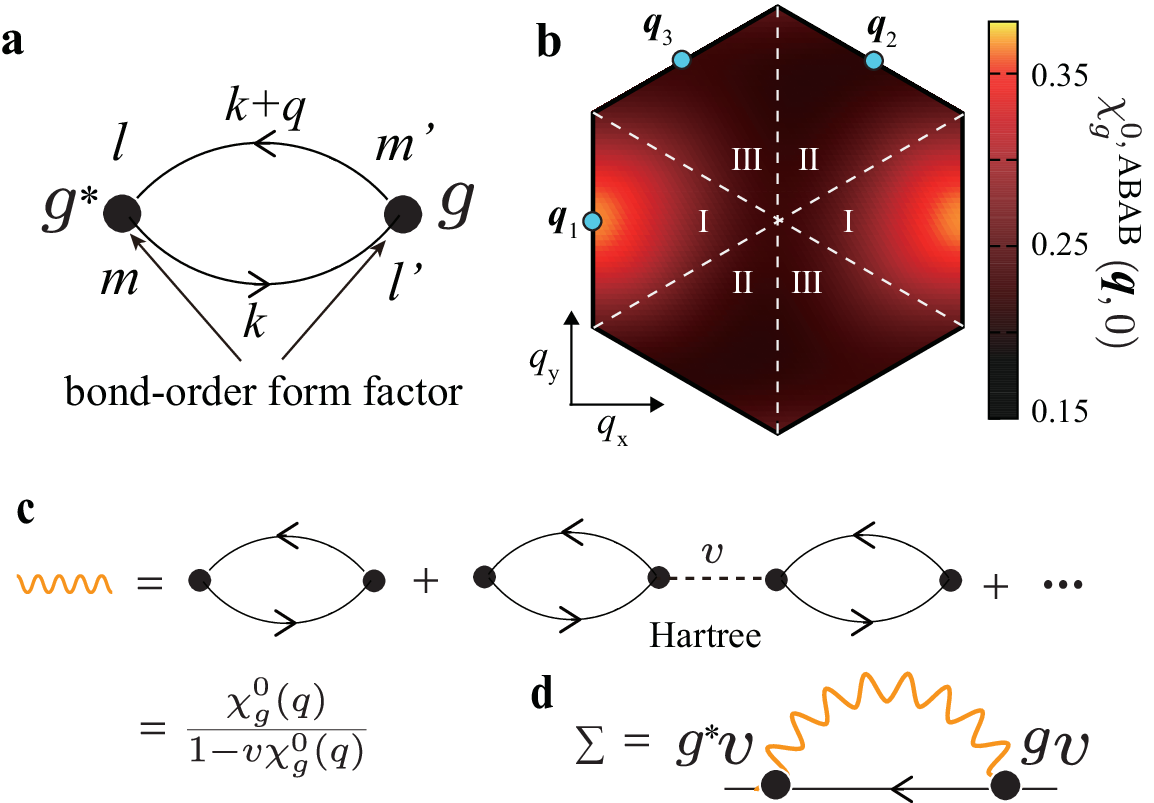}
\caption{
\color{black}
{\bf BO fluctuations and self-energy}. \
{\bf a}. Expressions of the bond-order (BO) irreducible susceptibility ${\chi}^{0,lmm'l'}_{g} (q)$.
{\bf b}. Obtained $\q$-dependence for ${\chi}^{0,\A\B\A\B}_{g} (q)$,
which takes the maximum at $\q=\q_1$.
{\bf c}. ${\hat \chi}_{g} (q)$ enlarged by the Hartree term of 
the electron-phonon interaction (\ref{eqn:elph1}).
[Note that ${\hat \chi}_{g} (q)$ is also enlarged by the Fock term of 
the off-site Coulomb interaction; see Supplementary Note 2-2.]
{\bf d}. Self-energy induced by the BO fluctuations.
\color{black}
}
\label{fig:fig1-1}
\end{figure}

The Fourier transform of the BO modulation,
$\delta t_{ij}^{\rm b}$, gives the even-parity BO form factor $g_\q^{lm}(\k)$
\cite{Tazai-rev2021,Kontani-AdvPhys}:
\begin{eqnarray}
g_{\q}^{lm}(\k) = \frac1N \sum_{i}^{{\rm sub}-l}\sum_{j}^{{\rm sub}-m}
\delta t_{ij}^{\rm b} e^{i\k\cdot(\r_i-\r_j)} e^{-i\q\cdot\r_j} ,
\label{eqn:form-r}
\end{eqnarray}
where $\q$ is the wavevector of the BO.
In this study, 
we use the simplified BO form factor due to the nearest sites
presented in Supplementary Note 2-1.
The form factor at $\q=\q_1$, $g_{\q_1}^{lm}$,
is nonzero only when $\{l,m\}=\{A,B\}$,
and we set $g_{\q}^{lm}=g_{\q_1}^{lm}$ when $\q$ is in region I
in Fig. \ref{fig:fig1-1} {\bf b}.
In the same way,
we set $g_{\q}^{lm}=g_{\q_2}^{lm}\ (g_{\q_3}^{lm})$ 
when $\q$ is in region II (III).
$g_{\q_2}^{lm}$ ($g_{\q_3}^{lm}$) is nonzero for 
$\{l,m\}=\{B,C\}$ ($\{C,A\}$). 
This treatment is justified 
because the BO fluctuations strongly develop only for 
$\q\approx \q_n$ in kagome metals.
Furthermore,
we use ${\bar f}_\q(\k)=(f_{\q_n}(\k)+f_{\q_n}(\k+\q-\q_n))/2$
for $\q\sim \q_n$ in the numerical study to improve the accuracy.
Both BO and cLC form factors are Hermite
$\delta t_{ij}^{lm}=(\delta t_{ji}^{ml})^*$,
which leads to the relation
$g_\q^{lm}(\k)=(g_{-\q}^{ml}(\k+\q))^*$
\cite{Kontani-AdvPhys}.


To express the development of the bond order and fluctuations
in kagome metals,
we introduce the following effective BO interaction:
%
\begin{eqnarray}
\hat{H}_{\rm int} &=& -\frac{1}{N} \sum_{\q} \frac{v}{2} \,\, \hat{O}^g_{\q} \,\, \hat{O}^g_{-\q} , 
\label{eqn:elph1} 
\end{eqnarray}
where 
$\hat{O}^g_{\q} \equiv \sum_{\k,l,m,\sigma} g_{\q}^{lm} (\k) c^{\dagger}_{\k+\q,l, \sigma} c_{\k,m,\sigma}$
is the BO operator \cite{Tazai-rev2021,Kontani-AdvPhys,Kontani-RPA}.
and $v$ is the effective interaction,
We assume that the form factor $g_{\q}^{lm} (\k)$ is 
normalized as $\max_{\k,l,m} |g_{\q}^{lm} (\k)|=1$ at each $\q$,
{\i.e.}, $|\delta t_{ij}^{\rm b}|\equiv1/2$ for the nearest sites.
Then, the maximum matrix element of BO interaction 
in Eq. (\ref{eqn:elph1}) is $v/2$.
The interaction (\ref{eqn:elph1}) 
would originate from the combination of
(i) the paramagnon-interference due to on-site $U$
\cite{Tazai-kagome}, 
(ii) the bond-stretching phonon
\cite{phonon-kagome}, and
(iii) the Fock term of off-site Coulomb interaction $V$ 
\cite{Neupert2021}.
In (i), Eq. (\ref{eqn:elph1}) is induced by the 
spin-fluctuation-mediated beyond-RPA processes,
whose diagrammatic expressions are shown in Figs. 3 (c) and (d)
in Ref. \cite{Tazai-kagome}.
This processes give rise to the nematic BO in Fe-based SCs
\cite{Kontani-AdvPhys}.
\textcolor{black}{
A great advantage of this theory \cite{Tazai-kagome} is that the function of the BO form factor and the BO wavevector are automatically optimized to maximize $T_{\rm BO}$. 
Based on this theory, the BO at $\q=\q_n$ ($n=1,2,3$) is robustly obtained based on the first principles multiorbital model for CsV$_3$Sb$_5$
\cite{Tazai-kagome}.
}
The effective parameter $v$ in Eq. (\ref{eqn:elph1}) is given as
$v_{\rm AL}\sim [g_{\rm back}+g_{\rm um}]/2$, which is about $1.5$
near the BO critical point ($\lambda_{\rm bond}\lesssim1$),
as we see in Fig. 3 (e) of Ref. \cite{Tazai-kagome}.
Thus, the value of $v$ given by the AL processes is comparable to 
that used in the present study.
In (ii),
$g_\q^{lm}(\k)$ is given by the hopping modulation due to the stretching mode
and $v=2\eta^2/\omega_{\rm D}$,
where $\eta$ is the electron-phonon ($e$-ph) coupling constant and
$\omega_{\rm D}$ is the phonon energy at $\q\approx\q_n$.
The BO interaction for the three vHS points model
was derived in Ref. \cite{Balents2021}.
In (iii), $v=2V$
as we explain in the Supplementary Note 2-2.
Thus, the effective interaction (\ref{eqn:elph1}) is general.
A possible driving forces of the BO have been discussed experimentally
\cite{Kohn,CDW-no-eph}.


Next, we study the susceptibility of the BO operator 
(per spin) defined as
%
\begin{eqnarray}
\chi_g (\q,\omega_l) & \equiv &\frac12\int^{\beta}_0 d \tau \langle \hat{O}_{\q}^g(\tau) \hat{O}_{-\q}^g(0) \rangle e^{i\omega_l \tau} 
\label{eqn:chi-BO-def} 
\end{eqnarray}
where $\omega_l$ is a boson Matsubara frequency.
$\hat{O}_{\q}^g(\tau)$ is the Heisenberg representation of the BO operator.
When $v=0$, $\chi_g (q)$ is equivalent to
the BO irreducible susceptibility $\chi^0_{g}(q)$
\cite{Tazai-rev2021,Kontani-AdvPhys}: 
\begin{eqnarray}
\chi^{0}_g (q) 
&=& \sum_{lmm'l'}  \chi^{0, lmm'l'}_{g} (q) ,
\label{eqn:chi0} \\
\chi^{0,lmm'l'}_g (\q,\omega_l) 
&=& \frac TN \sum_{\k,\e_n} g^{lm}_{\q} (\k)^* G_{lm'}(\k+\q,\e_n+\omega_l)
 \nonumber \\
&& \times  G_{l'm}(\k,\e_n) g^{m'l'}_{\q} (\k) ,
\label{eqn:chi0_2}
\end{eqnarray}
where $q\equiv (\q,\omega_l=2\pi Tl)$ and 
$\e_n$ is a fermion Matsubara frequency.
Equation (\ref{eqn:chi0_2}) contains two form factors, 
so it vanishes when $l=m$ or $l'=m'$.
Its diagrammatic expression is given in Fig. \ref{fig:fig1-1} {\bf a}.
The numerical result for $\chi^{\rm 0,ABAB}_g (\q,0)$ 
is shown in Fig. \ref{fig:fig1-1} {\bf b},
which exhibits the broad peak at the nesting vector 
between vHS-A and vHS-B; $\q=\q_1$.

The BO susceptibility in Eq. (\ref{eqn:chi-BO-def}) 
is strongly magnified by the Hartree term of Eq. (\ref{eqn:elph1})
because of the same form factors in both equations.
Its process is expressed in Fig. \ref{fig:fig1-1} {\bf c},
and its analytic expression is
\begin{eqnarray}
\chi_g (q) 
&=& \chi^0_{g}(q)/(1-v \chi^0_{g}(q)).
\label{eqn:chic}
\end{eqnarray}
where the notation $q\equiv (\q,\omega_l=2\pi Tl)$ is used.
Here, the relation $\chi_g (\q_n,0)\propto (1-\a_{\rm BO})^{-1}$ holds,
where $\a_{\rm BO}\equiv\max_\q v\chi_{g}^0 (\q)$ is the BO stoner factor.
$\chi_g (\q_n,0)$ diverges when $\a_{\rm BO}=1$.
In contrast,
the cLC susceptibility for the odd-parity cLC form factor,
$f_\q^{lm}(\k-\q/2)=-f_\q^{ml}(-\k-\q/2)$,
is unchanged by the Hartree term
because $g$ is even-parity.

The BO susceptibility is the largest in the Hartree-Fock (HF) approximation.
As we discuss in the Supplementary Note 3.
the BO and cLC susceptibilities at $\q=\q_n$ are 
${\tilde \chi}_{g}\propto (1-(v+v')\chi_g^0)^{-1}$ and
${\tilde \chi}_{\rm cLC}\propto (1-v''\chi_g^0)^{-1}$, respectively.
Here, $-v'\sim v''\sim0.3yv$ originates from the Fock term.
(The coefficient $y\ (\sim O(1))$ depends on the origin of BO fluctuations.
$y=1/2$ for $H_{\rm imt}$ in Eq. (\ref{eqn:elph1}).
The detailed discussion on $y$ will be presented later.)
Thus, both susceptibilities are enlarged,
while ${\tilde \chi}_{\rm cLC}<{\tilde \chi}_{g}$
within the HF approximation.
However, we discover that $-v'$ and $v''$ are further enlarged
by the Maki-Thompson (MT) vertex corrections.

\begin{figure}[htb]
\includegraphics[width=.99\linewidth]{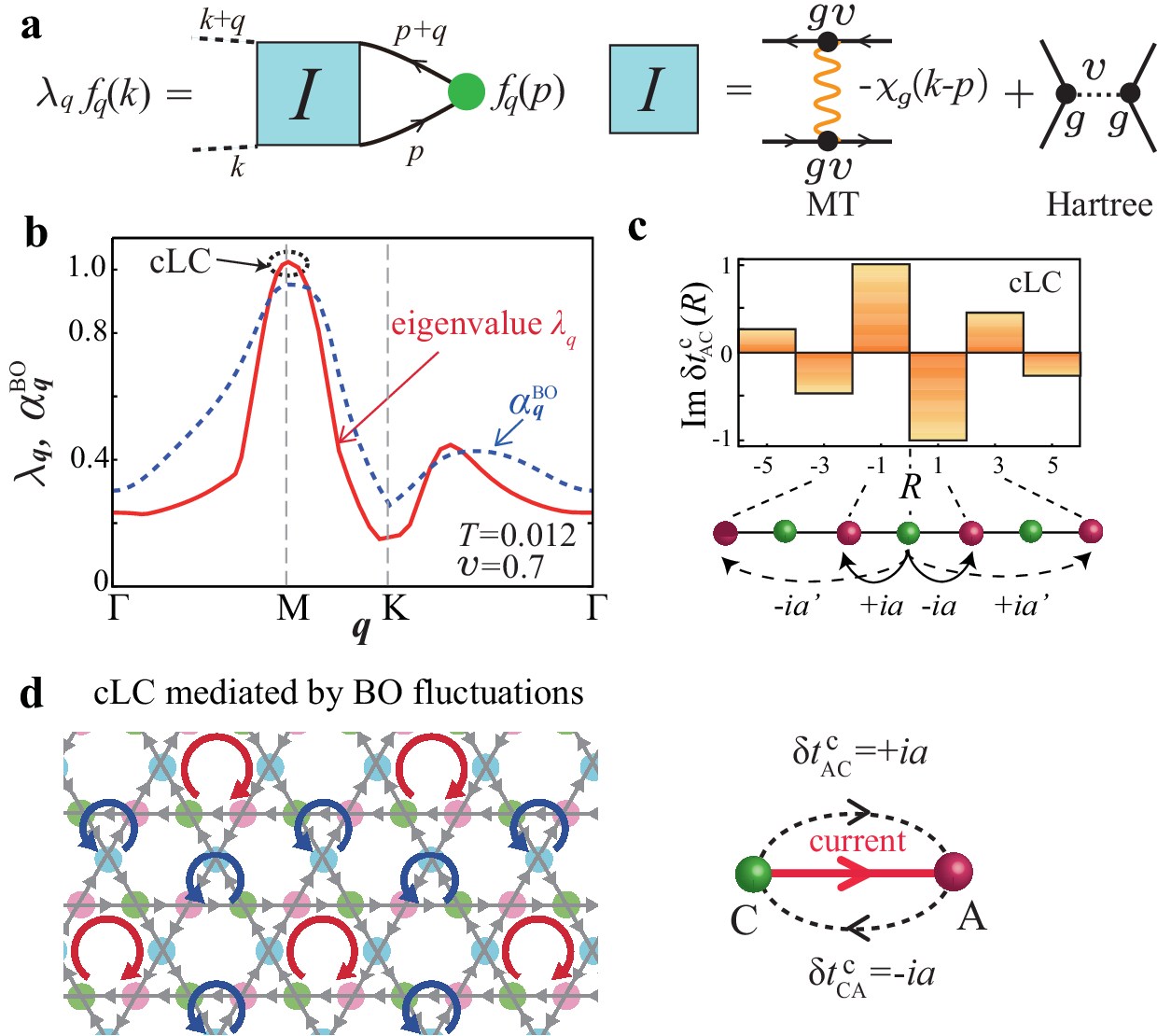}
\caption{
{\bf cLC order driven by BO fluctuation mechanism}. \
{\bf a}. Density-wave (DW) equation due to the single exchange term of the bond-order (BO) fluctuations.
{\bf b}. Eigenvalue of the DW equation $\lambda_\q$ (red-solid line)
and BO Stoner factor $\a_\q^{\rm BO}$ 
(blue-dashed line) for $v=0.7$ and $T=0.012$.
Both show peaks at $\q=\q_n$.
{\bf c}. Imaginary hopping modulation ${\rm Im}\delta t_{\rm AC}^{\rm c} (R)$.
Its triple-$\q$ order gives the cLC pattern in {\bf d}.
One can check that
the clock-wise (anti-clock-wise) loop currents on hexagons
(triangles) in {\bf d} are inverted and moved by ${\aa}_{\rm AC}$
under the sign change of $f_{\q_3}^{\rm AC}$.
}
\label{fig:fig2}
\end{figure}

The MT term describes the scattering of electrons 
due to the developed bosonic fluctuations.
This scattering process is important in metals near the quantum critical points.
For example, in nearly antiferromagnetic metals,
the $d$-wave SC transition is induced by the MT processes of spin fluctuations.
In kagome metals, the MT term
represents the strong inter vHS scattering of electrons
mediated by the abundant BO fluctuations; 
see Fig. \ref{fig:fig2} {\bf a}.
(The MT term also describes the $s$-wave SC state in kagome metals
 \cite{Tazai-kagome}.)
Here, we find that both ${\tilde \chi}_{g}$ and ${\tilde \chi}_{\rm cLC}$
are comparably enlarged due to the MT processes in the present theory.

To understand the BO+cLC phase diagram
and the energy scale of these orders accurately,
we have to include the self-energy that describes 
the quasiparticle properties.
We calculate the on-site self-energy due to BO fluctuations 
(see Eq. (\ref{eqn:Self}) in Methods).
The fluctuation-induced self-energy 
is essential to reproduce the 
$T$-dependence of various physical quantities,
as well-known in spin fluctuation theories
\cite{SCR,TPSC,Kontani-ROP}.
Here, we calculate $\chi_g(q)$ in Eq. (\ref{eqn:chic})
and $\Sigma_{m}(\e_n)$ in Eq. (\ref{eqn:Self}) self-consistently.

\vspace{5mm}
{\bf BO fluctuation-mediated cLC order}. \ 

Next, we discuss the cLC mechanism.
The HF approximation for the BO interaction (\ref{eqn:elph1})
does not lead to the cLC order, as we explain in the Supplementary Note 3.
(It is the same for off-site Coulomb interaction case; 
see Supplementary Note 2-1.)
Thus, the cLC order should be ascribed to the beyond-HF mechanism.
Here, we explain that the strong electron scattering 
between different vHS points due to the BO fluctuations, 
which are described as the MT processes, causes the odd-parity cLC order
$\delta t_{ij}^{\rm c}=-\delta t_{ji}^{\rm c}$.
(Note that the spin-fluctuation-exchange processes cause 
the cLC order in quasi-1D systems \cite{Tazai-cLC}.)
This process is generated by 
solving the following linearized DW equation
\cite{Tazai-Matsubara,Onari-FeSe,Tazai-cLC}:
\begin{eqnarray}
\lambda_{\q}f_\q^{L}(k)&=& \frac{T}{N}\sum_{p,M_1,M_2}
I_\q^{L,M_1}(k,p) 
\nonumber \\
& &\times \{ -G(p)G(p+\q) \}^{M_1,M_2} f_\q^{M_2}(p) ,
\label{eqn:DWeq}
\end{eqnarray}
where $L\equiv (l,l')$ and $M_i$ represent the pair of sublattice indices.
$I_\q^{L,M}(k,p)\propto -\chi_g(k-p)$ is given by the 
BO fluctuation scattering process shown in Fig. \ref{fig:fig2} {\bf a},
which is called the MT process.
The expression of $I_\q^{L,M}$ is given in
Eq. (\ref{eqn:DWkernel}) in Methods section.
Note that $T\sum_{n}\{-G(\p,\e_n)G(\p+\q,\e_n)\}>0$.

By solving the DW equation (\ref{eqn:DWeq}),
the optimized order parameter function is given as the 
eigenfunction $f_\q^L(k)$ for the maximum eigenvalue $\lambda_{\q}$.
$\max_\q\{\lambda_\q\}=1$ at the phase transition temperature.
Note that 
$f_\q^L(k)$ represents the symmetry-breaking part 
in the normal self-energy
$\Delta\Sigma(\k,\q)\sim\langle c_{\k+q\s}^\dagger c_{\k\s}\rangle$,
and DW equation is directly derived from the stationary condition
$\delta F[\Delta\Sigma]/\delta(\Delta\Sigma)=0$
\cite{Tazai-Matsubara}.
We can regard the DW equation (\ref{eqn:DWeq}) as 
the gap equation for the optimized particle-hole (p-h) condensation
\cite{Kontani-AdvPhys,Tazai-Matsubara}.


Note that the BO fluctuation-mediated interaction
for the self-energy (Eq. (\ref{eqn:Self})) 
and that for the kernel function (Eq. (\ref{eqn:DWkernel}))
have the same coefficient $y$, guaranteed by the Ward identity. 
The $y$ depends on the BO fluctuation mechanism:
$y=1/2$ for the BO interaction $v$ in Eq. (\ref{eqn:elph1})
that works only in charge-channel.
$y\approx1/2$ for the AL mechanism for the same reason \cite{Tazai-kagome}.
$y=2$ for the off-site $V$ that induced both charge-
and three spin-channel BO fluctuations
as we explain in Supplementary Note 2-2.
In kagome metals, both $v$ and $V$ coexist.
In this case, BO fluctuations in charge-channel 
dominate over those in spin-channel,
and therefore $y\gtrsim0.5$ is expected in real kagome metals.
Detailed explanation is given in Supplementary Note 2-3.
Because we are interested in a general argument,
we set $y=0.5\sim1$ as a model parameter below.
Note that the Aslaoazov-Larkin term is unimportant 
as we discuss in Supplementary Note 4.

Figure \ref{fig:fig2} {\bf b} shows the
largest eigenvalue of the DW equation $\lambda_{\q}$ (red line)
and BO Stoner factor $a_\q^{\rm BO}\equiv v\chi_g^0(\q)$ (blue line) 
as functions of $\q$, for $v=0.7$ and $T=0.012$.
They exhibit the maximum value at $\q=\q_n \ (n=1,2,3)$.
The corresponding solution of the DW equation is odd-parity:
$f_\q^{lm}(\k-\q/2)=-f_\q^{ml}(-\k-\q/2)$.
Then, the corresponding real-space hopping modulation is odd-parity
$\delta t_{ij}^{\rm c}=-\delta t_{ji}^{\rm c}$
and pure imaginary when $\delta t_{ij}^{\rm c}$ is Hermitian.
The obtained $\delta t_{\rm AC}^{\rm c} (R)\equiv \delta t_{i_\A j_\C}^{\rm c}$
for the cLC at $\q=\q_3$ along the A-C direction
is shown in Fig. \ref{fig:fig2} {\bf c}, where the odd integer
$R$ is defined as $\r_i^\C-\r_j^\A \equiv R{\aa}$.
In addition, the odd-parity relation 
$\delta t_{\rm AC}^{\rm c} (R)=-\delta t_{\rm CA}^{\rm c} (-R)$ is verified.
The obtained charge loop current pattern for the $3Q$ state
is depicted in Fig. \ref{fig:fig2} {\bf d}.


\begin{figure}[htb]
\includegraphics[width=.85\linewidth]{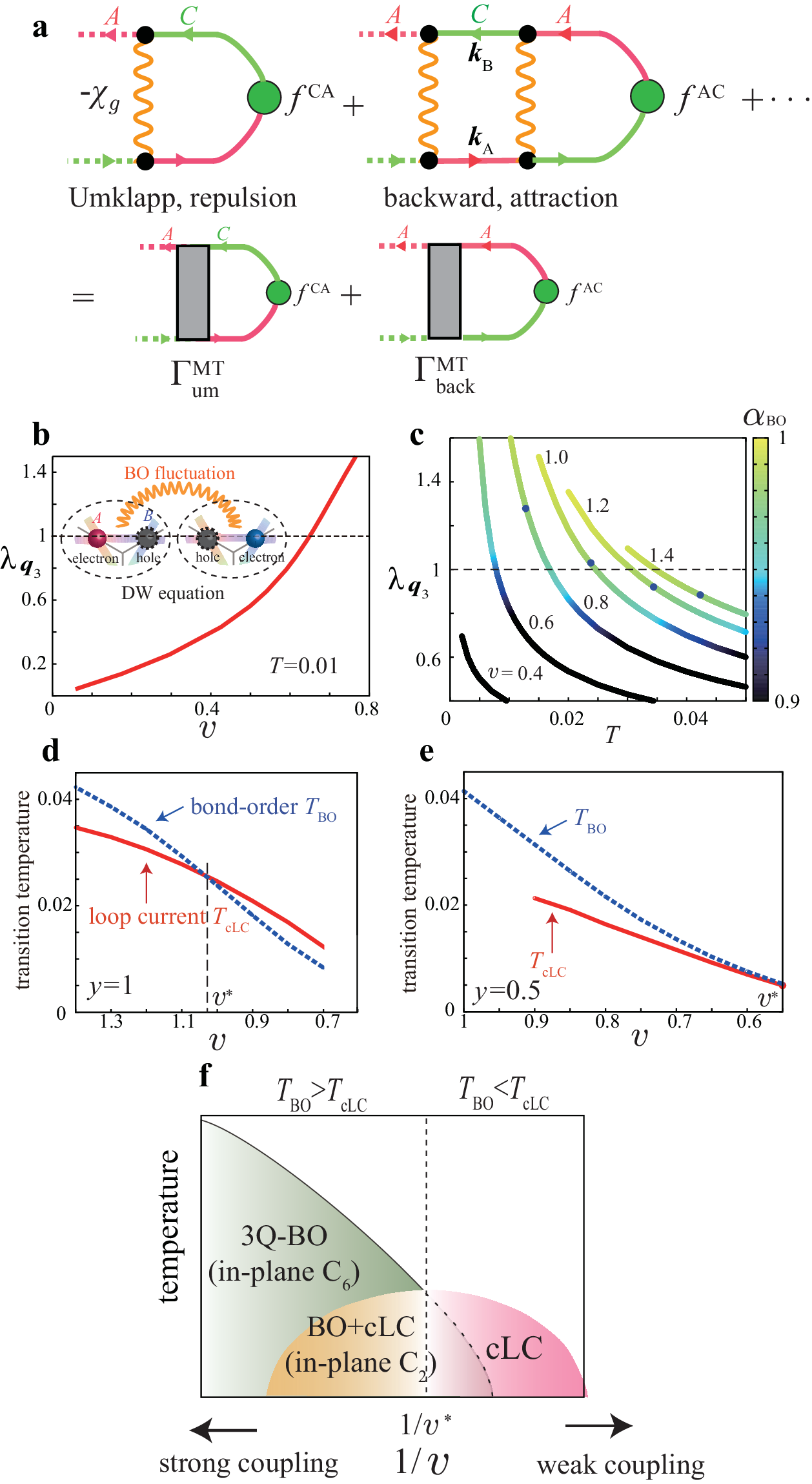}
\caption{
{\bf cLC and BO transition temperatures and predicted phase diagram}. \
\color{black}
{\bf a}. 
Series of Maki-Thompson (MT) processes produced in the density-wave (DW) equation.
Yellow wavy lines represent the bond order (BO) propagators.
The first-order and other odd-order terms give the repulsive
Umklapp scattering $\Gamma_{\rm um}^{\rm MT}<0$.
The second-order and other even-order terms give the attractive
backward scattering $\Gamma_{\rm back}^{\rm MT}>0$.
Both scatterings give the odd-parity charge loop current (cLC) order cooperatively.
{\bf b}. Obtained $v$-dependence of the eigenvalue of cLC 
$\lambda_{\q_3}$ at $y=1$.
{\bf c}. Obtained $T$-dependence of $\lambda_{\q_3}$ at $y=1$.
$\a_{\rm BO}$ is shown by the color of each line,
and the small black circle on each line represents $T_{\rm BO}$.
The relation $T_{\rm cLC}>T_{\rm BO}$
is satisfied in the weak-coupling region ($v<v^*$).
Obtained $T_{\rm cLC}$ and $T_{\rm BO}$ as functions of $v$
for {\bf d} $y=1$ and {\bf e} $y=0.5$.
{\bf f}. Schematic phase diagram in the present theory.
The nematic $3Q$ BO+cLC coexisting phase appears 
when $T_{\rm cLC}<T_{\rm BO}$.
\color{black}
}
\label{fig:fig3}
\end{figure}

Here, we discuss why the cLC order is mediated by the BO fluctuations.
Let us consider the infinite series of MT terms 
in Fig. \ref{fig:fig3} {\bf a},
which is equal to $f_{\q_3}^{\rm AC}(\lambda_{\q_3}^{-1}-1)^{-1}$
according to the DW equation (\ref{eqn:DWeq}).
The first term together with other odd-order MT terms 
in Fig. \ref{fig:fig3} {\bf a} give the 
repulsive Umklapp interaction, $\Gamma_{\rm um}^{\rm MT}<0$,
which leads to the odd-parity order $f_{\q_3}^{\A\C}=-f_{\q_3}^{\C\A}$.
In contrast, the second term together with other even-order MT terms
give the attractive backward interaction, $\Gamma_{\rm back}^{\rm MT} >0$,
which gives the attraction among the same $f_{\q_3}^{lm}$.
Therefore, all series of MT terms cooperatively induce 
the odd-parity current order form factor shown in Fig. \ref{fig:fig2} {\bf c}.
Figure \ref{fig:fig3} {\bf b} exhibits the
obtained $v$-dependence of the cLC eigenvalue
as a function of $T$ in the case of $y=1$.

Figure \ref{fig:fig3} {\bf c} exhibits
the $T$-dependence of $\lambda_{\q_3}$ for $v=0.4-1.4$ at $y=1$.
The cLC transition temperature 
$T_{\rm cLC}$ is given by the relation $\lambda_{\q_3}=1$.
The color on each line represents $\a_{\rm BO}$:
It is clearly seen that 
$\a_{\rm BO}$ at $T=T_{\rm cLC}$ monotonically increases with $v$.
In 2D systems, $\a_{\rm BO}$ asymptotically approaches 1 with $v$,
but never exceeds 1 due to the $\chi_g$-induced self-energy
\cite{Kino-Kontani,Kontani-ROP}.
Here, $T_{\rm BO}$ is defined as $\a_{\rm BO}=\a_{\rm BO}^*$ 
with $\a_{\rm BO}^*=0.985$, which is
shown as a small circle on each line in Fig. \ref{fig:fig3} {\bf c},
by considering the small inter-layer BO coupling $|v_\perp|\ (\ll v)$.
(Overall results are unchanged for $\a_{\rm BO}^*\sim0.99$.)
The three-dimensional (3D) BO appears when
$\chi_{g}^{\rm 3D}=\chi_{g}^{\rm 2D}/(1-|v_\perp|\chi_{g}^{\rm 2D})=\infty$,
that is, $|v_{\perp}|\sim(1-\a_{\rm BO})v$.
Similar method is frequently used in deriving $T_{\rm SDW}$
in spin fluctuation theories \cite{Kino-Kontani}.
When $v$ is small, the relation $T_{\rm cLC}>T_{\rm BO}$ holds,
which is natural because the MT term becomes large for $\a_{\rm BO}\lesssim1$.
With increasing $v$, however,
the opposite relation $T_{\rm cLC}<T_{\rm BO}$ is realized 
due to the large self-energy effect.

The obtained $T_{\rm cLC}$ and $T_{\rm BO}$ as functions of $v$
are shown in Figs. \ref{fig:fig3} {\bf d} $y=1$ and {\bf e} $y=0.5$.
In {\bf d}, $T_{\rm cLC}=T_{\rm BO}$ is realized at $v=v^*\approx1.03$,
and $T_{\rm cLC}/T_{\rm BO}>1$ is realized in the weak-coupling region $v<v^*$.
The opposite relation 
$T_{\rm cLC}/T_{\rm BO}<1$ is obtained in the strong-coupling region $v>v^*$
because the eigenvalue of DW equation (\ref{eqn:DWeq}) is suppressed 
by the large self-energy.
In {\bf e}, $T_{\rm cLC}=T_{\rm BO}$ at $v=v^*\approx0.55$.

Figures \ref{fig:fig3} {\bf d} and {\bf e} indicate that
both BO and cLC instabilities are comparable for $v^*\approx v$.
Based on the parity argument,
the BO (cLC) instability is given by $\Gamma_{\rm back}+(-)\Gamma_{\rm um}$.
Therefore, the relation $\Gamma_{\rm back}\gg|\Gamma_{\rm um}|$
should be satisfied for $v^*\approx v$.
In fact, the Hartree process gives positive 
$\Gamma_{\rm back}^{\rm H}=\Gamma_{\rm um}^{\rm H}\sim v/(1-v\chi_g^0)$,
so the Hartree and MT processes strengthen each other in 
$\Gamma_{\rm back}$ but cancel each other in $\Gamma_{\rm um}$.
This relation is verified by the parquet RG study in
Supplementary Note 5.


We discuss that 
Figs. \ref{fig:fig3} {\bf d} and {\bf e} naturally explain
the experimental $P$-$T$ phase diagram with $T_{\rm BO}$ and 
$T_2^* \ (\sim T_{\rm TRSB})$
given by $\mu$SR study \cite{muSR5-Rb} for $A$=Rb,
considering that $v/W_{\rm band}$ decreases with $P$.
A schematic BO+cLC phase diagram derived from the present theory 
is depicted in Fig. \ref{fig:fig3} {\bf f}.
(This schematic phase diagram is supported by 
the Ginzburg-Landau (GL) analysis in 
Supplemental Figs. 11 {\bf a}-{\bf c}.)
The suppression of the secondary order
due to the primary order is considered.)
The cLC phase is realized next to the BO phase 
because it is mediated by the BO fluctuations.
This cLC+BO phase diagram is reminiscent of the SC-SDW
phase diagram of spin-fluctuation-mediated superconductors, 
which has been reproduced by considering the self-energy
\cite{Kino-Kontani}.


\vspace{5mm}
{\bf $Z_3$-nematic state given by the cLC-BO coexistence}. \


\begin{figure}[htb]
\includegraphics[width=.99\linewidth]{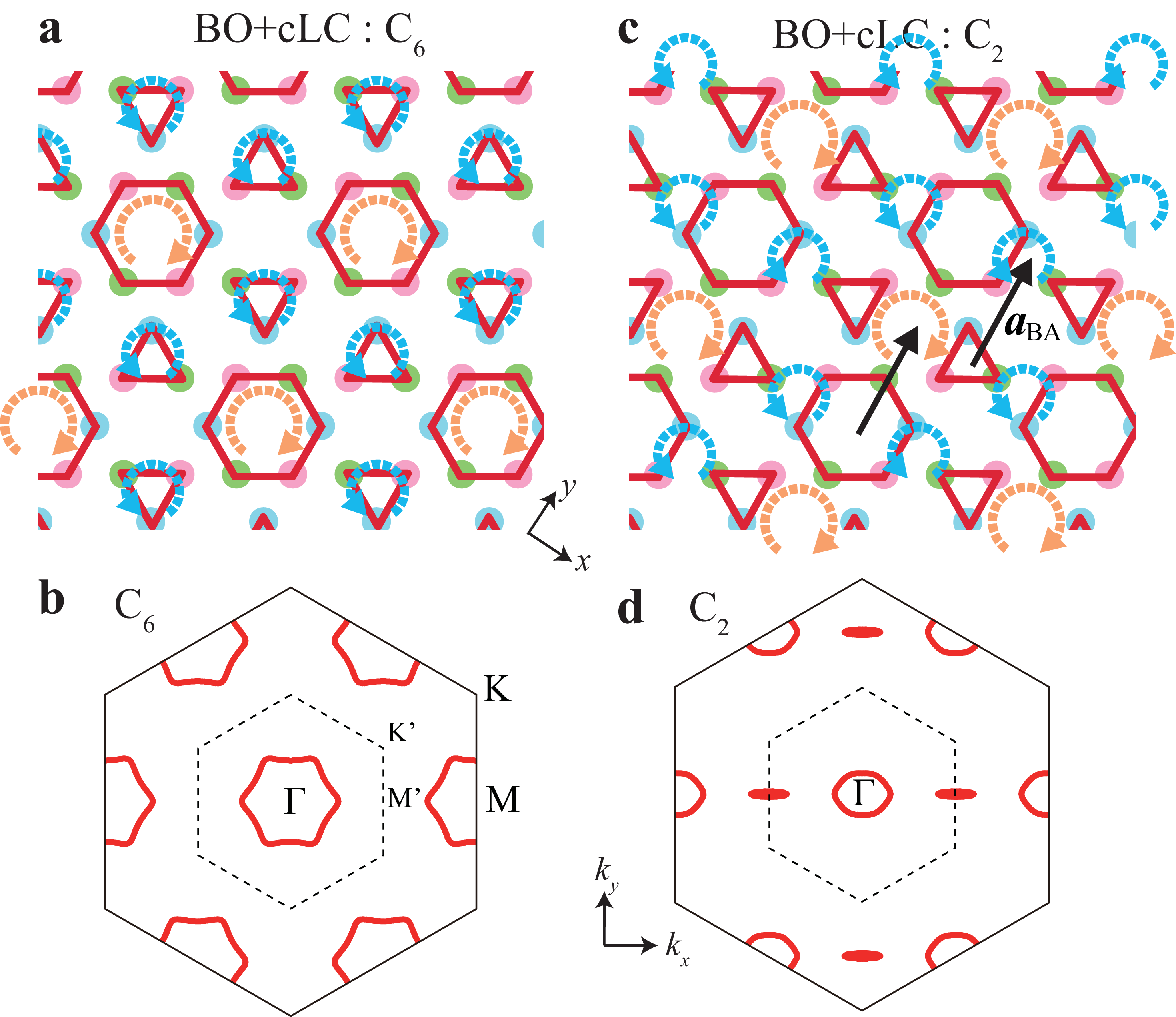}
\caption{
{\bf cLC+BO coexisting states with $C_6$ and $C_2$ symmetries}. \
{\bf a}. $C_6$-symmetric bond order (BO) and charge loop current (cLC) 
coexisting state in real space. 
{\bf b}. Its folded Fermi surface (FS). 
The folded Brillouin zone is shown by dotted lines.
{\bf c}. $C_2$-symmetric BO+cLC coexisting state.
The nematicity originates from the out-of-phase combination of bond and current orders.
{\bf d}. Its nematic FS.
Here, the director is parallel to ${\aa}_{\rm BA}$ because the cLC order in {\bf c} is shifted by ${\aa}_{\rm BA}$ from the cLC order in {\bf a}.
Thus, the $Z_3$ nematic state with different three directors is realized. 
Here, we use large $|\delta t_{ij}^{\rm b,c}| \ (=0.05)$ to exaggerate the nematicity.
}
\label{fig:fig4}
\end{figure}

To understand the cLC+BO coexisting states in Fig. \ref{fig:fig3} {\bf f},
the Ginzburg-Landau (GL) free energy analysis is very useful
\cite{Balents2021,Kennes-coexistence,Fernandes-coexistence}.
For example, the third-order GL term is 
$F^{(3)}= b_1 \phi_1\phi_2\phi_3
+b_2( \phi_1\eta_2\eta_3+\eta_1\phi_2\eta_3+ \eta_1\eta_2\phi_3 )$,
where the coefficients satisfy the relation $b_1\sim -b_2$,
and $(\phi_1,\phi_2,\phi_3)$ [$(\eta_1,\eta_2,\eta_3)$] 
is the magnitude of the BO [cLC] parameter
at $\q=\q_1$, $\q_2$, $\q_3$.
Here, we introduce the $3Q$ states 
${\bmphi}_1=(\phi/\sqrt{3})(1,1,1)$,
${\bmeta}_1=(\eta/\sqrt{3})(1,1,1)$, and
${\bmeta}_2=(\eta/\sqrt{3})(1,-1,-1)$.
The chiral center of ${\bmeta}_1$ coincides with 
the center of the BO ${\bmphi}_1$,
while the center of ${\bmeta}_2$ is shifted by ${\aa}_{\rm BA}$ from 
that of ${\bmphi}_1$.
Thus, the coexisting state $({\bmphi}_1,{\bmeta}_{1[2]})$
has the $C_{6[2]}$-symmetry as shown in Fig. \ref{fig:fig4} {\bf a} [{\bf c}],
and its FS in the folded Brillouin zone is in Fig. \ref{fig:fig4} {\bf b} [{\bf d}].
As we explain in Supplementary Note 6-1,
$F^{(3)}$ for the $C_2$-coexisting state is lower than 
that for the $C_6$-coexisting state
in the case of $|\phi|\gg|\eta|$ for fixed $|\phi|,|\eta|$.
(The optimized cLC order in the $C_2$-phase 
is ${\bmeta}_2'\propto(2,-1,-1)$; see Supplementary Note 6-1.)
Therefore, the BO+cLC $Z_3$ nematic state is realized when $v>v^*$.
\textcolor{black}{
This result is consistent with the recent observation of 
out-of-phase combination of bond charge order and loop currents
by STM measurement \cite{Madhavan}.
}
We comment that the nematic BO+cLC phase is obtained
when $T_{\rm BO}\gtrsim T_{\rm cLC}$ by minimizing the GL free energy 
$F[{\bmphi},{\bmeta}]$ exactly in Supplementary Note 6-2.

We also discuss the case of $v<v^*$,
where cLC is the primary order as shown in Fig. \ref{fig:fig3} {\bf f}.
The $C_6$ symmetry $3Q$ cLC order appears at $T=T_{\rm cLC}$
when $2d_{2,a}/d_{2,b}>1$ as we discuss in Supplementary Note 5,
where $d_{2,a}$ ($d_{2,b}$) is the GL coefficients of 
the $\eta_1^4$ ($\eta_1^2\eta_2^2$) term.
Note that the primary $3Q$ cLC order 
induces the secondary BO parameter even above $T_{\rm BO}$
through the $b_2$-term in $F^{(3)}$
\cite{Balents2021}.
\textcolor{black}{
In contrast, the $1Q$ cLC state is realized
when $2d_{2,a}/d_{2,b}$ is smaller than unity.
Thus, the electronic state becomes nematic at $T_{\rm cLC} \ (>T_{\rm BO})$.
In this case, there is no secondary BO component above $T_{\rm BO}$.
Recently, strong evidence of the emergence of the $1Q$ cLC state 
at $\sim130$K ($>T_{\rm BO}$) has been reported by the magnetic torque measurement
\cite{Asaba}.
}

The obtained nematic BO+cLC state is TRSB and two-dimensional.
Other possible nematic state is 
the shift-stacking of the $3Q$ BO layers,
each of which has $C_6$ symmetry.
The shift-stacking is caused by the $3Q$ state composed of the 
3D BO at $\q_n^{\rm 3D}$ with $q_{1,z}^{\rm 3D}=q_{2,z}^{\rm 3D}=\pi$ 
and $q_{3,z}^{\rm 3D}=0$ \cite{Balents2021}.
We stress that these two different nematic states
can be realized at different temperatures.

\vspace{5mm}
{\bf Giant AHE in cLC+BO state}. \

Next, we discuss the transport phenomena that originate from the cLC
\cite{Haldane,AHE-kagome-theory}.
Using the general expression of the intrinsic conductivity 
\cite{Kontani-Yamada,Kontani-ROP},
we calculate the Hall conductivity 
($\sigma_{xy}$ and $\sigma_{yx}$) due to the 
Fermi-surface contribution in the BO+cLC state.
The expression is
$\sigma_{\mu\nu}= \frac1N \sum_\k A_{\mu\nu}(\k)$,
%
where $A_{\mu\nu}(\k)= \frac{e^2}{\hbar}\frac1{\pi}{\rm Tr}\{ 
{\hat v}_{\k,\mu} {\hat G}_{\k}(i\gamma) {\hat v}_{\k,\nu} {\hat G}_{\k}(-i\gamma) \}$.
Here, ${\hat G}_\k(\e)= ((\e+\mu){\hat 1}-{\hat h}_\k)^{-1}$
is the Green function matrix, 
where ${\hat h}_\k$ is the $12\times12$ tight-binding model
with the $3Q$ BO and cLC order, 
and ${\hat v}_{\k,\mu} = d{\hat h}_\k/dk_\mu$ is the velocity operator.
$\gamma\ (>0)$ is the electron damping rate
that is given by the imaginary part of the self-energy.
We set $n=n_{\rm vHS}$ and $|\delta t_{ij}^{\rm b}|=|\delta t_{ij}^{\rm c}|=0.025$,
where the band hybridization gap due to the BO+cLC order is
about $\Delta\approx 2\sqrt{|\delta t_{ij}^{\rm b}|^2+|\delta t_{ij}^{\rm c}|^2}=0.07$.

\begin{figure}[htb]
\includegraphics[width=.99\linewidth]{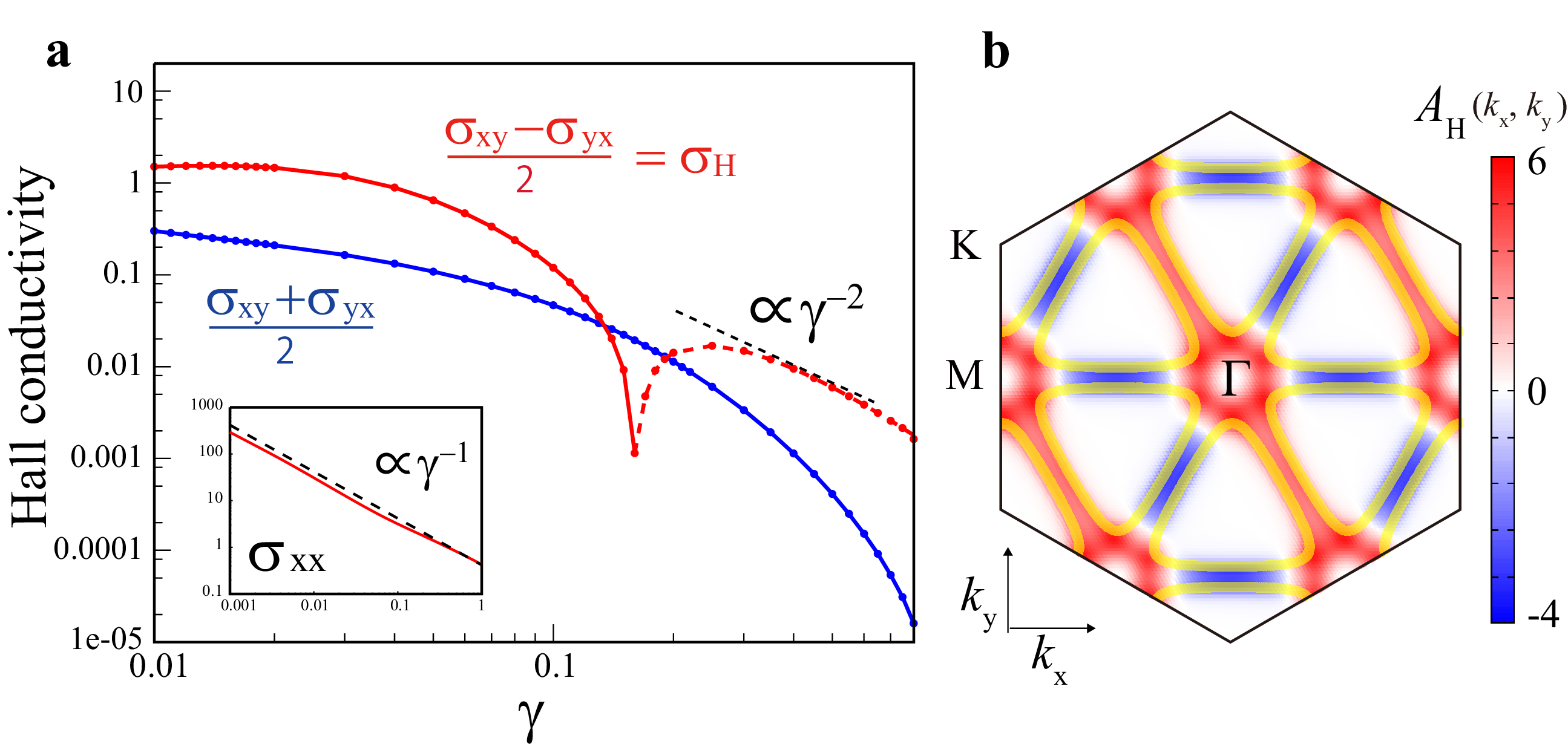}
\caption{
{\bf Giant AHE in nematic cLC+BO state}. \
{\bf a}. Anomalous Hall conductivity in the nematic BO+cLC state
($|\delta t_{ij}^{\rm b,c}|=0.025$)
as a function of the electron damping rate 
$\gamma \ \propto \tau^{-1}$.
The full (broken) line represents the positive (negative) value.
Thus, the Hall conductivity $\s_{\rm H}$ becomes large in the 
low-resistivity region ($\gamma\lesssim0.03$).
{\bf b}. $A_{\rm H}(\k)$ at $\gamma=0.05$:
$\frac1N \sum_\k A_{\rm H}(\k)=\s_{\rm H}$.
}
\label{fig:fig5}
\end{figure}

Figure \ref{fig:fig5} {\bf a} shows the obtained conductivities
in the nematic BO+cLC state,
in the unit of $\frac{e^2}{\hbar} \ (=2.4\times 10^{-4}\Omega^{-1})$.
When $\g\ll\Delta$, 
the Hall conductivity $\s_{\rm H}\equiv \frac12(\s_{xy}-\s_{yx})$ is almost constant,
and its magnitude is proportional to $|\delta t_{ij}^{\rm c}|$.
When $\g\gg\Delta$, in contrast,
$\s_{\rm H}$ decreases with $\gamma$ in proportion to $\gamma^{-2}$.
This crossover behavior is universal in the intrinsic Hall effect,
which was first revealed in heavy fermion systems,
and found to be universal in later studies
\cite{Kontani-ROP,Kontani-SHE-PRL,Kontani-Yamada,AHE-RMP}.
Note that $\frac12(\s_{xy}+\s_{yx})$ is nonzero
in the nematic state.
To understand the origin of the intrinsic Hall effect, 
we plot $A_{\rm H}(\k)\equiv (A_{xy}(\k)-A_{yx}(\k))/2$
at $\gamma=0.05$ in Fig. \ref{fig:fig5} {\bf b}:
It shows a large positive value
mainly around the vHS points,
due to the band-hybridization induced by the cLC order.
The obtained $\s_{\rm H}\sim 1$ corresponds to $4\times10^3 \Omega^{-1}{\rm cm}^{-1}$
because the interlayer spacing is $\sim0.6{\rm nm}$.
Thus, giant AHE $\s_{\rm H}\sim 10^2 \Omega^{-1}{\rm cm}^{-1}$
reported in Refs. \cite{AHE1,AHE2}
is understood in this theory.

\vspace{5mm}
{\bf Parquet RG theory, Field-induced cLC mechanism}. \

To verify the idea of the BO fluctuation-mediated cLC,
we perform the analysis of the parquet renormalization group (RG) 
formulation \cite{Balents2021,Chubukov-gr} 
and present the results in Supplementary Note 5. 
A great merit of the RG method is that both particle-particle 
and particle-hole channels are treated on the same footing. 
We find that both BO and cLC fluctuations cooperatively develop 
in Supplementary Fig. 6.
This result of the RG study strongly supports the validity of 
the DW equation analysis.


We comment on the complementary relationship between the present theory and the parquet RG theory. 
The latter theory solves a simplified 3-patch model in an unbiased way, leading to the development of both cLC order and BO, while the relationship between the two orders is not clear. 
On the other hand, the present theory focuses on the existence of the experimental BO phase and reveals that abundant BO quantum fluctuations lead to TRSB particle-hole condensation.
Thus, the concept of the BO fluctuation-mediated cLC has been verified based on different reliable theories.

\vspace{5mm}
{\bf Summary}. \

In summary,
we proposed a cLC mechanism 
mediated by the BO fluctuations in kagome metals.
This cLC mechanism is universal because 
it is independent of the origin of the BO.
The validity of the idea of the BO fluctuation-mediated cLC
has been confirmed by the parquet RG study in Supplementary Note 5. 
Furthermore, we revealed that novel $Z_3$ nematicity emerges
under the coexistence of the cLC and the BO
reported in Refs.
\cite{elastoresistance-kagome,birefringence-kagome,STM2}
in addition to the giant AHE
\cite{AHE1,AHE2}.
This theory presents a promising scenario 
for understanding the BO, the cLC and the nematicity 
in kagome metals in a unified way.

In the present study,
we focus on the pure-type band composed of $b_{3g}$-orbital. 
However, the impact of other $3d$-orbitals on the cLC order 
has also been studied in Refs. \cite{Fernandes-coexistence,Nat-g-ology}.
The extension of the present theory to multi-orbital models 
is a very important future issue.

Here, we shortly discuss several experimental evidences of 
the BO+cLC coexistence.
Recent transport measurement of highly symmetric fabricated 
CsV$_3$Sb$_5$ micro sample 
\cite{Moll-hz}
reveals that small magnetic field $h_z \ (<10{\rm T})$
or small strain gives rise to the nematic BO+cLC coexisting state
below $T_{\rm BO}$.
This finding is well explained by the recent GL theory under $h_z$
\cite{Mch2023}:
The current-bond-$h_z$ trilinear coupling caused by the orbital magnetization
gives rise to the sizable $h_z$-induced cLC order in the BO state.
This theory also explains $h_z$-induced 
enhancement of the cLC order observed by $\mu$SR measurements
\cite{muSR4-Cs,muSR2-K,muSR5-Rb}
and field-tuned chiral transport study 
\cite{eMChA}.
\textcolor{black}{
It is noteworthy that the nematic electronic state 
that supports the $C_2$ BO+cLC order in Fig. \ref{fig:fig4} {\bf c}
has been reported by recent STM measurement \cite{Madhavan}.
}

Finally, we comment on some interesting kagome metals other than $A$V$_3$Sb$_5$.
Double-layer kagome metal ScV$_6$Sn$_6$ shows $\sqrt{3}\times\sqrt{3}$ 
charge-density wave (CDW) \cite{166-CDW}.
It was proposed that the CDW originates from the 
flat phonon modes with Sn vibrations 
\cite{Bernevig-166,Bernevig-166-2}.
Interestingly, ScV$_6$Sn$_6$ also exhibits the spontaneous 
TRSB state \cite{166-mSR}.
The mechanism of the TRSB state in ScV$_6$Sn$_6$ 
is an interesting future problem.
(Note that the existence of the vHS points is not a 
requirement for the cLC order \cite{Tazai-cLC}.)
The GL free energy analysis was performed in Ref.
\cite{Thomale-GL}.
Recently, very weak but definite signal of the nematic electronic order 
has been observed in Ti-based kagome metal CsTi$_3$Bi$_5$
\cite{arXiv:2211.12264,arXiv:2211.16477}.
To explain the observed hidden nematicity,
the odd-parity BO without TRSB has been predicted theoretically
\cite{Ti-kagome}.



\subsection{Methods} 

\vspace{5mm}
{\bf Self-energy due to BO fluctuations}. \

To understand the BO+cLC phase diagram
and the energy scale of these orders accurately,
we have to include the self-energy that describes 
the quasiparticle properties.
We calculate the on-site self-energy due to BO fluctuations as
\begin{eqnarray}
\Sigma_{m}(\e_n)&=&\frac{T}{N}\sum_{\k,q,m'',m'''}G_{m'm''}(\k+\q,\e_n+\w_l) 
\nonumber \\
& &\times B_{mm,m''m'}(k,q),
\label{eqn:Self} \\
B_{mm,m''m'}(k,q)&=& g_\q^{m'm}(\k) g_{\q}^{m''m}(\k)^* \cdot 
y v(1+v\chi_g(q)) ,
\label{eqn:Self2}
\end{eqnarray}
which is shown in Fig. \ref{fig:fig1-1} {\bf d}.
Then, the Green function is given as
${\hat G}(k)=(i\e_n+\mu-{\hat h}(\k)-{\hat \Sigma}(\e_n))^{-1}$.
The effect of thermal fluctuations described by the self-energy 
is essential to reproduce the 
$T$-dependence of various physical quantities.
Here, $y=1/2$ when ${\hat H}_{\rm int}$ is given in Eq. (\ref{eqn:elph1}).
In the present numerical study,
we calculate $\chi_g(q)= \chi^0_{g}(q)/(1-v \chi^0_{g}(q))$ 
and $\Sigma_{m}(\e_n)$ in Eq. (\ref{eqn:Self}) self-consistently.

\vspace{5mm}
{\bf Kernel function of the DW equation}. \


The kernel function due to BO fluctuations
in Eq. (\ref{eqn:DWeq}) is given as
%
\begin{eqnarray}
I_{\q}^{ll',mm'}(k,p) &=&
-g_{\p- \k}^{m'l'}(\k) y v (1+v\chi_{g} (k-p)) 
g_{\k -\p}^{lm}(\p+\q)
\nonumber \\
& &+g_{\q}^{ll'}(\k) v g_{\q}^{mm'}(\p)^* ,
\label{eqn:DWkernel}
\end{eqnarray}
which is expressed in Fig. \ref{fig:fig2} {\bf a}
and Supplementary Fig. 4 {\bf a}.
The first term, the MT term,
is important when $\a_{\rm BO}\lesssim1$,
and its first term is the Fock term.
The second term, the Hartree term,
vanishes when the eigenfunction ${\hat f}_\q(k)$ 
is orthogonal to the BO form factor ${\hat g}_\q(k)$,
like the cLC order is.
Note that ${\hat B}(k,q)=-{\hat I}_{\bm 0}(k,k+q)$.
A more detailed discussion is presented in Supplementary Note 3.

{\bf Numerical Analysis}. \
In this study, we solved the eigenvalue equation with the kernel function (\ref{eqn:DWeq}) and the integral equations (\ref{eqn:Self}) and (\ref{eqn:Self2}) numerically, by dividing the Brillouin zone into $60\times60$ $\k$ meshes.
The number of $\k$ meshes is fine enough to achieve reliable numerical accuracy ($\sim1$\%) at the calculated temperatures ($T\sim0.01$).

\subsection{Acknowledgments}
We are grateful to S. Onari, A. Ogawa, Y. Matsuda, T. Shibauchi, K. Hashimoto,
and T. Asaba for fruitful discussions.












\clearpage
\newpage

\makeatletter
\renewcommand{\thefigure}{\arabic{figure}}
\renewcommand{\theequation}{\arabic{equation}}
\renewcommand{\figurename}{Supplementary Fig.}
\makeatother
\setcounter{figure}{0}
\setcounter{equation}{0}
\setcounter{page}{1}
\setcounter{section}{1}

\begin{widetext}
\begin{center}
{\bf \large 
[Supplementary Information] \\
\vspace{3mm}
{\large
Charge-loop current order and $Z_3$-nematicity mediated by bond order fluctuations in kagome metals 
}
}
\end{center}

\begin{center}
Rina Tazai$^1$, Youichi Yamakawa$^2$, and Hiroshi Kontani$^2$
\end{center}

\begin{center}
\textit{
$^1$Yukawa Institute for Theoretical Physics, Kyoto University,
Kyoto 606-8502, Japan \\
$^2$ Department of Physics, Nagoya University,
Furo-cho, Nagoya 464-8602, Japan. 
}
\end{center}

\end{widetext}

\subsection{Supplementary Note 1: Square-lattice kagome model}
\label{sec:SMA}

In this Supplementary Information,
we set the hopping integrals $t=-0.5$ eV and $t'=0$.
Hereafter, the unit of energy is eV unless otherwise noted.
In Ref. \cite{Tazai-kagomeS},
the present authors found that 
the paramagnon-interference theory naturally 
explains the bond order (BO) on the basis of the 
kagome-lattice Hubbard model.
The used lattice structure with the square unit cell and 
its Fermi surface (FS)
are shown in Supplementary Figs. \ref{fig:figS1} {\bf a} and {\bf b}, respectively.
The van-Hove singular point at $\k=\k_{X}$ ($X$=A,B,C)
is composed of the $X$-sublattice.
The form factor of the $3Q$ BO and that of the $3Q$ cLC in real space
are shown in Supplementary Figs. \ref{fig:figS1} {\bf c} and {\bf d}, respectively.
We stress that the
$D_{6h}$ point group symmetry of the original kagome lattice
is not harmed in the present DW equation solution 
using this square-lattice model.

\begin{figure}[htb]
\includegraphics[width=.99\linewidth]{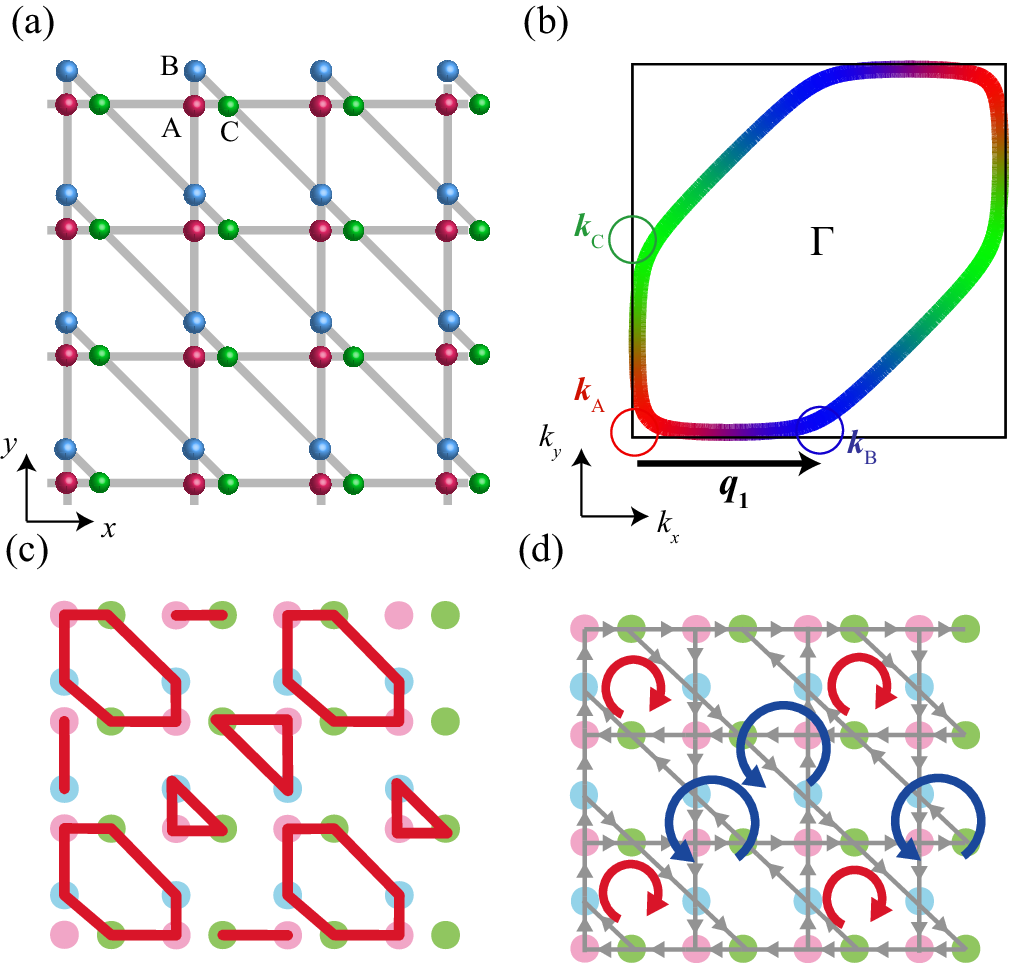}
\caption{
{\bf a}. Square-lattice kagome metal model
that is convenient for the numerical study.
{\bf b}. Obtained Fermi surface (FS) in the square Brillouin zone. 
The color represents the weight of the sublattice
(A = red, B = blue, C = green). 
{\bf c}. Form factor of the $3Q$ bond order (BO) in real space.
The red bonds represent the Tri-Hexagonal pattern.
{\bf d}. Form factor of the $3Q$ charge loop current (cLC).
The coexistence of {\bf c} and {\bf d} leads to the $C_6$ symmetry state.
}
\label{fig:figS1}
\end{figure}

\subsection{Supplementary Note 2: BO fluctuations due to Fock term by $V$}

Here, we discuss the important effect of the 
nearest-site Coulomb interaction $V$
in addition to the on-site one $U$.

\subsubsection{{Supplementary Note 2-1: HF approximation for the $U$-$V$ Hubbard model}}
\label{sec:SMB1}

The analysis of the kagome-lattice $U$-$V$ Hubbard model 
based on the mean-field theory is presented in Section SF of 
Ref. \cite{Tazai-kagomeS}.
The charge (spin) channel eigenvalue $\lambda^{c(s)}$ 
in the mean-field theory is given by 
solving the linearized DW equation with the 
Hartree-Fock (HF) kernel function made of $U$ and $V$.
Its diagrammatic expression for $V$
is shown in Supplementary Fig. \ref{fig:figS2} {\bf a}.

Supplementary Figure \ref{fig:figS2} {\bf b} shows the obtained 
several largest eigenvalues, $\lambda_{\rm SDW}^s$ 
and $\lambda_{X}^c$ ($X$=CDW, BO, cLC),
as functions of $V/U$ at $U=0.79$ 
\cite{Tazai-kagomeS}.
These eigenvalues linearly increase with respect to 
$U$ and $V$ at a fixed $V/U$.
When $V/U\ll 1$, a simple SDW order ($f=1$) at $\q\approx{\bm0}$ 
is realized for $U\sim1.6$.
It originates from the Hartree term of $U$.
When $V/U\gg 1$, on the other hand,
simple charge-density-wave (CDW) order ($f=1$) at $\q\approx \q_n$
is caused by the Hartree term of $V$.
In this model, $\q_1=(\pi,0)$, $\q_2=(\pi,\pi)$, $\q_3=(0,\pi)$.
For $V/U=0.4\sim0.65$, the BO is realized by the Fock term of $V$.
Note that the non-local BO is not suppressed by $U$,
while the simple CDW order due to the Hartree term of $V$ 
is strongly suppressed by $U$.
However, the cLC instability is smaller than other instabilities 
within the HF approximation.


The form factors of the BO between the nearest sites
in the square kagome-lattice model in Supplementary Fig. \ref{fig:figS1} {\bf a}
are given by
\begin{eqnarray}
b_{\rm AB}(\q)&=&(1-e^{-iq_y})/2, 
\label{eqn:BO1} \\
b_{\rm BC}(\q)&=&(1-e^{-iq_x+iq_y})/2, 
\label{eqn:BO2} \\
b_{\rm CA}(\q)&=&(1-e^{iq_x})/2, 
\label{eqn:BO3} 
\end{eqnarray}
which are normalized as $\max_\q\{b_{lm}(\q)\}=1$.
Here, $b_{lm}(\q)=b_{ml}(\q)^*$ and $b_{ll}(\q)=0$.

\subsubsection{{Supplementary Note 2-2: Effective interaction due to BO susceptibility}}
\label{sec:SMB2}

In Supplementary Fig. \ref{fig:figS2} {\bf b}, we found the development of 
the BO instability within the mean-field approximation.
Next, we derive the effective interaction mediated by 
the BO susceptibility.
The final result is given in Supplementary Eq. (\ref{eqn:I-V}),
which is essentially equivalent to  Eq. (\ref{eqn:DWkernel})
derived in the main text.

From now on, we derive Supplementary Eq. (\ref{eqn:I-V}).
First, we consider the
effective interaction due to the Fock term of $V$.
The Hamiltonian is
$H_V=\sum_{il,jm,\s\s'}V_{il,jm}c_{il,\s}^\dagger c_{il,\s} c_{jm,\s'}^\dagger c_{jm,\s'}$,
where $i,j$ represent the unit cell, $l,m=A,B,C$,
and $c_{il}$ is the electron annihilation operator.
(Here, we drop the spin indices for simplicity.)
$H_V$ is rewritten as
$H_V=\frac1N \sum_{\k\k'\q\s\s'}V_{lm}(\q)c_{\k+\q,l,\s}^\dagger c_{\k,l,\s} c_{\k',m,\s'}^\dagger c_{\k'+\q,m,\s'}$.
Here, 
$V_{lm}(\q)=\frac1N \sum_{i,j}V_{il,jm}e^{-i\q\cdot (\r_{i,l}-\r_{j,m})}$,
where $\r_{i,l}$ is the coordinate of the site $(i,l)$;
In Supplementary Fig. \ref{fig:figS1} {\bf a}, $\r_{i,l}=(i_x,i_y)$ is independent of $l$,
and $i_x, i_y$ are integer coordinates.
In the case of the nearest-site Coulomb interaction $V$,
$V_{lm}(\q)$ is expressed as
\begin{eqnarray}
V_{lm}(\q)&=&2V a_{lm}(\q).
\end{eqnarray}
Here,
\begin{eqnarray}
a_{\rm AB}(\q)&=&(1+e^{-iq_y})/2,\\
a_{\rm BC}(\q)&=&(1+e^{-iq_x+iq_y})/2, \\
a_{\rm CA}(\q)&=&(1+e^{iq_x})/2,
\end{eqnarray}
where $a_{lm}(\q)=a_{ml}(\q)^*$ and $a_{ll}(\q)=0$.

Considering the relation
$a_{lm}(\k-\k')=a_{lm}(\k)a_{lm}(\k')^*+b_{lm}(\k)b_{lm}(\k')^*$,
the Fock term $V_{lm}(\k-\k')$ in Supplementary Fig. \ref{fig:figS2} {\bf c} 
is expressed as
\begin{eqnarray}
\!\!\!\!\!
V_{lm}(\k-\k')&=&2V [a_{lm}(\k)a_{lm}(\k')^*+b_{lm}(\k)b_{lm}(\k')^*] 
\nonumber \\
& &=2V \sum_d^{a,b} d_{lm}(\k)d_{lm}(\k')^*
\label{eqn:Vab} 
\end{eqnarray}
which is expressed in Supplementary Fig. \ref{fig:figS2} {\bf c}.
Note that $a_{\rm AB}(\k)\approx a_{\rm AB}(\k')\approx 0$
and $b_{\rm AB}(\k)\approx b_{\rm AB}(\k')\approx 1$
for $(l,m)=({\rm A,B})$ and $(\k,\k')\approx (\k_{\rm A},\k_{\rm B})$.
Because $a_{lm}(\k)$ and $b_{lm}(\k)$ are orthogonal,
the form factor $g^{lm}_{\q}(\k)$ for the largest eigenvalue
of the DW equation is equal to $b_{lm}(\k)$
within the Fock approximation.

We next consider the second-order term with respect to $V$
shown in Supplementary Fig. \ref{fig:figS2} {\bf c}:
\begin{eqnarray}
\!\!\!\!\!
V_{lm,l'm'}^{(2)}(\k-\k';\q)&=& \frac{T}{N} \sum_{p}
V_{lm}(\k-\p)G_{ll'}(p+q)G_{m'm}(p)V_{l'm'}(\p-\k')
\nonumber\\
&=&(2V)^2 \sum_{d,d'}^{a,b}d_{lm}(\k)(d'_{l'm'}(\k'))^* \chi^{0dd'}_{lm,l'm'}(q),
\label{eqn:Vab-2nd} 
\end{eqnarray}
where
$\chi^{0dd'}_{lm,l'm'}(q)=
-\frac{T}{N}\sum_p (d_{lm}(\p))^*G_{ll'}(p+q)G_{m'm}(p)d'_{l'm'}(\p)$
($d,d'=a,b$).
In kagome metal model, $\chi^{0dd'}_{lm,l'm'}(q)$ takes large value
only for $d=d'$ and $\q\sim\q_m$ ($m=1,2,3$).

As we mentioned above,
we obtain $a_{\rm AB}(\k)\approx a_{\rm AB}(\k')\approx 0$
and $b_{\rm AB}(\k)\approx b_{\rm AB}(\k')\approx 1$
for $(\k,\k')\approx (\k_{\rm A},\k_{\rm B})$.
By considering these relations, we can drop $a_{lm}$ 
in Supplementary Eqs. (\ref{eqn:Vab}) and (\ref{eqn:Vab-2nd}).
Thus, the summation of the first and the second order terms is
given as
\begin{eqnarray}
2V b_{lm}(\k)b_{l'm'}(\k')^*[{\hat 1}+2V{\hat \chi}^{0b}(q)]_{lm,l'm'} ,
\end{eqnarray}
where
$\chi^{0b}_{lm,l'm'}(q)\equiv \chi^{0bb}_{lm,l'm'}(q)$.

Now, we consider all the ladder diagrams composed of the Fock terms
shown in Supplementary Fig. \ref{fig:figS2} {\bf d}.
It is obtained as
\begin{eqnarray}
W_{lm,l'm'}(k,k',q)\approx
2V b_{lm}(\k)b_{l'm'}(\k')^* [\hat{1}+2V\hat{\chi}^{b}(q)]_{lm,l'm'} ,
\label{eqn:W-V}
\end{eqnarray}
which takes sizable value when 
$\k\approx \k_m$, $\k'\approx \k_{m'}$, and
$\q\approx (\k_l-\k_m),\ (\k_{l'}-\k_{m'})$ (modulo original reciprocal vectors).
The BO susceptibility in Supplementary Eq. (\ref{eqn:W-V})
is given by the solution of 
${\hat\chi}^{b}(q)={\hat\chi}^{0b}(q)+(2V) {\hat\chi}^{0b}(q){\hat\chi}^{b}(q)$.
It is expressed as
\begin{eqnarray}
{\hat \chi}^{b}(q)
&=&{\hat \chi}^{0b}(q)({\hat 1}-2V{\hat \chi}^{0b}(q))^{-1} .
\label{eqn:chi-V}
\end{eqnarray}
Thus, ${\hat W}(k,k',q)$ is proportional to the 
BO susceptibility $\hat{\chi}^{b}(q)$.
(One can verify the relation
${\hat \chi}^{b}(q)={\hat \chi}^{0b}(q)+\frac{T^2}{N^2}\sum_{kk'}{\hat A}(k,q){\hat W}(k,k',q){\hat A}(k',q)$,
where $A_{ml,m'l'}(k,q)=G_{mm'}(k+q)G_{l'l}(k)$.)

By using Supplementary Eq. (\ref{eqn:W-V}),
the charge-channel MT term in the DW equation is 
\begin{eqnarray}
I_{ll'mm'}(k,k',q)=-2W_{m'l'ml}(k,k+q,k'-k),
\label{eqn:IW}
\end{eqnarray}
where the factor 2 comes from the summation of 
the parallel-spin ($\uparrow,\uparrow$) and 
the antiparallel-spin ($\uparrow,\downarrow$) ladder diagrams,
because the Pauli principle does not work on $V$.
It is expressed in Supplementary Fig. \ref{fig:figS2} {\bf d}.
Note that the present Fock term of the 
off-site Coulomb interaction corresponds to the 
Hartree term of the $e$-ph interaction in Eq. (\ref{eqn:elph1}) in the main text.

\begin{figure}[htb]
\includegraphics[width=.9\linewidth]{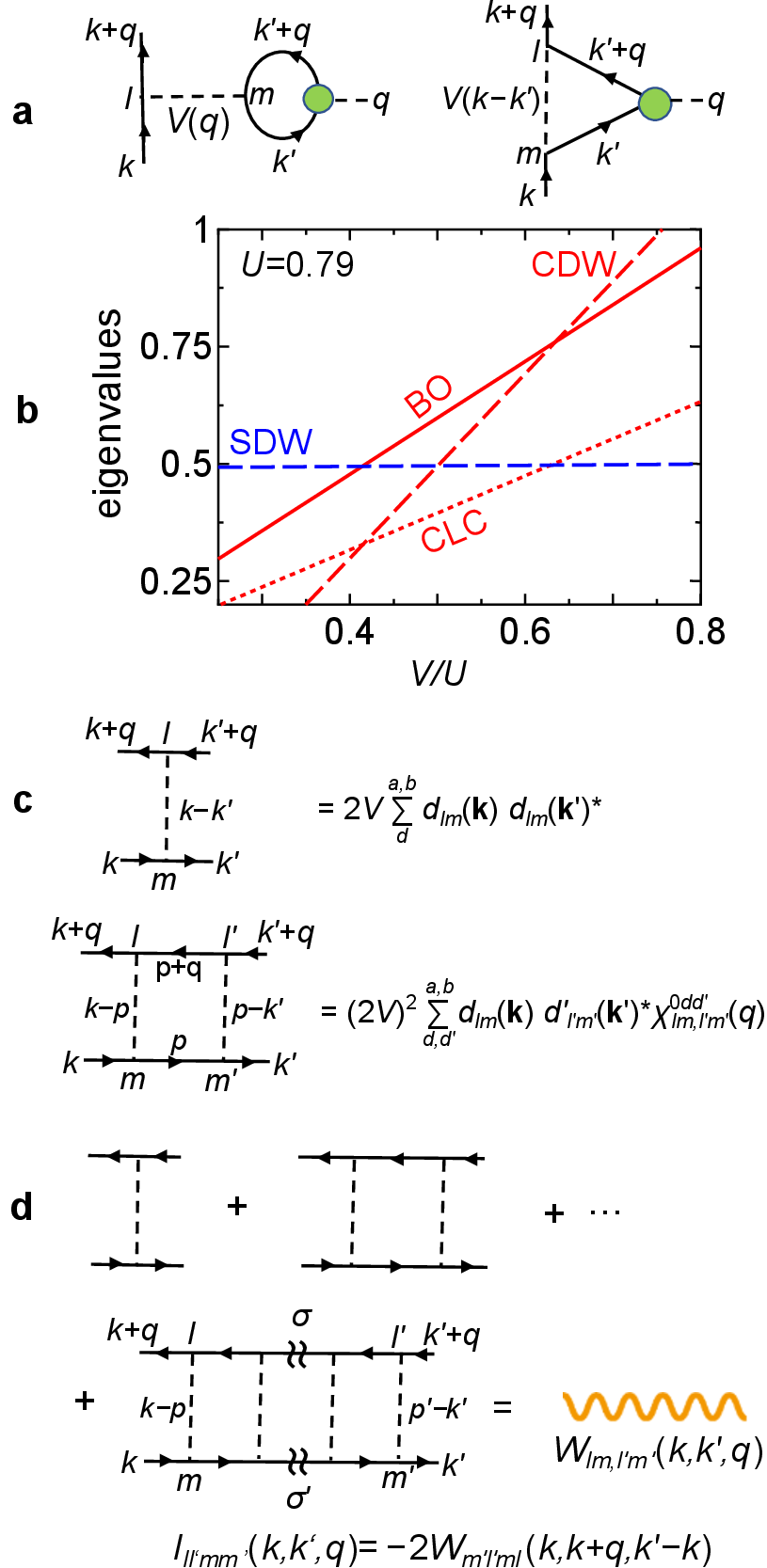}
\caption{
{\bf a}. 
Hartree and Fock terms with respect to
the on-site ($U$) and the nearest-site ($V$) 
Coulomb interactions in the density-wave (DW) equation.
Each green circle represents the form factor.
{\bf b}. Eigenvalues of the HF DW equation 
as a function of $V/U$:
$\lambda_{\rm SDW}^s$ and $\lambda_{X}^c$ ($X$=CDW, BO, cLC). 
Note that $\lambda^{\rm c}_{\rm BO} \approx\lambda^{\rm s}_{\rm BO}$;
see Ref. \cite{Tazai-kagomeS}.
{\bf c}. The first- and the second-order terms with respect 
to the Fock terms of $V$.
{\bf d} $W$ given by the summation of all ladder-type diagrams,
which is proportional to the BO susceptibility.
The present Fock term corresponds to the Hartree term of the 
$e$-ph interaction in Eq. (\ref{eqn:elph1}) in the main text.
}
\label{fig:figS2}
\end{figure}

As discussed above, $\chi_{lm,l'm'}^{0b}(q)$
is enlarged at $\q\approx \q_1$ only when $(lm,l'm')=({\rm AB,AB}),({\rm AB,BA})$.
(In the same way, it is enlarged
at $\q\approx \q_2$ only when $(lm,l'm')=({\rm BC,BC}),({\rm BC,CB})$.)
For this reason,
we can safely approximate Supplementary Eqs. (\ref{eqn:W-V}) and (\ref{eqn:chi-V})
as the $2\times2$ matrix expressions for $\q\approx \q_n$ ($n=1,2,3$).

Below, we explain that the relation 
$\chi^{b}_{\rm AB,AB}(\q_1) \approx \chi^{b}_{\rm AB,BA}(\q_1)$ holds.
In kagome metals,
the ratio in the irreducible susceptibility 
$\chi_{\rm AB,BA}^{0b}(\q_1)/\chi_{\rm AB,AB}^{0b}(\q_1)$ 
is just $\sim0.2$ 
because the Green function $G_{lm}(k)$ is nearly diagonal ($\propto \delta_{l,m}$).
Nonetheless of this fact,
the ratio in the BO susceptibility
$R\equiv \chi_{\rm AB,BA}^{b}(\q_1)/\chi_{\rm AB,AB}^{b}(\q_1)$ 
is of order unity,
when the BO Stoner factor 
$\a_{\rm BO}=2V(\chi_{\rm AB,AB}^{0b}(\q_1)+\chi_{\rm AB,BA}^{0b}(\q_1))$
is close to unity.
In fact, for $\q\approx\q_1$, the relation between the
BO susceptibility and its irreducible susceptibility is
$\hat{\chi}^{b}=\hat{\chi}^{b0}+2V\hat{\chi}^{b0}\hat{\chi}^{b}$,
where $\hat{\chi}^{b0}$ is the $2\times2$ matrix:
$\displaystyle
\hat{\chi}^{b0}=
\begin{pmatrix}
{\chi}^{b0}_{\rm AB,AB} & {\chi}^{b0}_{\rm AB,BA}  \\
{\chi}^{b0}_{\rm BA,AB} & {\chi}^{b0}_{\rm BA,BA} \\
\end{pmatrix} .
$
Then, the BO susceptibility for $\q\approx \q_1$
is obtained as 
\begin{eqnarray}
\chi^{b}_{\rm AB,AB}(q)&=& [a-2V(a^2-b^2)]/d,
\label{eqn:chib1} \\
\chi^{b}_{\rm AB,BA}(q)&=& b/d,
\label{eqn:chib2}
\end{eqnarray}
where $a\equiv {\chi}^{0b}_{\rm AB,AB}(q)$,
$b\equiv {\chi}^{0b}_{\rm AB,BA}(q)$,
and $d\equiv (1-(a+b)2V)(1-(a-b)2V)$.

\begin{figure}[htb]
\includegraphics[width=.8\linewidth]{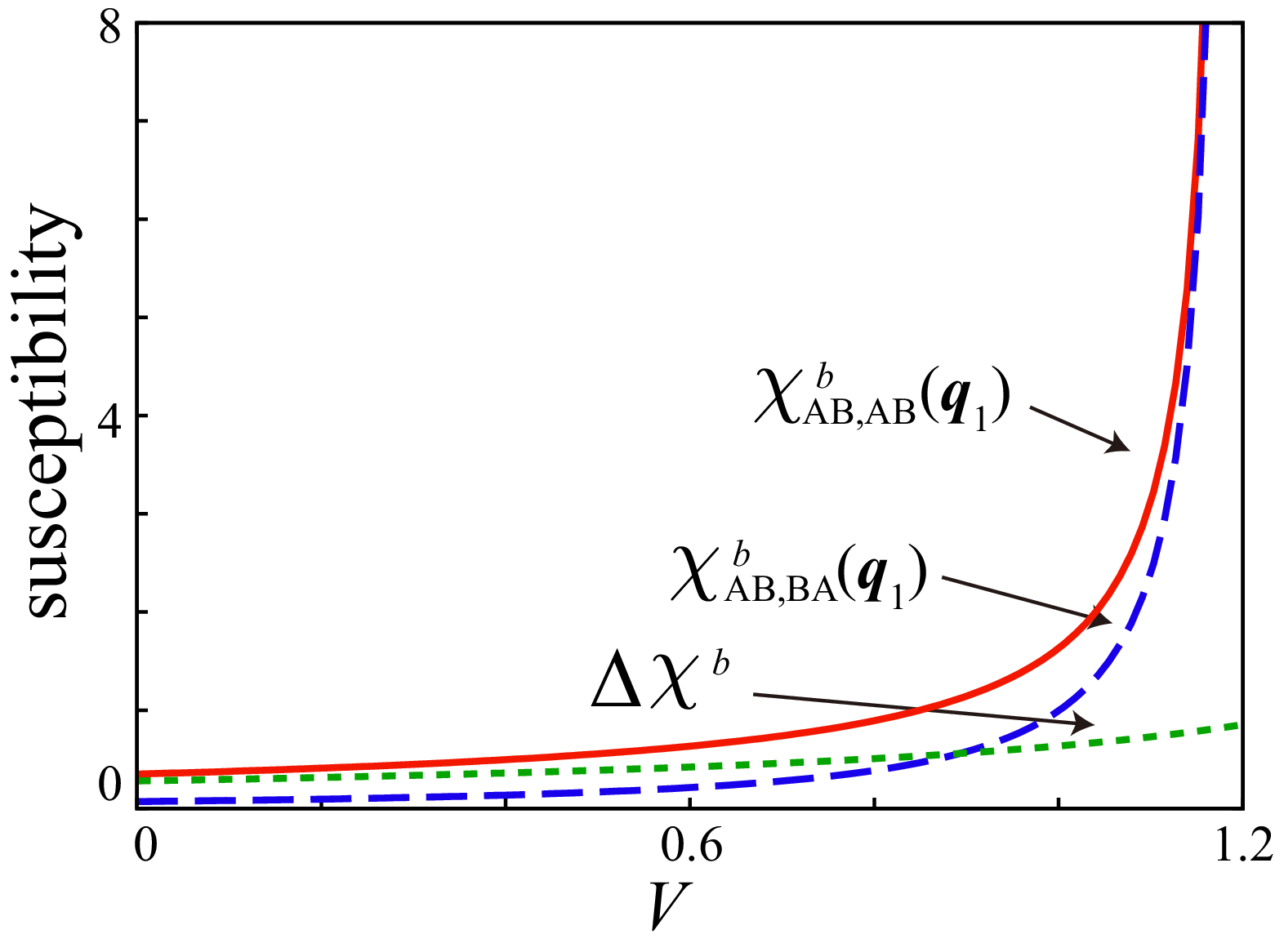}
\caption{
$\chi^{b}_{\rm AB,AB}$ and $\chi^{b}_{\rm AB,BA}$
given in Supplementary Eqs. (\ref{eqn:chib1}) and (\ref{eqn:chib2})
as functions of $V$.
We set $a=0.35$ and $b/a=0.2$.
$\Delta\chi^b\equiv \chi^{b}_{\rm AB,AB}-\chi^{b}_{\rm AB,BA}$
is also shown.
}
\label{fig:figS3}
\end{figure}

According to Supplementary Eqs. (\ref{eqn:chib1}) and (\ref{eqn:chib2}),
$R$ becomes $b/a (\ll1)$ when $V=0$.
In contrast, we obtain $R\approx1$ at $\a_{\rm BO}\approx1$ 
($\a_{\rm BO}=(a+b)2V$).
Supplementary Figure \ref{fig:figS3} shows the 
BO susceptibilities $\chi^{b}_{\rm AB,AB}$ and $\chi^{b}_{\rm AB,BA}$
as functions of $V$.
We set $a={\chi}^{b0}_{\rm AB,AB}(\q_1)=0.35$ and $b/a=0.2$.
Both $\chi^{b}_{\rm AB,AB}$ and $\chi^{b}_{\rm AB,BA}$ increase with $V$,
while $\Delta\chi^b\equiv \chi^{b}_{\rm AB,AB}-\chi^{b}_{\rm AB,BA}$
is almost constant.
This result means that the relation $R\approx1$ holds
near the BO-endpoint, around which the relation 
$\a_{\rm BO}\lesssim 1$ holds.

According to Supplementary Eqs. (\ref{eqn:W-V}) and (\ref{eqn:IW}),
in the case of $R\approx1$,
the MT kernel function for the DW equation is simply given as
\begin{eqnarray}
&& \!\!\!\!\!\!\!\!\!\!
I_{ll',mm'}(k,k',q)\approx -g_{k'-k}^{m'l'}(k)g_{k-k'}^{lm}(k'+q)
\nonumber \\
&&\ \ \ \ \ \ \ \ \ \ \ \ 
\times 2 (2V)^2\chi^{b}(k-k') ,
\label{eqn:I-V}
\end{eqnarray}
where $\chi^{b}(q)\equiv \chi^{b}_{\rm AB,AB}(q)+\chi^{b}_{\rm AB,BA}(q)
\approx \chi^{b0}_{\rm AB,AB}(q)/(1-\a_{\rm BO}(q))$ for $\q\approx \q_1$,
and $g^{lm}\approx b_{lm}$ is the normalized BO form factor
derived from the DW equation
\cite{Tazai-kagomeS}.
Supplementary Equation (\ref{eqn:I-V}) is equal to Eq. (\ref{eqn:DWkernel})
in the main text with $y=2$.
Note that both charge- and three spin-channel
BO fluctuations develop 
in the present off-site $V$ mechanism.

To summarize, the relation 
$\chi^{b}_{\rm AB,AB} \approx \chi^{b}_{\rm AB,BA}$
({\it i.e.}, $R\approx1$),
which is assumed in the cLC mechanism in the main text,
is well satisfied in the off-site $V$ mechanism.
Because this relation is also satisfied in the 
paramagnon-interference mechanism \cite{Tazai-kagomeS} 
and the phonon mechanism,
these three different BO mechanisms will cooperate.
We consider that the main mechanism of the BO in kagome metals
is the paramagnon interference mechanism \cite{Tazai-kagomeS},
and both the bond-stretching phonon mode
and the off-site Coulomb interaction 
will assist the BO formation.

\subsubsection{{Supplementary Note 2-3: The relation $y\gtrsim1/2$ when $v$ and $V$ coexist}}

We consider the MT kernel function in the presence of both 
the off-site Coulomb interaction $V$ 
and the BO interaction $v$ in Eq. (\ref{eqn:elph1}) in the main text.
In this case, $y\gtrsim1/2$ is expected to be realized.
Considering the $SU(2)$ symmetry in spin space,
the spin indices of the MT term in Supplementary Eq. (\ref{eqn:I-V}) are written as
\cite{Kontani-AdvPhys}
\begin{eqnarray}
I_{ll',mm'}^{\s\s',\rho\rho'}(k,k',q)
&=&\frac12 I_{ll',mm'}^{\rm c}(k,k',q)\delta_{\s,\s'}\delta_{\rho',\rho}
\nonumber \\
& &+\frac12 I_{ll',mm'}^{\rm s}(k,k',q){\bmsig}_{\s,\s'}\cdot{\bmsig}_{\rho',\rho},
\label{eqn:I-V-2}
\end{eqnarray}
where ${\bmsig}$ is the Pauli matrix vector.
${\hat I}^{\rm s(c)}$ is the spin (charge) channel interaction given as
$I_{ll'mm'}^{\rm s(c)}=-g_{k'-k}^{m'l'}(k)g_{k-k'}^{lm}(k'+q) (2V)^2\chi^{b,{\rm s(c)}}(k-k')$, 
where
$\displaystyle \chi^{b,{\rm s(c)}}(q)\approx \frac{\chi^{b0}_{\rm AB,AB}(q)}{1-\a^{\rm s(c)}(q)}$, and
$\a^{\rm s(c)}(q)$ is the spin (charge) channel Stoner factor.
Based on Supplementary Eq. (\ref{eqn:I-V-2}),
the MT kernel function for the charge-channel form factor is given as
${\hat I}^{\rm MT}=\frac12({\hat I}^{\rm c}+3{\hat I}^{\rm s})$
\cite{Kontani-AdvPhys}.
When $v=0$, $\a_{\rm s,c}(q)$ is equal to $\a_{\rm BO}(q)$
and therefore ${\hat I}^{\rm c}={\hat I}^{\rm s}$.
As a result,
${\hat I}^{\rm MT}=2 {\hat I}^{\rm c}$ for $v=0$, and therefore $y=2$.

Here, we consider the effect of the BO interaction $v$.
When $v>0$, the charge Stoner factor is magnified as
$\a_{\rm c}(q)=\a_{\rm BO}(q)+v\chi_g^0(\q)$,
where $\chi_g^0(\q)$ is the irreducible susceptibility
given by Eq. (\ref{eqn:chi0_2}) in the main text.
Thus, ${\hat I}^{\rm c}$ is enlarged by $v$.
In contrast, the spin Stoner factor and ${\hat I}^{\rm s}$ 
are unchanged by $v$.
Even if $v$ is small but finite ({\it e.g.}, $v\chi_g^0(\q)\sim0.1$), the relation ${\hat I}^{\rm c}\gg{\hat I}^{\rm s}$ will be realized near the charge-channel BO criticality.
Therefore, we obtain
${\hat I}^{\rm MT}\approx \frac12{\hat I}^{\rm c}$, 
which means that $y\gtrsim1/2$.
To summarize, the relation $y\gtrsim1/2$ is generally expected
when $V$ and $v$ (= AL interference mechanism and $e$-ph interaction) coexist.


\subsection{Supplementary Note 3: Self-consistent DW equation method: derivation of renormalized BO fluctuations}
\label{sec:SMC}
In the main text, 
we discuss the development of the BO susceptibility $\chi_g$
and its essential role in the cLC ordar.
Based on the BO interaction model in Eq. (\ref{eqn:elph1}) in the main text,
we find that the MT term due to $\chi_g$, $I^{\rm MT}\sim -\chi_g$, 
leads to the emergence of the cLC order.
On the other hand, $I^{\rm MT}$ also induces the 
renormalization of the $\chi_g$ itself.
Here, we explain that the latter effect does not change 
the results of the main text.
The kernel function composed of the Hartree and MT terms,
shown in Supplementary Fig. \ref{fig:figS11} {\bf a},
are respectively given as 
\begin{eqnarray}
{\tilde I}_{{\rm MT},\q}^{ll',mm'}(k,p) &=&
-g_{\p- \k}^{m'l'}(\k) g_{\k -\p}^{lm}(\p+\q) 
\nonumber \\
& &\times y {\tilde v} (1+{\tilde v}{\tilde \chi}_{g}(k-p) ),
\label{eqn:MT-S} 
\\
I_{{\rm H},\q}^{ll',mm'}(k,p) &=&
g_{\q}^{ll'}(k) v g_{\q}^{mm'}(p)^* ,
\label{eqn:H-S}
\end{eqnarray}
where ${\tilde \chi}_{g}(q)=\chi_g^0(q)/(1-{\tilde v}\chi_g^0(q))$
is the renormalized BO susceptibility.
In the main text, $v$ in the DW equation for the cLC form factor
and in the self-energy is considered as the renormalized ${\tilde v}$.

Here, by referring to the SCR theory \cite{SCRS},
we perform the self-consistent calculation of the 
renormalized BO susceptibility, ${\tilde \chi}_g(q)$,
based on the DW equation.
A natural self-consistency condition for the MT term 
composed of BO susceptibility is
\begin{eqnarray}
{\lambda}_{\q_1}={\tilde \a}_{\rm BO},
\label{eqn:selfconsist-S}
\end{eqnarray}
where the left-hand side is the eigenvalue of the BO type form factor,
and ${\tilde \a}_{\rm BO}\equiv {\tilde v}\chi_g^0(\q_1)$
is the renormalized BO Stoner factor.
Then, we can calculate the DW equation 
under the self-consistent condition of BO fluctuations.
${\tilde v}$ given by Supplementary Eq. (\ref{eqn:selfconsist-S})
is expressed as
\begin{eqnarray}
&&{\tilde v}=v+v' ,
\label{eqn:vvvS}
\\
&&v'=\frac{\frac{T^2}{N^2}\sum_{k,p,L,M}
A_{g'}^L(k+\q,-\q)I_{{\rm MT},\q}^{L,M}(k,p) A_{g'}^M(p,\q)}{(\chi_{g'}^0(\q))^2},
\nonumber \\
\\
&&{\hat A}_{g'}(p,\q)= {\hat G}(p+\q){\hat g}'_\q(p){\hat G}(p) ,
\label{eqn:selfconsist2-S}
\end{eqnarray}
at $\q=\q_1$, where $g'$ is the solution of the BO type form factor.
The kernel function $I_\q(k,p)$ is shown in Supplementary Fig. \ref{fig:figS11} {\bf a}.
It is composed of the Hartree and MT terms.
(Fock term is included in the MT term.)
Because $g'\approx g$,
${\tilde v}$ is simply derived from Supplementary Eq. (\ref{eqn:vvvS})
by setting $g'=g$.
By using ${\tilde v}$, we solve the DW equation 
and show the obtained $\lambda_{\rm BO} (={\tilde\a}_{\rm BO})$
and $\lambda_{\rm cLC}$ for $y=1$ and $T=0.01$
in Supplementary Fig. \ref{fig:figS11} {\bf b}. 

For small $v\ (\lesssim0.2)$, the DW equation solution 
in Supplementary Fig. \ref{fig:figS11} {\bf b}
corresponds to th HF approximation.
When ${f}_\q(k)$ is BO form factor,
the eigenvalue due to the Hartree term is equal to $\a_{\rm BO}=1.34v$.
However, it is reduced to $\lambda_{\rm BO}=1.34(v+v')$,
where $v'=-0.31yv$ is the Fock term contribution.
When ${f}_\q(k)$ is the cLC form factor,
the Hartree term vanishes, 
while the Fock term gives positive eigenvalue
$\lambda_{\rm cLC}=1.34v''$ with $v''=0.26yv$.
For large $v \ (\gtrsim1)$, the MT term becomes significant.
For this reason, $\lambda_{\rm BO}$ saturates while 
$\lambda_{\rm cLC}$ strongly increases, and
$\lambda_{\rm cLC}$ reaches unity at $v\approx1.35$.
This is because both $-v'$ and $v''$ are strongly 
enlarged by the MT term when ${\tilde \a}_{\rm BO}\lesssim1$.

Supplementary Figures \ref{fig:figS11} {\bf c} and {\bf d} show the obtained 
$T_{\rm cLC}$ and $T_{\rm BO}$
by solving the present DW equation,
in the case of {\bf c} $y=1$ and {\bf d} $y=0.5$.
Here, $T_{\rm cLC}$ is defined by the condition
$\lambda_{\q_1}=1$ for the cLC type form factor.
Also, $T_{\rm BO}$ is defined as $\lambda_{\q_1}=0.985$ 
for the BO type form factor.
The overall $v$-dependences of the order parameters 
are similar to those in Figs. \ref{fig:fig3} {\bf d} and {\bf e} 
in the main text.
By introducing the self-energy, 
both $T_{\rm cLC}$ and $T_{\rm BO}$ will be suppressed,
and the relation $T_{\rm cLC}<T_{\rm BO}$ will be realized 
in the strong coupling region.

Supplementary Figure \ref{fig:figS11} {\bf e} presents the
beyond-RPA processes $\Delta\chi_w(q)$ in the present study. 
The total susceptibility is 
$\chi_w^{\rm tot}(q) = \chi_w^{\rm RPA}(q)+\Delta\chi_w(q)$,
All these diagrams are generated by solving the DW equation.
For $w=g$ (=BO form factor), 
$\chi_g^{\rm RPA}(q)$ is large and positive, 
while $\Delta\chi_g(q)$ takes negative values due to the MT terms. 
For $w=f$ (=cLC form factor), 
$\chi_f^{\rm RPA}(q)$ is very small, 
while $\Delta\chi_f(q)$ takes positive values, 
which becomes significant when $\a_{\rm BO}\sim1$. 
Therefore, the cLC susceptibility 
$\chi_f^{\rm tot}(q)$ develops as large as the BO susceptibility 
in the present theory.

In the main text, 
we solve the DW equation with including the self-energy,
while we solve the DW equation under the
self-consistent condition in Supplementary Eq. (\ref{eqn:selfconsist-S}) in this section.
We find that two different DW equation analyses with MT-type kernel function
produce essentially equivalent numerical results.
The present self-consistent DW equation analysis
strongly supports the reliability of the numerical study in the main text.

\begin{figure}[htb]
\includegraphics[width=.99\linewidth]{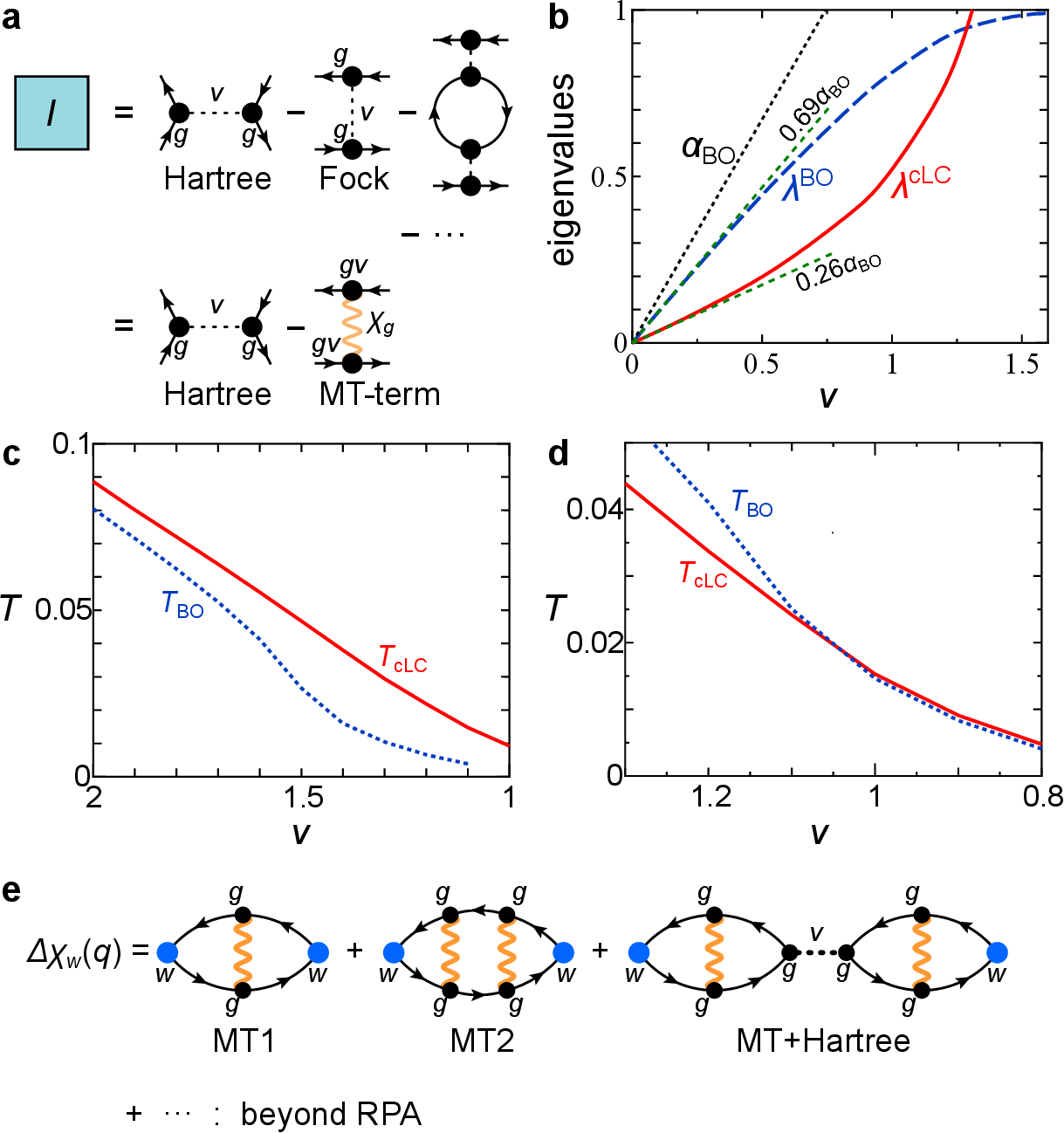}
\caption{
{\bf a}. Diagrammatic expression of the kernel function. 
{\bf b}. $\lambda_{\rm BO}$, $\lambda_{\rm cLC}$, and $\a_{\rm BO}$
given by the present self-consistent density-wave (DW) equation method,
for $y=1$ and $T=0.01$.
{\bf c}.,{\bf d}. Obtained $T_{\rm cLC}$ and $T_{\rm BO}$ as functions of $v$
derived from the self-consistent DW equation method,
in the cases of {\bf c} $y=1.0$ and {\bf d} $y=0.5$.
The self-energy correction is dropped in this calculation.
{\bf e}. Beyond-RPA processes $\Delta\chi_w(q)$
generated by solving the DW equation.
}
\label{fig:figS11}
\end{figure}

\subsection{Supplementary Note 4: Effects of AL-type VCs}
\label{sec:SMD}
Here, we examine the role of the Aslamazov-Larkin (AL)
vertex corrections (VCs) due to the interference between two 
bosonic susceptibilities ($\chi^{\rm boson}$)
shown in Supplementary Fig. \ref{fig:figs4}.
The AL terms are significant for the even-parity 
order parameter in the paramagnon-interference mechanism.
This mechanism is responsible for the BO and 
the orbital order in Fe-based superconductors
\cite{Onari-SCVCS,Yamakawa-FeSeS,Kontani-AdvPhysS}, 
high-$T_c$ cuprates
\cite{Tazai-MatsubaraS,Tsuchiizu4S,Yamakawa-CuS}, 
and kagome metals
\cite{Tazai-kagomeS}.
In contrast, the AL term is unimportant for the 
odd-parity order parameter, and instead, 
the MT term is significant 
for the current order in the frustrated Hubbard models
\cite{Tazai-cLCS}
and non-Fermi liquid transport phenomena
\cite{Kontani-ROPS}.

Here, we explain that 
the AL terms due to the bond susceptibilities,
which were neglected in the main text,
are unimportant in the present cLC mechanism in kagome metals.
Supplementary Figure \ref{fig:figs4} exhibits the VC 
for $f_{\q_1}^{\rm AB}(\k)$ at $\k\approx \k_\A$.
These terms are almost canceled for the odd-parity cLC order 
because of the relation $f_{\q_1}^{\rm AB}(\k)=-f_{\q_1}^{\rm BA}(-\k-\q)$.
In fact, 
the order parameter in the real space satisfies the relation 
$\delta t_{ij}={\cal P}\delta t_{ji}$,
where ${\cal P}=+1\ (-1)$ for the even (odd) parity order 
and $l,m=\A,\B,\C$.
Then, its Fourier transform gives the form factor:
\begin{eqnarray}
f^{lm}_\q(\k) &=& \frac1N \sum_{i}^{{\rm sub}-l} \sum_{j}^{{\rm sub}-m}
 \delta t_{ij} e^{-\k\cdot({\r}_i-{\r}_j)}e^{-\q\cdot{\r}_j}
\nonumber \\
&=& \frac 1N \sum_{i}^{{\rm sub}-l} \sum_{j}^{{\rm sub}-m}
 ({\cal P} \delta t_{ji})
 e^{-\k\cdot({\r}_i-{\r}_j)}e^{-\q\cdot{\r}_j}
\nonumber \\
&=&{\cal P} f^{ml}_\q(-\k-\q) .
\end{eqnarray}
%

\begin{figure}[htb]
\includegraphics[width=.85\linewidth]{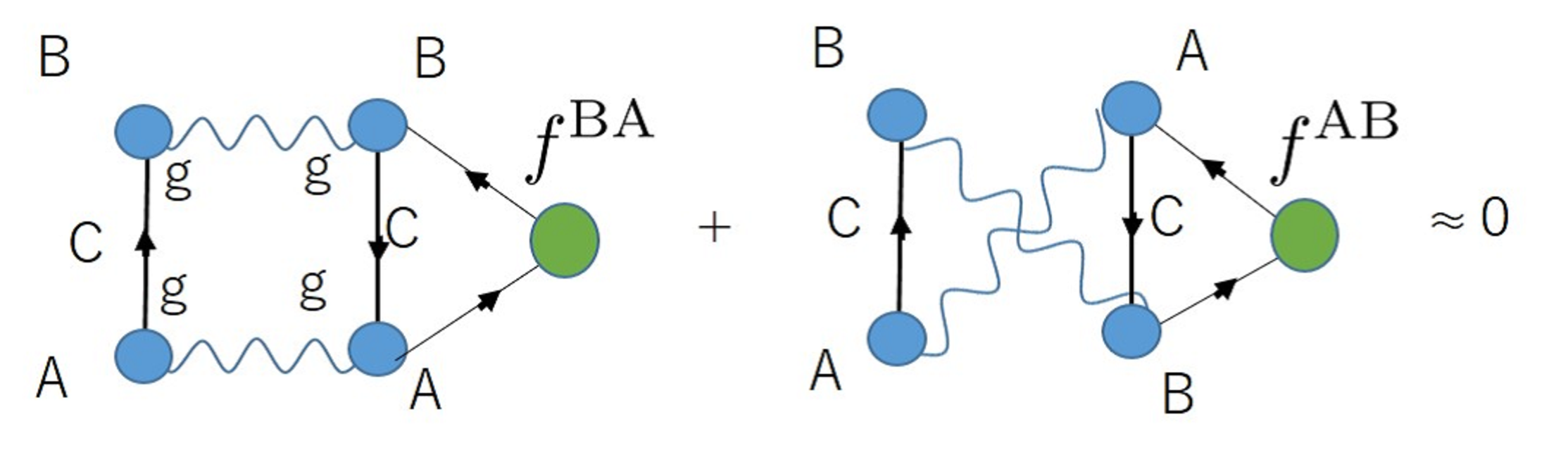}
\caption{
Two Aslamazov-Larkin (AL) terms for the charge loop current (cLC) order at $\q=\q_1$.
These terms are almost canceled 
for the cLC order with $f_{\q_1}^{\rm AB}(\k)=-f_{\q_1}^{\rm BA}(-\k-\q_1)$.
}
\label{fig:figs4}
\end{figure}

In addition, it is verified that 
each AL term in Supplementary Fig. \ref{fig:figs4} is small 
because the momentum summation is restricted by four $g$'s.
(Note that $|g_\q^{lm}(\k)|\le1$.)

\subsection{Supplementary Note 5: Parquet RG theory for kagome metals: Derivation of cLC and BO instabilities}
In the main text, 
we revealed the BO fluctuation-mediated cLC mechanism in kagome metals
based on the DW equation method.
The BO fluctuations in the MT term of the DW equation
causes the scattering between three van-Hove singularity (vHS) points.
It is found that the cLC and BO fluctuations 
develop cooperatively, as demonstrated in Figs. \ref{fig:fig3} {\bf c}-{\bf f} in the main text and Supplementary Fig. \ref{fig:figS11} {\bf b}-{\bf d}.

\begin{figure}[htb]
\includegraphics[width=.99\linewidth]{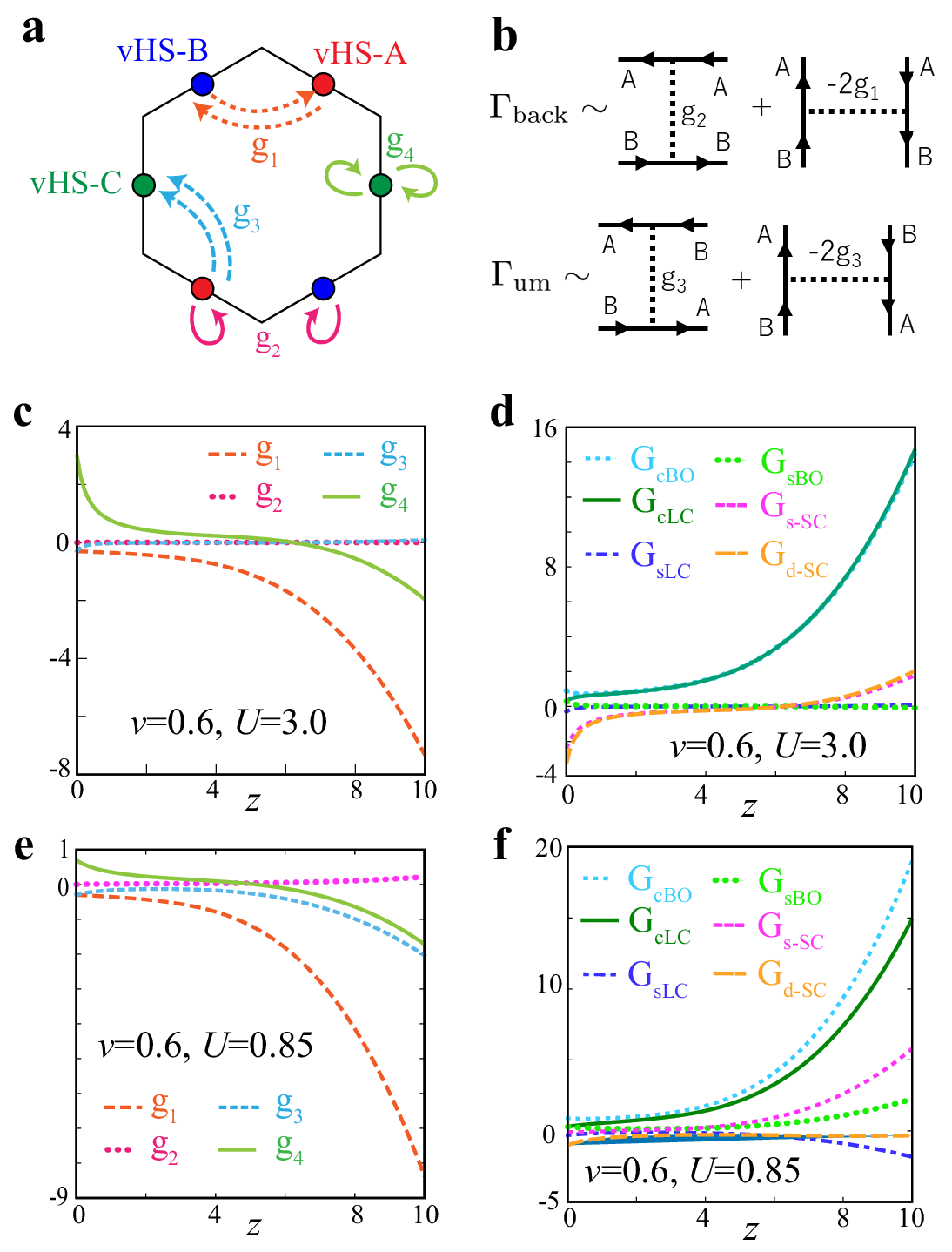}
\caption{
{\bf a}. Scattering processes between three vHS points; $g_i$ ($i=1\sim4$).
{\bf b}. Umklapp and backward terms in the kernel of the 
charge-channel density-wave (DW) equation.
$\Gamma_{\rm um}=-g_3$ and $\Gamma_{\rm back}=g_2-2g_1$.
$\Gamma_{\rm back}+(-)\Gamma_{\rm um}$ gives the cBO (cLC) instability.
{\bf c}. Obtained interactions $g_i(z)$ and
{\bf d}. instability for the channel $X$, $G_X$,
($X$=cBO, cLC, sBO, sLC, $s$-SC and $d$-SC)
as function of $y$, in the case of $v=0.6$ and $U=3$.
In this case, 
strong increment of $G_{\rm cBO}\approx G_{\rm cLC}$ are obtained.
{\bf e}. Obtained $g_i(z)$ and
{\bf f}. $G_X$ in the case of $v=0.6$ and $U=0.7$.
In this case, the relations $G_{\rm cBO}\gtrsim G_{\rm cLC}$ 
and $G_{\rm cLC}\sim 2.5G_{s{\rm SC}}$ are obtained.
}
\label{fig:RG-flow}
\end{figure}

The aim of this section is to verify the idea of the 
BO fluctuation-mediated cLC based on a different 
reliable theoretical framework.
Here, we study the kagome lattice model based on the 
parquet renormalization group (RG) formulation 
\cite{Balents2021S,Chubukov-grS}.
Supplementary Figure \ref{fig:RG-flow} {\bf a} represents the 
four scattering processes between three vHS points.
$g_1$ is the backward scattering, 
$g_3$ is the Umklapp scattering,
and $g_2$ and $g_4$ are the forward scatterings.
A great merit of the RG method
is that both particle-particle and particle-hole channels 
are treated on the same footing,
while the details of the shape of the FS are dropped.
Using $g_i$'s, $\Gamma_{\rm um}$ and $\Gamma_{\rm back}$ 
included in the kernel of the DW equation,
which are introduced in the main text,
are expressed in Supplementary Fig. \ref{fig:RG-flow} {\bf b}.
The BO (cLC) instability is given by $\Gamma_{\rm back}+(-)\Gamma_{\rm um}$.

Here, we solve the following parquet RG equation 
for the three vHS points model
to obtain the renormalized $g_i$ ($i=1\sim4$)
due to the electron correlation
\cite{Balents2021S,Chubukov-grS}:
\begin{eqnarray}
\frac{g_1}{dz}&=& 2d g_1(g_2-g_1),
\\
\frac{g_2}{dz}&=& 2d (g_2^2+g_3^2),
\\
\frac{g_3}{dz}&=& -g_3^2-2g_3g_4+2dg_3(2g_2-g_1),
\\
\frac{g_4}{dz}&=& -2g_3^2-g_4^2,
\end{eqnarray}
where 
$z=\Pi^{\rm 0,AAAA}({\bm0},E)\sim {\rm ln}^2(E_0/E)$
and $d=d\chi^{\rm 0,ABBA}(\q_1,E)/dz$.
Here, $\Pi^{\rm 0,AAAA}({\bm0},E)$ is the Cooper channel bubble 
with the low-energy (high-energy) cutoff $E$ $(E_0)$,
and we consider the effect of the vHS at $E_{\rm F}$.
\textcolor{black}{
Note that $E$ corresponds to $\sim T$, so $T\approx E_0\exp(-z^{1/2})$.
When $E_0=0.2$eV, $z=10$ corresponds to $T\approx 100$K.
The parameter $d$ is bounded $0<d<1/2$, and 
$d=1/2$ corresponds to the perfect nesting.
Hereafter, we set $d=1/4$ because the nesting of the real FS is not perfect.
In this study, strong cLC and BO instabilities are robustly obtained for $d=1/4 \sim 1/2$.
}

The BO interaction in Eq. (\ref{eqn:elph1}) in the main text
gives the following initial values at $z=0$ ({\it i.e.}, $E=E_0$):
$g_1^0=g_3^0=-v/2$, $g_2^0=0$.
They correspond to $y=1/2$ in the main text.
Similar sets of $\{g_i^0\}$ were discussed in Ref. \cite{Balents2021S}.
In addition, we include the on-site Coulomb interaction $g_4=U$.
\textcolor{black}{
Supplementary Figure \ref{fig:RG-flow} {\bf c} shows the obtained flows of $g_i$'s
as functions of $z \ (\ge0)$ for $v=0.6$ and $U=3$.
}
In this case, the renormalized $g_1$ takes a large negative value,
while other $g_i$'s approach zero.
In this case, the interaction for the $X$-channel susceptibility, $G_X$,
is shown in Supplementary Fig. \ref{fig:RG-flow} {\bf d}.
$G_X$ is given as 
\cite{Balents2021S,Chubukov-grS}
$G_{\rm cBO}=-2g_1+g_2-g_3$,
$G_{\rm cLC}=-2g_1+g_2+g_3$,
$G_{\rm sBO}=g_2+g_3$,
$G_{\rm sLC}=g_2-g_3$,
$G_{s-{\rm SC}}=-2g_3-g_4$, and
$G_{d-{\rm SC}}=g_3-g_4$.
Here, $X=$c(s)BO: charge (spin) bond order, 
$X=$c(s)LC: charge (spin) loop-current,
$X=s(d)$-SC: $s(d)$-wave SC.
Thus, both $G_{\rm cBO}$ and $G_{\rm cLC}$ strongly develop.
The relation $G_{\rm cBO}\approx G_{\rm cLC}$ is obtained
because $g_3$ is irrelevant.
Therefore, we find $T_{\rm cBC}\approx T_{\rm cLC}$.


\textcolor{black}{
Supplementary Figure \ref{fig:RG-flow} {\bf e} shows the obtained flow of $g_i$
for $v=0.6$ and $U=0.7$.
}
In this case, both $|g_3|$ and $|g_1|$ are enlarged.
($|g_1|>|g_3|$ is satisfied.)
The corresponding instability $G_X$
is shown in Supplementary Fig. \ref{fig:RG-flow} {\bf f}.
Thus, both $G_{\rm cBO}$ and $G_{\rm cLC}$ are strongly enlarged,
while the relation $G_{\rm cBO}\gtrsim G_{\rm cLC}\gtrsim G_{s-{\rm SC}}$
is obtained.
Therefore, we find $T_{\rm cBO}\gtrsim T_{\rm cLC}\gtrsim T_{s-{\rm SC}}$.

Note that the CDW instability 
for the onsite operator $n_{\rm A}+s n_{\rm B}$ ($s=\pm1$)
at $\q=\q_1$ is $G_{{\rm CDW},s}=s(g_1-2g_2)-g_4$.
Also, the SDW instability 
for the onsite operator $m_{\rm A}+s m_{\rm B}$ 
($m_{\rm A}=n_{{\rm A}\uparrow}-n_{{\rm A}\downarrow}$)
at $\q=\q_1$ is $G_{{\rm SDW},s}=sg_1+g_4$.
Both quantities are smaller than $G_{\rm cBO}$ and $G_{\rm cLC}$
after the renormalization, as verified in 
Supplementary Figs. \ref{fig:RG-flow} {\bf d} and {\bf f}.

To summarize, the strong cLC instability derived from the present RG study 
strongly indicates the validity of the 
BO fluctuation-mediated cLC mechanism in kagome metals.
Considering the initial condition
$G_{\rm cLC}^0=G_{\rm cBO}^0/3=v/2$,
the strong cLC instability originates from the beyond-RPA effect.
The comparing study between the diagrammatic method
and the RG method will lead to further interesting discoveries.
On the other hand, parquet RG method cannot derive quantitative results, 
such as the long-range components of the order parameters
obtained from the DW equations.

\subsection{Supplementary Note 6: Stability of the nematic BO+cLC state}

\subsubsection{Supplementary Note 6-1: Analytic discussion}
\label{sec:SME1}
In the main text, we explained that the coexistence of the 
$3Q$ BO and the $3Q$ cLC leads to one $C_6$ state 
and three nematic ($C_2$) states.  
Here, we discuss the stability of these four states
based on the Ginzburg-Landau (GL) theory.
It is found that the nematic states are expected to emerge
when $T_{\rm BO}>T_{\rm cLC}$ by considering the third-order GL terms,
which play essential roles in kagome metals.

Here, we assume that the BO form factor $g^{lm}$
is composed of the nearest-neighbor components,
as we did in the main text.
Then, the BO form factors
$g_{\q_1}^{\rm AB}(\k)$, $g_{\q_2}^{\rm BC}$ and $g_{\q_3}^{\rm CA}$ are given by 
Supplementary Eqs. (\ref{eqn:BO1})-(\ref{eqn:BO3}), respectively.
The relation $g_{\q}^{lm}(\k)=\{g_{\q}^{ml}(\k)\}^*$ is satisfied.
As for the cLC form factor $f^{lm}$, we use the solution of the DW equation
obtained in the main text.
The obtained $f^{lm}$ contains long-range components as shown in 
Fig. \ref{fig:fig2} {\bf c} in the main text.
In fact, its $\k$-space expression in Supplementary Figs. \ref{fig:figS8} {\bf a} and {\bf b}
is very different from Supplementary Eq. (\ref{eqn:BO1}).
In addition, the frequency dependence $f^{lm}$ shown in  
Supplementary Fig. \ref{fig:figS8} {\bf c} is very strong,
by reflecting drastic frequency dependence of the MT term
($\propto -\chi_g(q)$).
The BO [cLC] parameter at $\q=\q_1$ is $\phi_{1}g_{\q}^{lm}$ 
[$\eta_{1}f_{\q}^{lm}$] with $lm={\rm AB}$ or ${\rm BA}$.

Here, we introduce the vector representation
${\bmphi}\equiv (\phi_1,\phi_2,\phi_3)$,
and define 
${\bmphi}_1\equiv(\phi,\phi,\phi)/\sqrt{3}$ and
${\bmphi}_2\equiv(-\phi,\phi,\phi)/\sqrt{3}$.
When ${\bmphi}={\bmphi}_1$,
we obtain the Tri-Hexagonal (Star-of-David) pattern 
for $\phi>0$ ($\phi<0$) shown in Supplementary Fig. \ref{fig:figS1} {\bf c}.
(The free energy for ${\bmphi}$ is different from that 
for $-{\bmphi}$ due to the third-order term.)
When ${\bmphi}={\bmphi}_2$,
we obtain the Tri-Hexagonal pattern for $\phi<0$,
while it is displaced by ${\aa}_{\rm AB}$ from Supplementary Fig. \ref{fig:figS1} {\bf c}.

We also introduce the notation 
${\bmeta}\equiv (\eta_1,\eta_2,\eta_3)$,
and define 
${\bmeta}_1\equiv(\eta,\eta,\eta)/\sqrt{3}$ and
${\bmeta}_2\equiv(-\eta,\eta,\eta)/\sqrt{3}$.
The cLC pattern in Supplementary Fig. \ref{fig:figS1} {\bf d}
is given by ${\bmeta}={\bmeta}_1$ with $\eta>0$,
and the direction of the cLC is reversed when $\eta<0$.
(The free energy is unchanged by ${\bmeta}\rightarrow-{\bmeta}$.)
When ${\bmeta}={\bmeta}_2$,
the cLC pattern is given by the parallel shift
of Supplementary Fig. \ref{fig:figS1} {\bf d} by ${\aa}_{\rm AB}$.

Hereafter, we construct the GL free energy up to the fourth-order terms:
\begin{eqnarray}
F&=&F^{(2)}+F^{(3)}+F^{(4)} .
\end{eqnarray}
The second-order term is
\begin{eqnarray}
F^{(2)}&=& a_1|{\bmphi}|^2 + a_2|{\bmeta}|^2 ,
\label{eqn:F2} \\
a_{\rm 1}&=&-\chi_{g}^0(\q_1) +I_{\rm b}^{-1},
\label{eqn:a1}\\
a_{\rm 2}&=&-\chi_{f}^0(\q_1) +I_{\rm c}^{-1},
\label{eqn:a2}
\end{eqnarray}
where $I_{\rm b(c)} \ (>0)$ is the effective interaction 
for the BO (cLC order)
\cite{Balents2021S,Tazai-MatsubaraS}.
($a_{\rm 1,2}\le0$ corresponds to $\lambda_\q\ge1$ 
in the DW equation as proved in Ref. \cite{Tazai-MatsubaraS}.)
Because $\chi_{g}^0(\q_1)\approx \chi_{f}^0(\q_1)$,
the relation $T_{\rm BO}>T_{\rm cLC}$ would be realized
when $I_{\rm b} >I_{\rm c}$.
We note that, in the present theory, 
both $I_{\rm b}$ and $I_{\rm c}$ originate from the 
electron correlations,
and therefore they exhibit strong $T$-dependences.

The third-order term and the forth-order term 
for general ${\bmphi}$ and ${\bmeta}$ are given as
\begin{eqnarray}
F^{(3)}&=& b_1 \phi_1\phi_2\phi_3
+b_2( \phi_1\eta_2\eta_3+\eta_1\phi_2\eta_3+ \eta_1\eta_2\phi_3 ),
\label{eqn:F3} \\
F^{(4)}&=& 
d_{1,a}(\phi_1^4+\phi_2^4+\phi_3^4)+d_{1,b}(\phi_1^2\phi_2^2+\phi_2^2\phi_3^2+\phi_3^2\phi_1^2)
\nonumber \\
& &+d_{2,a}(\eta_1^4+\eta_2^4+\eta_3^4)+d_{2,b}(\eta_1^2\eta_2^2+\eta_2^2\eta_3^2+\eta_3^2\eta_1^2)
\nonumber \\
& &+2d_{3,a}(\phi_1^2\eta_1^2+\phi_2^2\eta_2^2+\phi_3^2\eta_3^2)
\nonumber \\
& &+d_{3,b}(\phi_1^2\eta_2^2+\phi_2^2\eta_3^2+\phi_3^2\eta_1^2+\phi_2^2\eta_1^2+\phi_3^2\eta_2^2+\phi_1^2\eta_3^2) .
\nonumber \\
\label{eqn:F4} 
\end{eqnarray}
%
Note that the term $\sim\phi_1\phi_2\eta_1\eta_2$ is absent.
In the absence of the cLC order,
the third-order free energy term is given by the first term of 
Supplementary Eq. (\ref{eqn:F3}).
The diagrammatic expression for the coefficient $b_1$ 
is given in Supplementary Fig. \ref{fig:figS6} {\bf a},
and its analytic expression is found in Refs. \cite{HirataS,Balents2021S}.
When the BO and the cLC coexist, we obtain the additional cross-terms 
given by the second term of Supplementary Eq. (\ref{eqn:F3}).
The diagrammatic expression for $b_2$ 
is given in Supplementary Fig. \ref{fig:figS6} {\bf b}.
The relation $b_1=-b_2$ holds when the relation
$f_{\q_2}^{\rm BC}(\k_{\rm C})f_{\q_3}^{\rm CA}(\k_{\rm A})=-g_{\q_2}^{\rm BC}(\k_{\rm C})g_{\q_3}^{\rm CA}(\k_{\rm A})$ holds.
Note that the minimum of the GL free energy does not diverge
due to the positive fourth-order term.

\begin{figure}[htb]
\includegraphics[width=.8\linewidth]{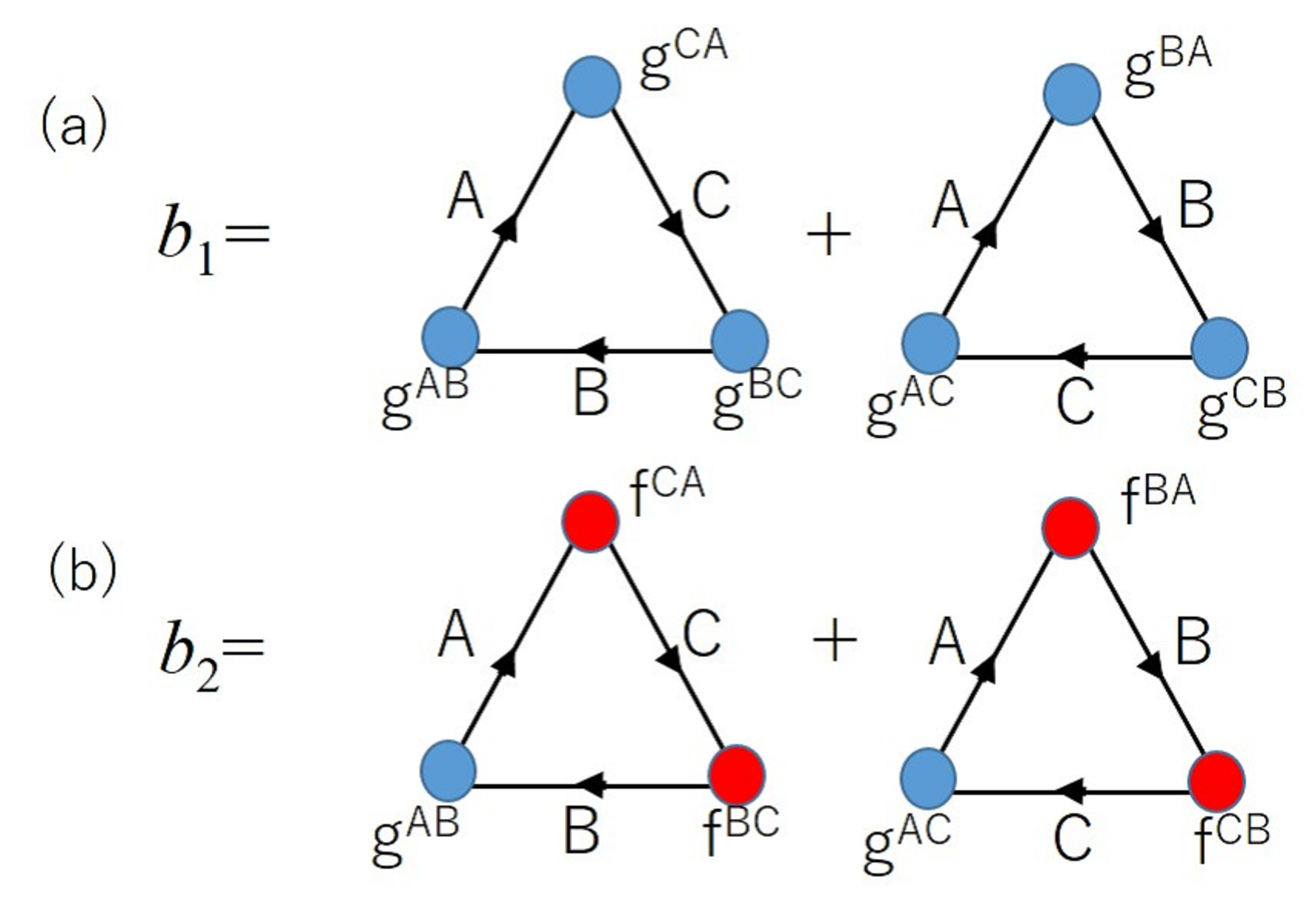}
\caption{
Diagrams for the third-order Ginzburg-Landau terms $b_1$ and $b_2$.
The relation $b_1=-b_2$ holds when $f_{\q}^{lm}(\k)=\pm ig_{\q}^{lm}(\k)$.
}
\label{fig:figS6}
\end{figure}

Now, we set ${\bmphi}={\bmphi}_1$ because the $3Q$ BO 
is stabilized by the third-order GL free energy 
($\propto \phi_1\phi_2\phi_3$), and it is actually observed experimentally.
When ${\bmeta}={\bmeta}_{1[2]}$,
the BO+cLC state is $C_6$ [$C_2$] symmetric
as shown in Fig. \ref{fig:fig4} {\bf a} [{\bf b}] in the main text.
Hereafter, we explain that the $C_2$-symmetric BO+cLC state
is realized due to ${\bmeta}\ne{\bmeta}_1$,
based on both analytic and numerical studies.

\begin{figure}[htb]
\includegraphics[width=.99\linewidth]{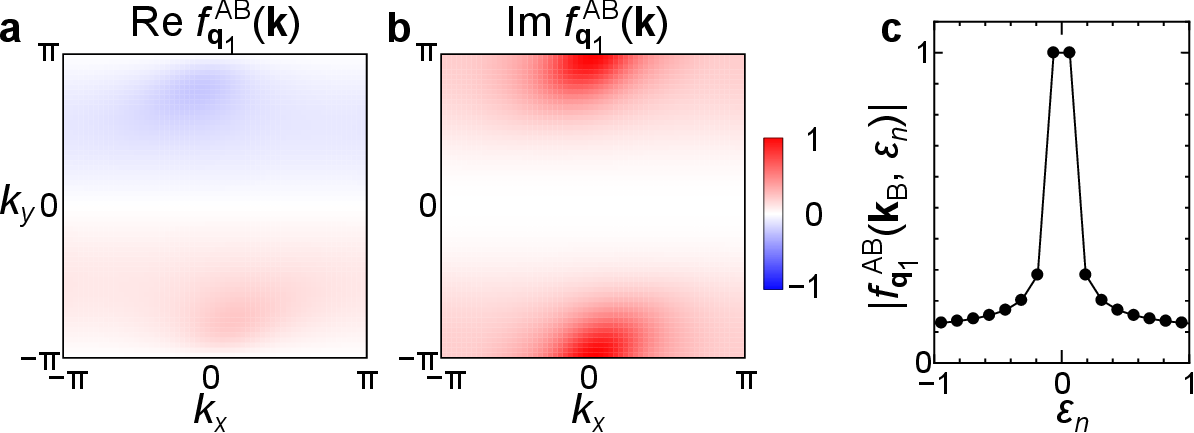}
\caption{
{\bf a}. Real part and {\bf b}. imaginary part of the 
cLC form factor $f_{\q_1}^{\rm AB}(\k)$
in the square kagome-lattice model.
$f_{\q_1}^{\rm AB}(\k)$ is very different from 
the nearest-neighbor BO form factor
$-ig_{\q_1}^{\rm AB}(\k)= \frac12\sin k_y-i\frac12(1-\cos k_y)$,
and its strong $\k$-dependence means the existence of 
large cLC order between distant sites.
This means that the inner-product 
$|\frac1N \sum_\k [if_{\q}^{\rm AB}(\k)]^*g_{\q}^{\rm AB}(\k)|$ is small,
indicating that the cLC and the BO are nearly independent.
{\bf c}. $\e_n$-dependence of $|f_{\q_1}^{\rm AB}(\k_B,e_n)|$,
where $y=1$, $v=1.0$ and $T=0.02$.
}
\label{fig:figS8}
\end{figure}

Here, we discuss the possible $3Q$ BO+cLC states 
in the case of $T_{\rm BO}>T_{\rm cLC}$.
When ${\bmphi}={\bmphi}_1$, the third-order term is
\begin{eqnarray}
F^{(3)'}&=&\frac{b_1}{3\sqrt{3}}\phi^3 
\nonumber \\
& &+\frac{b_2}{2\sqrt{3}}\phi[(\eta_1+\eta_2+\eta_3)^2-|{\bmeta}|^2] .
\label{eqn:F3d}
\end{eqnarray}
Now, we minimize Supplementary Eq. (\ref{eqn:F3d}) under the constraints
$|{\bmeta}|=$const. and $|{\bmphi}|=$const.,
where the 2nd order GL free energy is constant.
In the case of $|{\bmeta}|\gg|{\bmphi}|$ ($T_{\rm cLC}\gg T_{\rm BO}$),
$b_2\phi$ becomes negative and ${\bmeta}={\bmeta}_1$
to get the energy gain from the second term in Supplementary Eq. (\ref{eqn:F3d}).
(The first term is small and positive because of $b_1b_2<0$.)
Thus, the coexisting state has $C_6$ symmetry.

In the case of $|{\bmeta}|\ll|{\bmphi}|$ ($T_{\rm cLC}\ll T_{\rm BO}$),
$b_1\phi$ becomes negative 
to get the energy gain from the first term in Supplementary Eq. (\ref{eqn:F3d}).
Then, the second term is minimized when $\eta_1+\eta_2+\eta_3=0$
for a fixed $|{\bmeta}|$.
For example, ${\bmeta}'\propto(\eta,\eta,-2\eta)$
or ${\bmeta}''\propto(\eta,-\eta,0)$.
In both cases, the coexisting state has $C_2$ symmetry.
The nematic BO+cLC state with ${\bmphi}_1$ and ${\bmeta}'$ is
depicted in Fig. \ref{fig:fig4} {\bf c} in the main text.
Note that the total free energy is unchanged 
under $\eta_1+\eta_2+\eta_3=0$ within the 4th-order GL terms,
while this degeneracy is lifted by the 6th-order GL terms.
Thus, the $Z_3$ nematic BO+cLC state by ${\bmeta}'$ or ${\bmeta}''$
is expected to be realized for $T<T_{\rm cLC}<T_{\rm BO}$,
as shown in a schematic phase diagram in Fig. \ref{fig:fig3} {\bf f}
in the main text.



Finally, we discuss the fourth-order GL term in Supplementary Eq. (\ref{eqn:F4}).
The coefficients $d_{l,a}$ and $d_{l,b}$ ($l=1,2,3$) are positive,
and they are given by the closed Feynman diagrams
made of four form factors and four $G$'s
\cite{Balents2021S}.
The relations $d_{1,a}=d_{2,a}=d_{3,a}$ and 
$d_{1,b}=d_{2,b}=d_{3,b}$ hold in the case of
$f_{\q}^{lm}(\k)\approx\pm ig_{\q}^{lm}(\k)$.
For any
${\bmphi}={\bmphi}_\a$ and ${\bmeta}={\bmeta}_\b$ ($\a,\b=1,2$),
the fourth-order GL term is expressed as
\begin{eqnarray}
F^{(4)'}= d_1 \phi^4 + d_2 \eta^4 + 2d_3 \phi^2\eta^2 ,
\label{eqn:F4d}
\end{eqnarray}
where the coefficients are $d_1=(d_{1,a}+d_{1,b})/3$, 
$d_2=(d_{2,a}+d_{2,b})/3$, and $d_3=(d_{3,a}+d_{3,b})/3$.
However, the relation $f_{\q}^{lm}(\k)\approx\pm ig_{\q}^{lm}(\k)$
is not satisfied
because the mechanisms of the cLC and the BO are different in the present theory.
Therefore, all the coefficients $b_l$, $d_{m,a}$, $d_{m,b}$ depend on 
$l=1,2$ and $m=1,2,3$.
In the successive section, we discuss 
the stability of the nematic BO+cLC state based on the numerical study.

\subsubsection{Supplementary Note 6-2: Numerical study}
\label{sec:SME2}
Here, we calculate the 3rd- and 4th-order GL coefficients
based on the diagrammatic method.
We normalize the form factors as
$\max_{\k} |g_{\q_1}^{\rm AB} (\k)|=\max_{\k,l,m} |f_{\q_1}^{\rm AB} (\k)|=1$ 
which is realized at $\k=\k_{\rm A}$ in the present DW equation analysis.
Then, the BO parameter $\phi_m g_{\q_m}(\k)$ at $\q=\q_m$ ($m=1\sim3$) 
gives the hybridization gap $\Delta_{\rm BO}\sim|\phi|$
in the folded band at $\Gamma$ point.
[In the same way, $\Delta_{\rm cLC}\sim|\eta|$ 
for the cLC order $\eta_m f_{\q_m}(\k)$.] 
Thus, the present normalization rule for $g$ and $f$ 
is physically reasonable and convenient. 
Note that the normalization rule does not influence the
physical quantities, because the change of $H_{\rm int}$
due to $g\rightarrow r\cdot g$ is absorbed by $v\rightarrow v/r^2$.

The 3rd order GL parameters per unit cell are given as
\begin{eqnarray}
b_1&=& 3I_{123}^{ggg}+3I_{132}^{ggg},
\\
b_2&=& 3I_{123}^{gff}+3I_{132}^{gff},
\end{eqnarray}
where
\begin{eqnarray}
I_{lmn}^{xyz}&=&-\frac{T}{3N}\sum_{k,\s} {\rm Tr}{\hat x}_{q_l}(k+q_n+q_m){\hat G}(k+q_n+q_m)
\nonumber \\
&\times&
{\hat y}_{q_m}(k+q_n){\hat G}(k+q_n){\hat z}_{q_n}(k){\hat G}(k) ,
\label{eqn:I3}
\end{eqnarray}
where $x,y,z$ is $f$ or $g$, and $l,m,n$ is 1, 2, or 3.
The relation $q_l+q_m+q_n=0$ should be satisfied.
Here, ${\hat G}(k)$ is the $3\times3$ matrix expression of the 
Green function with the self-energy given in the main text.

The 4th order GL parameters per unit cell are given as
\begin{eqnarray}
d_{1,a}&=& I_{1111}^{gggg},
\\
d_{1,b}&=& 2I_{1212}^{gggg}+4I_{1122}^{gggg},
\\
d_{2,a}&=& I_{1111}^{ffff},
\\
d_{2,b}&=& 2I_{1212}^{ffff}+4I_{1122}^{ffff},
\\
d_{3,a}&=& \frac12(2I_{1111}^{fgfg}+4I_{1111}^{ffgg}),
\\
d_{3,b}&=& 2I_{1212}^{fgfg}+4I_{1122}^{ffgg},
\end{eqnarray}
where
\begin{eqnarray}
I_{hlmn}^{wxyz}&=&\frac{T}{4N}\sum_{k,\s} {\rm Tr}
{\hat w}_{q_h}(k+q_n+q_m+q_l){\hat G}(k+q_n+q_m+q_l)
\nonumber \\
&\times&{\hat x}_{q_l}(k+q_n+q_m){\hat G}(k+q_n+q_m)
\nonumber \\
&\times&{\hat y}_{q_m}(k+q_n)
{\hat G}(k+q_n){\hat z}_{q_n}(k){\hat G}(k) ,
\label{eqn:I4}
\end{eqnarray}
where $w,x,y,z$ is $f$ or $g$, and $h,l,m,n$ is 1, 2, or 3.
The relation $q_h+q_l+q_m+q_n=0$ should be satisfied.
The diagrammatic expression of $I_{lmn}^{xyz}$ and $I_{hlmn}^{wxyz}$
are depicted in Supplementary Fig. \ref{fig:GL0}.

Note that these GL coefficients depend on the 
Green functions (=bandstructure) and the form factors $g$ and $f$.
Thus, the GL coefficients depend on the BO/cLC mechanisms indirectly, 
just through the form factors.

\begin{figure}[htb]
\includegraphics[width=.8\linewidth]{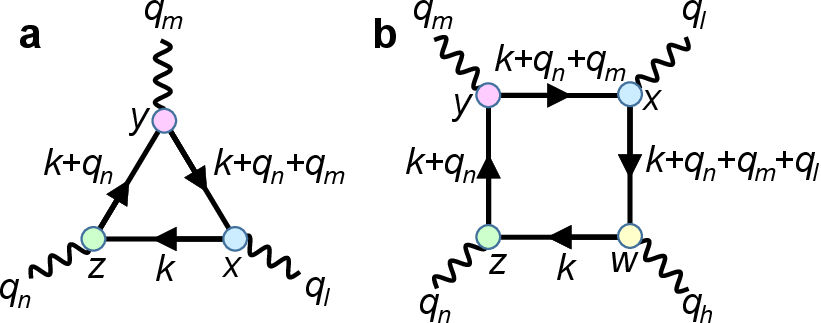}
\caption{
Diagrammatic expressions of Ginzburg-Landau parameters:
{\bf a}. Third-order term.
{\bf b}. Fourth-order term.
}
\label{fig:GL0}
\end{figure}

The numerical results of the 3rd and 4th GL parameters
are shown in Supplementary Figs. \ref{fig:GL1} {\bf a} and {\bf b}, respectively,
for $v=0.6$ and $y=1$.
We find that $b_1$ and $b_2$ have different signs,
consistently with the discussion in the previous section.
The obtained GL coefficients exhibit moderate and monotonic 
$T$-dependences for a wide $T$ range
because the self-energy suppresses unrealistic
singular behaviors of GL coefficients at low temperatures.
As we show in Supplementary Fig. \ref{fig:GL1} {\bf c},
the obtained ratio $r=2d_{1,a}/d_{1,b}$ is larger than 1.
This result means that experimentally observed $3Q$ BO state 
is realized irrespective of the size of $b_1$.
Also, the relation $r'=2d_{2,a}/d_{2,b}>1$ means that
the $3Q$ cLC state is realized.

In addition,
the obtained GL coefficients satisfy the relations
$R=d_{1,a}d_{2,a}/d_{3,a}^2>1$ and $R'=d_{1,b}d_{2,b}/d_{3,b}^2>1$
as shown in Supplementary Fig. \ref{fig:GL1} {\bf d}.
These relations indicate the smallness of 
the competition between cLC and BO 
described by $d_{3,a}$ and $d_{3,b}$ terms.
(Note that $R=R'=1$ when the BO and cLC form factors are
composed of only the nearest bonds \cite{Balents2021S}.)
Therefore, we can expect the coexistence of 
the BO and cLC orders by analyzing the GL free energy.
The relations $R,R'>1$ originates from the 
smallness of the inner-product 
$|\frac1N \sum_\k [if_{\q}^{ml}(\k)]^*g_{\q}^{lm}(\k)|$ 
(see Supplementary Fig. \ref{fig:figS8} {\bf a} and {\bf b})
and the strong frequency dependence of the cLC form factor
shown in Supplementary Fig. \ref{fig:figS8} {\bf c}.
This is not surprising because the driving forces of 
the BO and the cLC are different in the present theory.

\begin{figure}[htb]
\includegraphics[width=.9\linewidth]{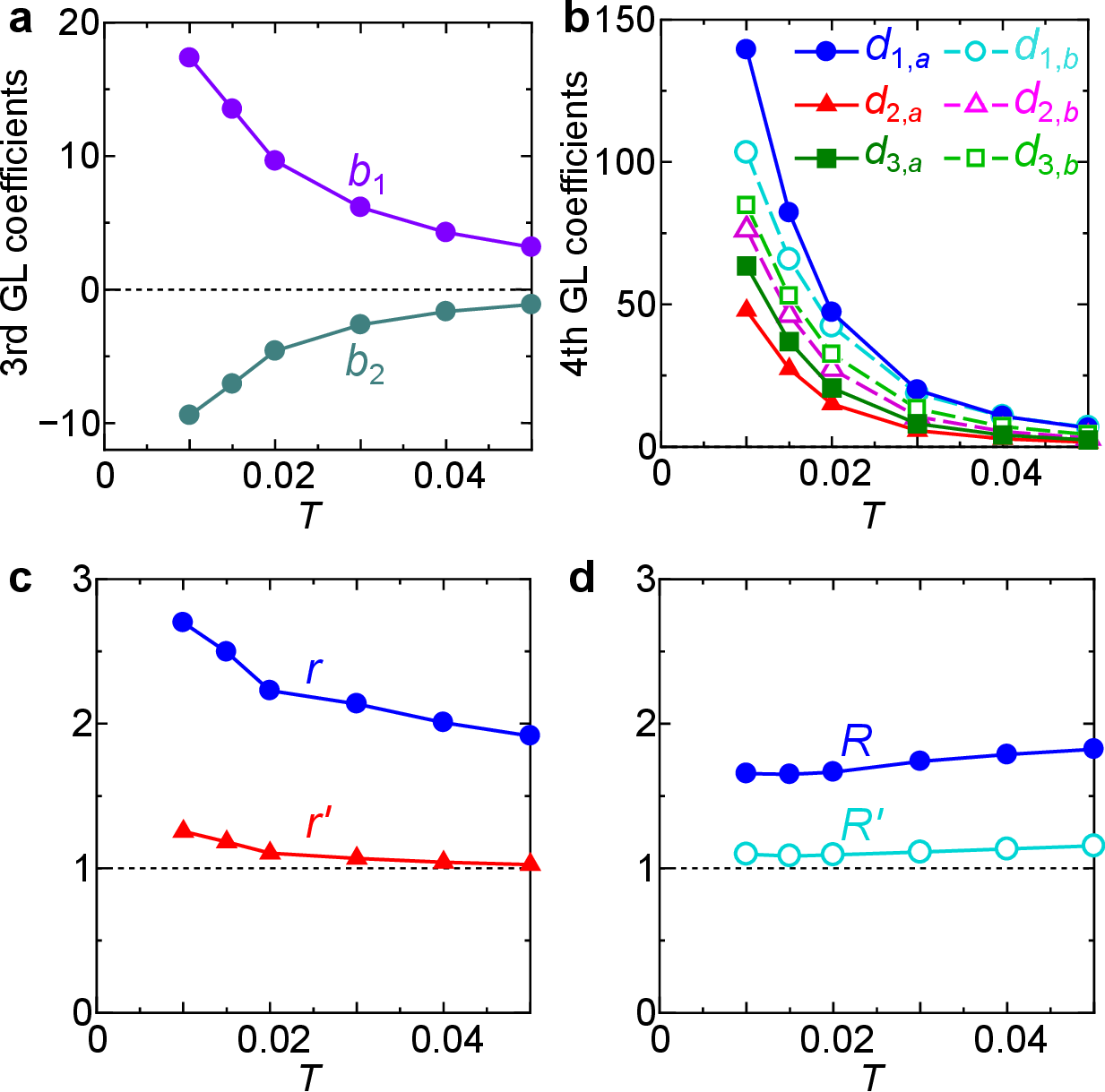}
\caption{
{\bf a}.,{\bf b}. Numerical results of the Ginzburg-Landau parameters 
for $v=0.6$ as functions of $T$.
The relation $b_1 b_2<0$ is verified.
{\bf c}. Ratio $r=2d_{1,a}/d_{1,b}$ (for BO)
and ratio $r'=2d_{2,a}/d_{2,b}$ (for cLC).
The $3Q$ state is stable when $r$ ($r'$) is larger than unity.
{\bf d}. Ratio $R=d_{1,a}d_{2,a}/d_{3,a}^2$ 
and ratio $R'=d_{1,b}d_{2,b}/d_{3,b}^2$.
The $C_2$ symmetric BO+cLC coexisting state is 
energetically stable when $R,R'>1$.
}
\label{fig:GL1}
\end{figure}

\begin{figure}[htb]
\includegraphics[width=.9\linewidth]{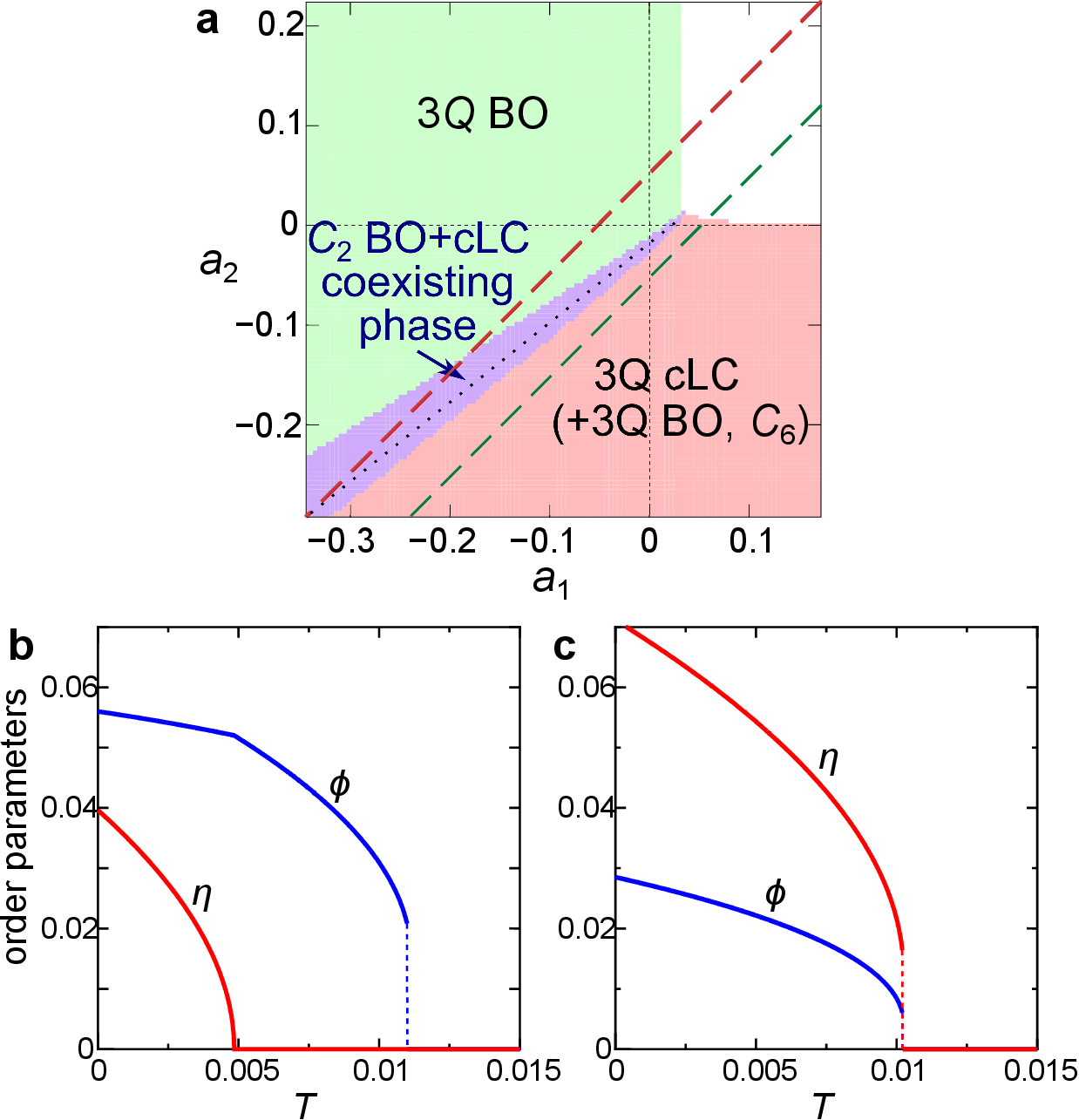}
\caption{
{\bf a}. Obtained phase diagram using the Ginzburg-Landau coefficients
in Supplementary Figs. \ref{fig:GL1} {\bf a} and {\bf b} at $T=0.01$.
The horizontal (vertical) axis is $a_1$ ($a_2$).
The purple region represents the $C_2$ symmetric BO+cLC phase.
The dotted line is the phase boundary without optimizing the 
$C_2$ BO and cLC order parameters.
{\bf b}.,{\bf c}. Obtained order parameters along {\bf b} the red broken line
and {\bf c} the green broken line in {\bf a}.
}
\label{fig:GL2}
\end{figure}

Supplementary Figure \ref{fig:GL2} {\bf a} exhibits the obtained phase diagram
derived from the GL free energy, by using the GL coefficients
in Supplementary Figs. \ref{fig:GL1} {\bf a} and {\bf b} at $T=0.01$.
The horizontal (vertical) axis is the second-order GL parameter for 
the BO $a_1$ (cLC $a_2$).
Each phase is determined by minimizing the GL free energy 
$F[{\bmphi},{\bmeta}]$ with respect to the
following $13^2$ patterns
$({\bmphi},{\bmeta})=({\bmphi}_m,{\bmeta}_n)$ with $m,n=1\sim13$
exactly numerically:
\begin{eqnarray*}
&&3Q \ {\rm BO}: \ \ {\bmphi}_{1}=\phi(1,1,1)/\sqrt{3}, 
\\
&&\ \ \ \ \ \ \ \ \ \ \ \ \ \ {\bmphi}_{2}=\phi(-1,1,1)/\sqrt{3}, \ \ {\bmphi}_{3}, \ {\bmphi}_{4}={\rm cycl.} ,
\\
&&2Q \ {\rm BO}: \ \ {\bmphi}_5=\phi(1,1,0)/\sqrt{2}, \ \ {\bmphi}_6, \ {\bmphi}_7={\rm cycl.} ,
\\
&&\ \ \ \ \ \ \ \ \ \ \ \ \ \ {\bmphi}_8=\phi(1,-1,0)/\sqrt{2}, \ \ {\bmphi}_9, \ {\bmphi}_{10}={\rm cycl.} ,
\\
&&1Q \ {\rm BO}: \ \ {\bmphi}_{11}=\phi(1,0,0), \ \ {\bmphi}_{12}, \ {\bmphi}_{13}={\rm cycl.} ,
\end{eqnarray*}
and ${\bmeta}_n={\bmphi}_n|_{\phi\rightarrow\eta}$ for the cLC.

The obtained $C_2$ symmetric BO+cLC phase with ${\bmphi}={\bmphi}_1$ 
and ${\bmeta}=(\eta,-\eta,0)/\sqrt{2}$
(or equivalently ${\bmeta}=(\eta,\eta,-2\eta)/\sqrt{6}$; see E-1)
is shown by the purple region.
This $C_2$ coexisting phase is always realized 
when $T_{\rm BO}\gtrsim T_{\rm cLC}$.
Also, the green (red) region corresponds to $3Q$ BO (cLC) phase.
In the red region,
the secondary $3Q$ BO appears through the $b_2$-terms
even for $a_1>0$, and $|{\bmphi}|$ 
becomes comparable to $|{\bmeta}|$ when $a_1<0$.
The coexisting state is $C_6$ symmetry.
\cite{Balents2021S}.

In deriving the phase boundary between 
the $C_2$ and $C_6$ coexisting phases in Supplementary Fig. \ref{fig:GL2} {\bf a}, 
we optimized three parameters $\{\phi,\phi',\eta\}$ numerically 
by considering additional 14th order parameter 
${\bmphi}_{14}\sim(\phi,\phi,\phi')$ 
and ${\bmeta}\sim(\eta,-\eta,0)$.
The dotted line in Supplementary Fig. \ref{fig:GL2} {\bf a}
is the phase boundary by setting $\phi=\phi'$.

Supplementary Figure \ref{fig:GL2} {\bf b}
shows the obtained order parameters along the red broken line in {\bf a}.
Here, we set 2nd order GL parameters as functions of $T$:
$a_1=A_1(T/T_{\rm BO}^0-1)$,
$a_2=A_2(T/T_{\rm cLC}^0-1)$, where
$A_1=A_2=0.17$, $T_{\rm BO}^0=0.01$ and $T_{\rm cLC}^0=0.85T_{\rm BO}^0$.
Here, $A_{1(2)}=0$ at $T=T_{\rm BO(cLC)}^0$.
The GL parameters at $T=0.01$ in Supplementary Fig. \ref{fig:GL1} are used.
We find that
$|{\bmphi}|$ exhibits the 1st order transition at 
$T_{\rm BO}\approx 1.1T_{\rm BO}^0$,
and $|{\bmeta}|$ appears as the 2nd-order transition at 
$T_{\rm cLC}\approx 0.6T_{\rm cLC}^0\approx 0.5T_{\rm BO}^0$.
This result corresponds to the region $v>v^*$
in Fig. \ref{fig:fig3} {\bf f} in the main text.
By performing careful numerical analyses, 
we find that the $C_2$ coexisting region 
(=purple region in Supplementary Figure \ref{fig:GL2} {\bf a}) appears
when the relations $R>1$ and $R'>1$ are satisfied.

Supplementary Figure \ref{fig:GL2} {\bf c}
shows the order parameters along the green broken line in {\bf a}.
The primary $3Q$ cLC order induces the secondary $3Q$ BO 
due to the 3rd order GL terms.
The symmetry of this coexisting phase is $C_6$ \cite{Balents2021S}.

\begin{thebibliography}{999}
\bibitem{kagome-exp1}
B. R. Ortiz, L. C. Gomes, J. R. Morey, M. Winiarski, M. Bordelon, J. S. Mangum, I. W. H. Oswald, J. A. Rodriguez-Rivera, J. R. Neilson, S. D. Wilson, E. Ertekin, T. M. McQueen, and E. S. Toberer,
{\it New kagome prototype materials: discovery of ${{KV}}_{3}{{Sb}}_{5},{{RbV}}_{3}{{Sb}}_{5}$, and ${{CsV}}_{3}{{Sb}}_{5}$},
{Phys. Rev. Materials} {\bf 3}, 094407 (2019).

\bibitem{kagome-exp2}
B. R. Ortiz, S. M. L. Teicher, Y. Hu, J. L. Zuo, P. M. Sarte, E. C. Schueller, A. M. M. Abeykoon, M. J. Krogstad, S. Rosenkranz, R. Osborn, R. Seshadri, L. Balents, J. He, and S. D. Wilson,
{\it ${Cs}{{V}}_{3}{{Sb}}_{5}$: A ${\mathbb{Z}}_{2}$ Topological Kagome Metal with a Superconducting Ground State},
{Phys. Rev. Lett.} {\bf 125}, 247002 (2020).

\bibitem{kagome-P-Tc1}
F. H. Yu, D. H. Ma, W. Z. Zhuo, S. Q. Liu, X. K. Wen, B. Lei, J. J. Ying, and X. H. Chen,
{\it Unusual competition of superconductivity and charge-density-wave state in a compressed topological kagome metal},
{Nat. Commun.} {\bf 12}, 3645 (2021).



\bibitem{STM1}
Y.-X. Jiang, J.-X. Yin, M. M. Denner, N. Shumiya, B. R. Ortiz, G. Xu, Z. Guguchia, J. He, M. S. Hossain, X. Liu, J. Ruff, L. Kautzsch, S. S. Zhang, G. Chang, I. Belopolski, Q. Zhang, T. A. Cochran, D. Multer, M. Litskevich, Z.-J. Cheng, X. P. Yang, Z. Wang, R. Thomale, T. Neupert, S. D. Wilson, and M. Z. Hasan,
{\it Unconventional chiral charge order in kagome superconductor KV$_{3}$Sb$_{5}$},
{Nat. Mater.} {\bf 20}, 1353 (2021).

\bibitem{STM2}
H. Li, H. Zhao, B. R. Ortiz, T. Park, M. Ye, L. Balents, Z. Wang, S. D. Wilson, and I. Zeljkovic,
{\it Rotation symmetry breaking in the normal state of a kagome superconductor KV$_{3}$Sb$_{5}$},
{Nat. Phys.} {\bf 18}, 265 (2022).

\bibitem{Thomale2013}
M. L. Kiesel, C. Platt, and R. Thomale,
{\it Unconventional Fermi Surface Instabilities in the Kagome Hubbard Model},
{Phys. Rev. Lett.} {\bf 110}, 126405 (2013).

\bibitem{SMFRG}
W.-S. Wang, Z.-Z. Li, Y.-Y. Xiang, and Q.-H. Wang,
{\it Competing electronic orders on kagome lattices at van Hove filling},
{Phys. Rev. B} {\bf 87}, 115135 (2013).

\bibitem{Thomale2021}
X. Wu, T. Schwemmer, T. M\"uller, A. Consiglio, G. Sangiovanni, D. Di Sante, Y. Iqbal, W. Hanke, A. P. Schnyder, M. M. Denner, M. H. Fischer, T. Neupert, and R. Thomale,
{\it Nature of Unconventional Pairing in the Kagome Superconductors $A{{V}}_{3}{{Sb}}_{5}$ ($A={K},{Rb},{Cs}$)},
{Phys. Rev. Lett.} {\bf 127}, 177001 (2021).

\bibitem{Neupert2021}
M. M. Denner, R. Thomale, and T. Neupert,
{\it Analysis of Charge Order in the Kagome Metal $A{{V}}_{3}{{Sb}}_{5}$ ($A={K},{Rb},{Cs}$)},
{Phys. Rev. Lett.} {\bf 127}, 217601 (2021).

\bibitem{Balents2021}
T. Park, M. Ye, and L. Balents,
{\it Electronic instabilities of kagome metals: Saddle points and Landau theory},
{Phys. Rev. B} {\bf 104}, 035142 (2021).

\bibitem{Nandkishore}
Y.-P. Lin and R. M. Nandkishore,
{\it Complex charge density waves at Van Hove singularity on hexagonal lattices: Haldane-model phase diagram and potential realization in the kagome metals AV$_{3}$Sb$_{5}$ (A = K, Rb, Cs)},
Phys. Rev. B {\bf 104}, 045122 (2021).

\bibitem{Tazai-kagome}
R. Tazai, Y. Yamakawa, S. Onari, and H. Kontani,
{\it Mechanism of exotic density-wave and beyond-Migdal unconventional superconductivity in kagome metal AV$_{3}$Sb$_{5}$ (A = K, Rb, Cs)},
{Sci. Adv.} {\bf 8}, eabl4108 (2022).

\bibitem{Roppongi}
M. Roppongi, K. Ishihara, Y. Tanaka, K. Ogawa, K. Okada, S. Liu, K. Mukasa, Y. Mizukami, Y. Uwatoko, R. Grasset, M. Konczykowski, B. R. Ortiz, S. D. Wilson, K. Hashimoto, and T. Shibauchi,
{\it Bulk evidence of anisotropic s-wave pairing with no sign change in the kagome superconductor CsV$_3$Sb$_5$},
Nat. Commun. {\bf 14}, 667 (2023).

\bibitem{SC2}
W. Zhang, X. Liu, L. Wang, C. Wai T., Z. Wang, S. T. Lam, W. Wang, J. Xie, X. Zhou, Y. Zhao, S. Wang, J. Tallon, K. T. Lai, and S. K. Goh,
{\it Nodeless superconductivity in kagome metal CsV$_3$Sb$_5$ with and without time reversal symmetry breaking},
Nano Lett., {\bf 23}, 872 (2023).

\bibitem{muSR5-Rb}
Z. Guguchia, C. Mielke III, D. Das, R. Gupta, J.-X. Yin, H. Liu, Q. Yin, M.H. Christensen, Z. Tu, C. Gong, N. Shumiya, Ts. Gamsakhurdashvili, M. Elender, Pengcheng Dai, A. Amato, Y. Shi, H.C. Lei, R.M. Fernandes, M.Z. Hasan, H. Luetkens, and R. Khasanov,
{\it Tunable nodal kagome superconductivity in charge ordered RbV$_3$Sb$_5$},
Nat. Commun. {\bf 14}, 153 (2023).

\bibitem{muSR3-Cs}
L. Yu, C. Wang, Y. Zhang, M. Sander, S. Ni, Z. Lu, S. Ma, Z. Wang, Z. Zhao, H. Chen, K. Jiang, Y. Zhang, H. Yang, F. Zhou, X. Dong, S. L. Johnson, M. J. Graf, J. Hu, H.-J. Gao, and Z. Zhao,
{\it Evidence of a hidden flux phase in the topological kagome metal CsV$_3$Sb$_5$},
arXiv:2107.10714 (available at https://arxiv.org/abs/2107.10714).

\bibitem{muSR2-K}
C. Mielke, D. Das, J.-X. Yin, H. Liu, R. Gupta, Y.-X. Jiang, M. Medarde, X. Wu, H. C. Lei, J. Chang, P. Dai, Q. Si, H. Miao, R. Thomale, T. Neupert, Y. Shi, R. Khasanov, M. Z. Hasan, H. Luetkens, and Z. Guguchia,
{\it Time-reversal symmetry-breaking charge order in a kagome superconductor},
{Nature} {\bf 602}, 245 (2022).

\bibitem{muSR4-Cs}
R. Khasanov, D. Das, R. Gupta, C. Mielke, M. Elender, Q. Yin, Z. Tu, C. Gong, H. Lei, E. T. Ritz, R. M. Fernandes, T. Birol, Z. Guguchia, and H. Luetkens,
{\it Time-reversal symmetry broken by charge order in CsV$_3$Sb$_5$}, Phys. Rev. Research {\bf 4}, 023244 (2022).

\bibitem{birefringence-kagome}
Y. Xu, Z. Ni, Y. Liu, B. R. Ortiz, S. D. Wilson, B. Yan, L. Balents, and L. Wu,
{\it Universal three-state nematicity and magneto-optical Kerr effect in the charge density waves in AV$_3$Sb$_5$ (A=Cs, Rb, K)},
Nat. Phys. {\bf 18}, 1470 (2022).

\bibitem{eMChA}
C. Guo, C. Putzke, S. Konyzheva, X. Huang, M. Gutierrez-Amigo, I. Errea, D. Chen, M. G. Vergniory, C. Felser, M. H. Fischer, T. Neupert, and P. J. W. Moll,
{\it Switchable chiral transport in charge-ordered Kagome metal CsV$_3$Sb$_5$},
Nature {\bf 611}, 461 (2022).

\bibitem{Asaba}
T. Asaba, A. Onishi, Y. Kageyama, T. Kiyosue, K. Ohtsuka, S. Suetsugu, Y. Kohsaka, T. Gaggl, Y. Kasahara, H. Murayama, K. Hashimoto, R. Tazai, H. Kontani, B. R. Ortiz, S. D. Wilson, Q. Li, H.-H. Wen, T. Shibauchi, and Y. Matsuda,
{\it Evidence for an odd-parity nematic phase above the charge density wave transition in kagome metal CsV$_3$Sb$_5$},
arXiv:2309.16985 (available at https://arxiv.org/abs/2309.16985).
(to be published in Nat. Phys.)

\bibitem{Kapitulnik}
D. R. Saykin, C. Farhang, E. D. Kountz, D. Chen, B. R. Ortiz, C. Shekhar, C. Felser, S. D. Wilson, R. Thomale, J. Xia, and A. Kapitulnik,
{\it High Resolution Polar Kerr Effect Studies of CsV$_3$Sb$_5$: Tests for Time Reversal Symmetry Breaking Below the Charge Order Transition},
Phys. Rev. Lett. {\bf 131}, 016901 (2023).

\bibitem{nematic-SC2}
Y. Xiang, Q. Li, Y. Li, W. Xie, H. Yang, Z. Wang, Y. Yao, and H.-H. Wen,
{\it Twofold symmetry of c-axis resistivity in topological kagome superconductor CsV3Sb5 with in-plane rotating magnetic field},
{Nat. Commun.} {\bf 12}, 6727 (2021).

\bibitem{Haldane}
F. D. M. Haldane,
{\it Model for a Quantum Hall Effect without Landau Levels: Condensed-Matter Realization of the "Parity Anomaly"},
{Phys. Rev. Lett.} {\bf 61}, 2015 (1988).

\bibitem{AHE1}
S.-Y. Yang, Y. Wang, B. R. Ortiz, D. Liu, J. Gayles, E. Derunova, R. Gonzalez-Hernandez, L. $\check{S}$mejkal, Y. Chen, S. S. P. Parkin, S. D. Wilson, E. S. Toberer, T. McQueen, and M. N. Ali,
{\it Giant, unconventional anomalous Hall effect in the metallic frustrated magnet candidate, KV$_{3}$Sb$_{5}$},
{Sci. Adv.} {\bf 6}, eabb6003 (2020).

\bibitem{AHE2}
F. H. Yu, T. Wu, Z. Y. Wang, B. Lei, W. Z. Zhuo, J. J. Ying, and X. H. Chen,
{\it Concurrence of anomalous Hall effect and charge density wave in a superconducting topological kagome metal},
{Phys. Rev. B} {\bf 104}, L041103 (2021).

\bibitem{elastoresistance-kagome}
L. Nie, K. Sun, W. Ma, D. Song, L. Zheng, Z. Liang, P. Wu, F. Yu, J. Li, M. Shan, D. Zhao, S. Li, B. Kang, Z. Wu, Y. Zhou, K. Liu, Z. Xiang, J. Ying, Z. Wang, T. Wu, and X. Chen,
{\it Charge-density-wave-driven electronic nematicity in a kagome superconductor},
{Nature} {\bf 604}, 59 (2022).

\bibitem{nematic-SC1}
S. Ni, S. Ma, Y. Zhang, J. Yuan, H. Yang, Z. Lu, N. Wang, J. Sun, Z. Zhao, D. Li, S. Liu, H. Zhang, H. Chen, K. Jin, J. Cheng, L. Yu, F. Zhou, X. Dong, J. Hu, H.-J. Gao, and Z. Zhao,
{\it Anisotropic Superconducting Properties of Kagome Metal {CsV}$_3$Sb$_5$},
{Chin. Phys. Lett.} {\bf 38}, 057403 (2021).

\bibitem{Zhou-cLC-RPA}
J.-Wei D., Z. Wang, S. Zhou,
{\it Loop-current charge density wave driven by long-range Coulomb repulsion on the kagome lattice},
Phys. Rev. B {\bf 107}, 045127 (2023).

\bibitem{Fernandes-coexistence}
M. H. Christensen, T. Birol, B. M. Andersen, R. M. Fernandes,
{\it Loop currents in AV3Sb5 kagome metals: multipolar and toroidal magnetic orders},
Phys. Rev. B {\bf 106}, 144504 (2022).

\bibitem{Kennes-coexistence}
F. Grandi, A. Consiglio, M. A. Sentef, R. Thomale, D. M. Kennes,
{\it Theory of nematic charge orders in kagome metals},
Phys. Rev. B {\bf 107}, 155131 (2023).

\bibitem{Varma}
C. M. Varma,
{\it Non-Fermi-liquid states and pairing instability of a general model of copper oxide metals},
{Phys. Rev. B} {\bf 55}, 14554 (1997).

\bibitem{Nersesyan}
A. A. Nersesyan, G. I. Japaridze, and I. G. Kimeridze,
{\it Low-temperature magnetic properties of a two-dimensional spin nematic state},
{J. Phys.: Condens. Matter} {\bf 3}, 3353 (1991).

\bibitem{Fradkin-rev2012}
E. Fradkin and S. A. Kivelson,
{\it Ineluctable complexity},
{Nat. Phys.} {\bf 8}, 864 (2012).

\bibitem{Davis-rev2013}
J. C. S. Davis and D.-H. Lee,
{\it Concepts relating magnetic interactions, intertwined electronic orders, and strongly correlated superconductivity},
{Proc. Natl. Acad. Sci. U.S.A.} {\bf 110}, 17623 (2013).


\bibitem{Onari-SCVC}
S. Onari and H. Kontani,
{\it Self-consistent Vertex Correction Analysis for Iron-based Superconductors: Mechanism of Coulomb Interaction-Driven Orbital Fluctuations},
{Phys. Rev. Lett.} {\bf 109}, 137001 (2012).

\bibitem{Tsuchiizu1}
M. Tsuchiizu, Y. Ohno, S. Onari, and H. Kontani,
{\it Orbital Nematic Instability in the Two-Orbital Hubbard Model: Renormalization-Group + Constrained RPA Analysis},
{Phys. Rev. Lett.} {\bf 111}, 057003 (2013).

\bibitem{Tsuchiizu4}
M. Tsuchiizu, K. Kawaguchi, Y. Yamakawa, and H. Kontani,
{\it Multistage electronic nematic transitions in cuprate superconductors: A functional-renormalization-group analysis},
{Phys. Rev. B} {\bf 97}, 165131 (2018).

\bibitem{Yamakawa-Cu}
Y. Yamakawa and H. Kontani,
{\it Spin-Fluctuation-Driven Nematic Charge-Density Wave in Cuprate Superconductors: Impact of Aslamazov-Larkin Vertex Corrections},
{Phys. Rev. Lett.} {\bf 114}, 257001 (2015).

\bibitem{Yamakawa-FeSe}
Y. Yamakawa, S. Onari, and H. Kontani,
{\it Nematicity and Magnetism in FeSe and Other Families of Fe-Based Superconductors},
{Phys. Rev. X} {\bf 6}, 021032 (2016).

\bibitem{Onari-FeSe}
S. Onari, Y. Yamakawa, and H. Kontani,
{\it Sign-Reversing Orbital Polarization in the Nematic Phase of FeSe due to the ${C}_{2}$ Symmetry Breaking in the Self-Energy},
{Phys. Rev. Lett.} {\bf 116}, 227001 (2016).

\bibitem{Chubukov-PRX2016}
A. V. Chubukov, M. Khodas, and R. M. Fernandes,
{\it Magnetism, Superconductivity, and Spontaneous Orbital Order in Iron-Based Superconductors: Which Comes First and Why?},
{Phys. Rev. X} {\bf 6}, 041045 (2016).

\bibitem{Fernandes-rev2018}
R. M. Fernandes, P. P. Orth, and J. Schmalian,
{\it Intertwined Vestigial Order in Quantum Materials: Nematicity and Beyond},
{Annu. Rev. Condens. Matter Phys.} {\bf 10}, 133 (2019).

\bibitem{Onari-TBG}
S. Onari and H. Kontani,
{\it SU(4) $\text{Valley}+\text{Spin}$ Fluctuation Interference Mechanism for Nematic Order in Magic-Angle Twisted Bilayer Graphene: The Impact of Vertex Corrections},
{Phys. Rev. Lett.} {\bf 128}, 066401 (2022).

\bibitem{Kontani-AdvPhys}
H. Kontani, R. Tazai, Y. Yamakawa, and S. Onari,
{\it Unconventional density waves and superconductivities in Fe-based superconductors and other strongly correlated electron systems},
Adv. Phys. {\bf 70}, 355 (2021).


\bibitem{PDW-theory}
S. Zhou and Z. Wang,
{\it Chern Fermi pocket, topological pair density wave, and charge-4e and charge-6e superconductivity in kagome superconductors},
Nat. Commun. {\bf 13}, 7288 (2022).

\bibitem{Raghu-PDW}
Y.-M. Wu, R. Thomale, S. Raghu,
{\it Sublattice Interference promotes Pair Density Wave order in Kagome Metals},
Phys. Rev. B {\bf 108}, L081117 (2023).

\bibitem{Wu-6e}
Z. Pan, C. Lu, F. Yang, C. Wu,
{\it Frustrated superconductivity and charge-6e ordering},
arXiv:2209.13745 (available at https://arxiv.org/abs/2209.13745).

\bibitem{X-ray160K}
Q. Chen, D. Chen, W. Schnelle, C. Felser, and B. D. Gaulin,
{\it Charge Density Wave Order and Fluctuations above $T_{CDW}$ and below
Superconducting $T_c$ in the Kagome Metal CsV$_3$Sb$_5$},
Phys. Rev. Lett. {\bf 129}, 056401 (2022).

\bibitem{He-BOfluc}
K. Yang, W. Xia, X. Mi, L. Zhang, Y. Gan, A. Wang, Y. Chai, X. Zhou, X. Yang, Y. Guo, M. He,
{\it Charge fluctuations above $T_{CDW}$ revealed by glasslike thermal},
Phys. Rev. B {\bf 107}, 184506 (2023).

\bibitem{Tazai-Matsubara}
R. Tazai, S. Matsubara, Y. Yamakawa, S. Onari, and H. Kontani,
{\it A Rigorous Formalism of Unconventional Symmetry Breaking in Fermi Liquid Theory and Its Application to Nematicity in FeSe},
Phys. Rev. B {\bf 107}, 035137 (2023).

\bibitem{Frantz}
H.-M. Guo and M. Franz,
{\it Topological insulator on the kagome lattice},
Phys. Rev. B {\bf 80}, 113102 (2009).

\bibitem{ARPES-VHS}
Y. Hu, X. Wu, B. R. Ortiz, S. Ju, X. Han, J. Ma, N. C. Plumb, M. Radovic, R. Thomale, S. D. Wilson, A. P. Schnyder, and M. Shi,
{\it Rich nature of Van Hove singularities in Kagome superconductor CsV3Sb5},
{Nature} {\bf 13}, 2220 (2022).

\bibitem{ARPES-band}
Y. Luo, S. Peng, S. M. L. Teicher, L. Huai, Y. Hu, B. R. Ortiz, Z. Wei, J. Shen, Z. Ou, B. Wang, Y. Miao, M. Guo, M. Shi, S. D. Wilson, and J.-F. He,
{\it Distinct band reconstructions in kagome superconductor CsV$_3$Sb$_5$},
Phys. Rev. B {\bf 105}, L241111 (2022).

\bibitem{ARPES-CDWgap}
K. Nakayama, Y. Li, T. Kato, M. Liu, Z. Wang, T. Takahashi, Y. Yao, and T. Sato,
{\it Multiple energy scales and anisotropic energy gap in the charge-density-wave phase of the kagome superconductor ${{CsV}}_{3}{{Sb}}_{5}$},
{Phys. Rev. B} {\bf 104}, L161112 (2021).

\bibitem{ARPES-Lifshitz}
Z. Liu, N. Zhao, Q. Yin, C. Gong, Z. Tu, M. Li, W. Song, Z. Liu, D. Shen, Y. Huang, K. Liu, H. Lei, and S. Wang,
{\it Charge-Density-Wave-Induced Bands Renormalization and Energy Gaps in a Kagome Superconductor ${{RbV}}_{3}{{Sb}}_{5}$},
{Phys. Rev. X} {\bf 11}, 041010 (2021).

\bibitem{ARPES-CDWgap2}
Z. Wang, S. Ma, Y. Zhang, H. Yang, Z. Zhao, Y. Ou, Y. Zhu, S. Ni, Z. Lu, H. Chen, K. Jiang, L. Yu, Y. Zhang, X. Dong, J. Hu, H.-J. Gao, and Z. Zhao,
{\it Distinctive momentum dependent charge-density-wave gap observed in CsV$_3$Sb$_5$ superconductor with topological Kagome lattice},
arXiv:2104.05556 (available at https://arxiv.org/abs/2104.05556).


\bibitem{Tazai-rev2021}
R. Tazai, Y. Yamakawa, M. Tsuchiizu, and H. Kontani,
{\it $d$- and $p$-wave Quantum Liquid Crystal Orders in Cuprate Superconductors, \ifmmode {\kappa}\else ${\kappa}$\fi{}-(BEDT-TTF)$_2$X, and Coupled Chain Hubbard Models: Functional-renormalization-group Analysis},
{J. Phys. Soc. Jpn.} {\bf 90}, 111012 (2021).

\bibitem{Kontani-RPA}
H. Kontani and S. Onari,
{\it Orbital-Fluctuation-Mediated Superconductivity in Iron Pnictides: Analysis of the Five-Orbital Hubbard-Holstein Model},
{Phys. Rev. Lett.} {\bf 104}, 157001 (2010).

\bibitem{phonon-kagome}
H. Tan, Y. Liu, Z. Wang, and B. Yan,
{\it Charge Density Waves and Electronic Properties of Superconducting Kagome Metals},
{Phys. Rev. Lett.} {\bf 127}, 046401 (2021).

\bibitem{Kohn}
H. Li, T. T. Zhang, T. Yilmaz, Y. Y. Pai, C. E. Marvinney, A. Said, Q. W. Yin, C. S. Gong, Z. J. Tu, E. Vescovo, C. S. Nelson, R. G. Moore, S. Murakami, H. C. Lei, H. N. Lee, B. J. Lawrie, and H. Miao,
{\it Observation of Unconventional Charge Density Wave without Acoustic Phonon Anomaly in Kagome Superconductors ${A{V}}_{3}{{Sb}}_{5}$ ($A={Rb}$, Cs)},
{Phys. Rev. X} {\bf 11}, 031050 (2021).

\bibitem{CDW-no-eph}
Z. X. Wang, Q. Wu, Q. W. Yin, C. S. Gong, Z. J. Tu, T. Lin, Q. M. Liu, L. Y. Shi, S. J. Zhang, D. Wu, H. C. Lei, T. Dong, and N. L. Wang,
{\it Unconventional charge density wave and photoinduced lattice symmetry change in the kagome metal ${{CsV}}_{3}{{Sb}}_{5}$ probed by time-resolved spectroscopy},
{Phys. Rev. B} {\bf 104}, 165110 (2021).

\bibitem{SCR}
T. Moriya and K. Ueda,
{\it Spin fluctuations and high temperature superconductivity},
Adv. Phys. {\bf 49}, 555 (2000).

\bibitem{TPSC}
Y. Vilk, and A.-M.S. Tremblay, 
{\it Non-perturbative many-body approach to the Hubbard model and single-particle pseudogap},
J. Phys I (France) {\bf 7}, 1309 (1997).

\bibitem{Kontani-ROP}
H. Kontani,
{\it Anomalous transport phenomena in Fermi liquids with strong magnetic fluctuations},
{Rep. Prog. Phys.} {\bf 71}, 026501 (2008).

\bibitem{Tazai-cLC}
R. Tazai, Y. Yamakawa, and H. Kontani,
{\it Emergence of charge loop current in the geometrically frustrated Hubbard model: A functional renormalization group study},
{Phys. Rev. B} {\bf 103}, L161112 (2021).

\bibitem{Kino-Kontani}
H. Kino and H. Kontani,
{\it Phase Diagram of Superconductivity on the Anisotropic Triangular Lattice Hubbard Model: An Effective Model of $\kappa$-(BEDT-TTF) Salts},
J. Phys. Soc. Jpn. {\bf 67}, 3691 (1998);
M. Kitatani, N. Tsuji, and H. Aoki,
{\it FLEX+DMFT approach to the $d$-wave superconducting phase diagram of the two-dimensional Hubbard model},
Phys. Rev. B {\bf 92}, 085104 (2015).



\bibitem{Madhavan}
Y. Xing, S. Bae, E. Ritz, F. Yang, T. Birol, A. N. C. Salinas, B. R. Ortiz, S. D. Wilson, Z. Wang, R. M. Fernandes, and V. Madhavan,
{\it Optical Manipulation of the Charge Density Wave state in RbV$_3$Sb$_5$},
arXiv:2308.04128 (available at https://arxiv.org/abs/2308.04128).

\bibitem{AHE-kagome-theory}
X. Feng, K. Jiang, Z. Wang, and J. Hu,
{\it Chiral flux phase in the Kagome superconductor AV$_3$Sb$_5$},
{Sci. Bull.} {\bf 66}, 1384 (2021).


\bibitem{Kontani-SHE-PRL}
H. Kontani, T. Tanaka, D. S. Hirashima, K. Yamada, and J. Inoue,
{\it Giant Orbital Hall Effect in Transition Metals: Origin of Large Spin and Anomalous Hall Effects},
{Phys. Rev. Lett.} {\bf 102}, 016601 (2009).

\bibitem{Kontani-Yamada}
H. Kontani, T. Tanaka, and K. Yamada,
{\it Intrinsic anomalous Hall effect in ferromagnetic metals studied by the multi-$d$-orbital tight-binding model},
{Phys. Rev. B} {\bf 75}, 184416 (2007).


\bibitem{AHE-RMP}
N. Nagaosa, J. Sinova, S. Onoda, A. H. MacDonald, and N. P. Ong,
{\it Anomalous Hall effect},
{Rev. Mod. Phys.} {\bf 82}, 1539 (2010).

\bibitem{Chubukov-gr}
R. Nandkishore, L. S. Levitov, A. V. Chubukov,
{\it Chiral superconductivity from repulsive interactions in doped graphene},
Nat. Phys. {\bf 8}, 158 (2012).

\bibitem{Nat-g-ology}
H. D. Scammell, J. Ingham, T. Li, and O. P. Sushkov,
{\it Chiral excitonic order from twofold van Hove singularities in kagome metals},
Nat. Commun. {\bf 14}, 605 (2023)

\bibitem{Moll-hz}
C. Guo, G. Wagner, C. Putzke, D. Chen, K. Wang, L. Zhang, M. G. Amigo, I. Errea, M. G. Vergniory, C. Felser, M. H. Fischer, T. Neupert, and P. J. W. Moll,
{\it Correlated order at the tipping point in the kagome metal CsV$_3$Sb$_5$},
arXiv:2304.00972 (available at https://arxiv.org/abs/2304.00972).

\bibitem{Mch2023}
R. Tazai, Y. Yamakawa and H. Kontani, 
{\it Drastic magnetic-field-induced chiral current order and emergent current-bond-field interplay in kagome metal AV$_3$Sb$_5$ (A=Cs,Rb,K)},
arXiv:2303.00623 (available at https://arxiv.org/abs/2303.00623).

\bibitem{166-CDW}
H. W. S. Arachchige, W. R. Meier, M. Marshall, T. Matsuoka, R. Xue, M. A. McGuire, R. P. Hermann, H. Cao, and D. Mandrus,
{\it Charge Density Wave in Kagome Lattice Intermetallic ScV$_6$Sn$_6$},
Phys. Rev. Lett. {\bf 129}, 216402 (2022).

\bibitem{Bernevig-166}
H. Hu, Y. Jiang, D. Calug, X. Feng, D. Subires, M. G. Vergniory, C. Felser, S. Blanco-Canosa, and B. A. Bernevig,
{\it Kagome Materials I: SG 191, ScV$_6$Sn$_6$. Flat Phonon Soft Modes and Unconventional CDW Formation: Microscopic and Effective Theory},
arXiv:2305.15469 (available at https://arxiv.org/abs/2305.15469).

\bibitem{Bernevig-166-2}
A. Korshunov, H. Hu, D. Subires, Y. Jiang, D. Calugaru, X. Feng, A. Rajapitamahuni, C. Yi, S. Roychowdhury, M. G. Vergniory, J. Strempfer, C. Shekhar, E. Vescovo, D. Chernyshov, A. H. Said, A. Bosak, C. Felser, B. Andrei Bernevig, S. Blanco-Canosa,
{\it Softening of a flat phonon mode in the kagome ScV$_6$Sn$_6$},
arXiv:2304.09173 (available at https://arxiv.org/abs/2304.09173).

\bibitem{166-mSR}
Z. Guguchia, D.J. Gawryluk, Soohyeon Shin, Z. Hao, C. Mielke III, D. Das, I. Plokhikh, L. Liborio, K. Shenton, Y. Hu, V. Sazgari, M. Medarde, H. Deng, Y. Cai, C. Chen, Y. Jiang, A. Amato, M. Shi, M.Z. Hasan, J.-X. Yin, R. Khasanov, E. Pomjakushina, and H. Luetkens,
{\it Hidden magnetism uncovered in charge ordered bilayer kagome material ScV$_6$Sn$_6$},
arXiv:2304.06436 (available at https://arxiv.org/abs/2304.06436).

\bibitem{Thomale-GL}
F. Grandi, A. Consiglio, M. A. Sentef, R. Thomale, and D. M. Kennes,
{\it Theory of nematic charge orders in kagome metals},
Phys. Rev. B {\bf 107}, 155131 (2023)

\bibitem{arXiv:2211.12264}
H. Yang, Y. Ye, Z. Zhao, J. Liu, X.-W. Yi, Y. Zhang, J. Shi, J.-Y. You, Z. Huang, B. Wang, J. Wang, H. Guo, X. Lin, C. Shen, W. Zhou, H. Chen, X. Dong, G. Su, Z. Wang, H.-J. Gao,
{\it Superconductivity and orbital-selective nematic order in a new titanium-based kagome metal CsTi$_3$Bi$_5$},
arXiv:2211.12264 (available at https://arxiv.org/abs/2211.12264).

\bibitem{arXiv:2211.16477}
H. Li, S. Cheng, B. R. Ortiz, H. Tan, D. Werhahn, K. Zeng, D. Jorhendt, B. Yan, Z. Wang, S. D. Wilson, and I. Zeljkovic,
{\it Electronic nematicity in the absence of charge density waves in a new titanium-based kagome metal},
Nat. Phys. (2023). https://doi.org/10.1038/s41567-023-02176-3

\bibitem{Ti-kagome}
J. Huang, Y. Yamakawa, R. Tazai, and H. Kontani,
{\it Odd-parity intra-unit-cell bond-order and induced nematicity in kagome metal CsTi$_3$Bi$_5$ driven by quantum interference mechanism},
arXiv:2305.18093 (available at https://arxiv.org/abs/2305.18093).







\end{thebibliography}

\begin{thebibliography}{99}

\bibitem{Tazai-kagomeS}
R. Tazai, Y. Yamakawa, S. Onari, and H. Kontani,
{\it Mechanism of exotic density-wave and beyond-Migdal unconventional superconductivity in kagome metal AV$_{3}$Sb$_{5}$ (A = K, Rb, Cs)},
{Sci. Adv.} {\bf 8}, eabl4108 (2022).

\bibitem{Onari-SCVCS}
S. Onari and H. Kontani,
{\it Self-consistent Vertex Correction Analysis for Iron-based Superconductors: Mechanism of Coulomb Interaction-Driven Orbital Fluctuations},
{Phys. Rev. Lett.} {\bf 109}, 137001 (2012).

\bibitem{Yamakawa-FeSeS}
Y. Yamakawa, S. Onari, and H. Kontani,
{\it Nematicity and Magnetism in FeSe and Other Families of Fe-Based Superconductors},
{Phys. Rev. X} {\bf 6}, 021032 (2016).

\bibitem{Onari-FeSeS}
S. Onari, Y. Yamakawa, and H. Kontani,
{\it Sign-Reversing Orbital Polarization in the Nematic Phase of FeSe due to the ${C}_{2}$ Symmetry Breaking in the Self-Energy},
{Phys. Rev. Lett.} {\bf 116}, 227001 (2016).

\bibitem{Tsuchiizu4S}
M. Tsuchiizu, K. Kawaguchi, Y. Yamakawa, and H. Kontani,
{\it Multistage electronic nematic transitions in cuprate superconductors: A functional-renormalization-group analysis},
{Phys. Rev. B} {\bf 97}, 165131 (2018).

\bibitem{Yamakawa-CuS}
Y. Yamakawa and H. Kontani,
{\it Spin-Fluctuation-Driven Nematic Charge-Density Wave in Cuprate Superconductors: Impact of Aslamazov-Larkin Vertex Corrections},
{Phys. Rev. Lett.} {\bf 114}, 257001 (2015).

\bibitem{Tazai-cLCS}
R. Tazai, Y. Yamakawa, and H. Kontani,
{\it Emergence of charge loop current in the geometrically frustrated Hubbard model: A functional renormalization group study},
{Phys. Rev. B} {\bf 103}, L161112 (2021).

\bibitem{Kontani-ROPS}
H. Kontani,
{\it Anomalous transport phenomena in Fermi liquids with strong magnetic fluctuations},
{Rep. Prog. Phys.} {\bf 71}, 026501 (2008).

\bibitem{Kontani-AdvPhysS}
H. Kontani, R. Tazai, Y. Yamakawa, and S. Onari,
{\it Unconventional density waves and superconductivities in Fe-based superconductors and other strongly correlated electron systems},
Adv. Phys. {\bf 70}, 355 (2021).

\bibitem{SCRS}
T. Moriya and K. Ueda,
{\it Spin fluctuations and high temperature superconductivity},
Adv. Phys. {\bf 49}, 555 (2000).

\bibitem{Tazai-MatsubaraS}
R. Tazai, S. Matsubara, Y. Yamakawa, S. Onari, and H. Kontani,
{\it A Rigorous Formalism of Unconventional Symmetry Breaking in Fermi Liquid Theory and Its Application to Nematicity in FeSe},
Phys. Rev. B {\bf 107}, 035137 (2023).

\bibitem{HirataS}
T. Hirata, Y. Yamakawa, S. Onari, and H. Kontani,
{\it Unconventional orbital charge density wave mechanism in the transition metal dichalcogenide $1T\ensuremath{-}{{TaS}}_{2}$},
{Phys. Rev. Research} {\bf 3}, L032053 (2021).

\bibitem{Balents2021S}
T. Park, M. Ye, and L. Balents,
{\it Electronic instabilities of kagome metals: Saddle points and Landau theory},
{Phys. Rev. B} {\bf 104}, 035142 (2021).

\bibitem{Chubukov-grS}
R. Nandkishore, L. S. Levitov, and A. V. Chubukov,
{\it Chiral superconductivity from repulsive interactions in doped graphene},
Nat. Phys. {\bf 8}, 158 (2012).

\end{thebibliography}
\end{document}